\documentclass{emulateapj}
\usepackage{apjfonts}
\journalinfo{ApJ Supplement, in press (v.\  171, August 2007)}

\shorttitle{CORONAL HEATING AND SOLAR WIND ACCELERATION}
\shortauthors{CRANMER, VAN BALLEGOOIJEN, \& EDGAR}

\begin{document}

\title{Self-consistent Coronal Heating and Solar Wind Acceleration
from Anisotropic Magnetohydrodynamic Turbulence}

\author{Steven R. Cranmer,
Adriaan A. van Ballegooijen,
and
Richard J. Edgar}
\affil{Harvard-Smithsonian Center for Astrophysics,
60 Garden Street, Cambridge, MA 02138 \\
Submitted 2006 December 15; \, accepted 2007 March 11}
\email{scranmer@cfa.harvard.edu,
avanballegooijen@cfa.harvard.edu,
redgar@cfa.harvard.edu}

\begin{abstract}
We present a series of models for the plasma properties along
open magnetic flux tubes rooted in solar coronal holes,
streamers, and active regions.
These models represent the first self-consistent solutions that
combine: (1) chromospheric heating driven by an empirically
guided acoustic wave spectrum, (2) coronal heating from
Alfv\'{e}n waves that have been partially reflected, then damped
by anisotropic turbulent cascade, and (3) solar wind acceleration
from gradients of gas pressure, acoustic wave pressure, and
Alfv\'{e}n wave pressure.
The only input parameters are the photospheric lower boundary
conditions for the waves and the radial dependence of the
background magnetic field along the flux tube.
We have not included multifluid or collisionless effects (e.g.,
preferential ion heating) which are not yet fully understood.
For a single choice for the photospheric wave properties, our
models produce a realistic range of slow and fast solar wind
conditions by varying only the coronal magnetic field.
Specifically, a two-dimensional model of coronal holes and
streamers at solar minimum reproduces the latitudinal
bifurcation of slow and fast streams seen by {\em Ulysses.}
The radial gradient of the Alfv\'{e}n speed affects where the
waves are reflected and damped, and thus whether energy
is deposited below or above the Parker critical point.
As predicted by earlier studies, a larger coronal ``expansion
factor'' gives rise to a slower and denser wind, higher
temperature at the coronal base, less intense Alfv\'{e}n waves
at 1 AU, and correlative trends for commonly measured ratios
of ion charge states and FIP-sensitive abundances that are in
general agreement with observations.
These models offer supporting evidence for the idea that
coronal heating and solar wind acceleration (in open magnetic
flux tubes) can occur as a result of wave dissipation and
turbulent cascade.
\end{abstract}


\section{Introduction}

One of the most persistent problems in solar physics has been
the unambiguous identification of the mechanisms that heat the
Sun's corona and accelerate the solar wind.
Many processes have been proposed for converting some fraction of
the mechanical energy in subphotospheric convective motions to
heat, but it has proved very difficult to make distinguishing
comparisons between the predictions of these competing ideas and
specific observations.
We are entering an era, though, where both the models and the
measurements are improving to the point of soon being able to
eliminate many of the candidate theories.
For example, it seems increasingly clear that closed loops in the
low corona are heated by small-scale, intermittent magnetic
reconnection that is driven by the continual stressing of their
magnetic footpoints (see recent reviews by Longcope 2004;
Gudiksen 2005; Aschwanden 2006; Klimchuk 2006).

In this paper we model the coronal heating along open field lines
that reach into interplanetary space.
We construct a self-consistent model of the photosphere,
chromosphere, corona, and solar wind that is driven mainly by
magnetohydrodynamic (MHD) turbulence and is free of arbitrary
``heating functions'' that have been used in the past.
This work continues earlier studies
(Cranmer \& van Ballegooijen 2005; Cranmer 2005a) which used
prescribed empirical descriptions of the density and flow velocity
along an open flux tube in order to compute the rates of turbulent
heating and acceleration.
The main goal of this paper is to provide a possible
explanation for the origin and properties of fast and slow
solar wind streams.
Another goal, though, is to illustrate how the particular
description of MHD turbulence can be applied to more advanced
modeling efforts---specifically those that attempt to reproduce
the full three-dimensional and time-dependent nature of the
corona and heliosphere (e.g., Riley et al.\  2001;
Roussev et al.\  2003; T\'{o}th et al.\  2005;
Usmanov \& Goldstein 2006).

Coronal heating and solar wind acceleration have been known to be
closely linked since the initial contributions of Parker (1958).
However, nearly all subsequent theoretical attempts to model both
processes together have made limiting assumptions about either the
heating (i.e., ad~hoc energy input rates) or the acceleration
(i.e., a prescribed mass flux).
It has been realized over the past few decades that it is key
to resolve the full chromosphere-corona transition region in
order to produce a model with an internally consistent radial
dependence of pressure, density, and flow speed (see
Hammer 1982; Hansteen \& Leer 1995; Hansteen et al.\  1997;
Lie-Svendsen \& Esser 2005).
To our knowledge, the only solar wind models---other than the ones
presented in this paper---that contain {\em both} a first-principles
approach to coronal heating and a self-consistent chromosphere,
transition region, and mass flux are those of
Suzuki \& Inutsuka (2005, 2006).\footnote{%
Other models can be included if some of the above conditions are
relaxed.
For example, Hu et al.\  (2000) and Li (2002, 2003) considered
wave-driven coronal heating with a lower boundary within the
transition region (i.e., temperatures ranging between
$6 \times 10^{4}$ and $8 \times 10^{5}$ K).
See {\S}~2 below for a comparison of the relevant physical
assumptions and numerical approaches.}

The study of MHD turbulence as a potential source of heating
for the solar wind goes back to Coleman (1968) and
Jokipii \& Davis (1969).
Hollweg (1986) extended these ideas down into the corona and
laid the foundations for the cascade-driven heating rates used
by Isenberg (1990), Li et al.\  (1999), Matthaeus et al.\  (1999),
Dmitruk et al.\  (2001, 2002), and this paper.
These ideas are highly complementary to theoretical models of
wave dissipation via ion cyclotron resonance (e.g.,
Hollweg \& Isenberg 2002; Cranmer 2000, 2001, 2002) that have
been invoked to explain the observed preferential heating of heavy
ions in the extended corona (Kohl et al.\  1997, 1998, 2006).
In the cascade paradigm, the energy in large-scale Alfv\'{e}nic
fluctuations must eventually be dissipated in small-scale
(short wavelength or high frequency) collisionless kinetic modes.
The results from this paper can thus be used as initial conditions
for detailed models of the anisotropic turbulent cascade and
minor ion heating.

The remainder of this paper is organized as follows.
In {\S}~2 we present an overview of the general principles
involved in our modeling effort (i.e., why certain physical
processes are included or excluded).
{\S}~3 gives the conservation equations that are solved and the
adopted prescriptions for basic ingredients such as radiative
heating/cooling and conduction.
The detailed treatment of waves is described in the following
three sections: acoustic waves and shock heating in {\S}~4,
Alfv\'{e}n waves and turbulent cascade in {\S}~5, and
the ponderomotive wave pressure acceleration due to both types
of fluctuations in {\S}~6.
The numerical methods used to solve the equations and the
newly developed computer code, called ZEPHYR, are described
in {\S}~7.
We then present a comprehensive set of solutions for the solar
wind emerging from coronal holes, helmet streamers, and active
regions ({\S}~8).
A summary of the major results of this paper, together with
a discussion of the implications for understanding the winds
of other stars, is given in {\S}~9.

\section{Model Philosophy}

We consider the {\em one-dimensional} variation of plasma
parameters along a radially pointed magnetic flux tube rooted
in the solar photosphere and open to interplanetary space.
Some effects of the multi-dimensional magnetic field in the
photosphere and chromosphere are included as explicit superradial
expansion of the flux tube (see also
Cranmer \& van Ballegooijen 2005), but otherwise we ignore
the effects of neighboring closed flux tubes.
We assume that most of the plasma that eventually becomes the
time-steady solar wind originates in small (100 km sized)
intergranular flux concentrations that are concentrated most
densely in the supergranular network.

The models we construct are {\em time-independent} solutions to
the hydrodynamic conservation equations.
This may be a restrictive simplification since all observed layers
of the solar atmosphere are intrinsically dynamic over a wide
range of time scales.
It is evident, for example, that the transition region is
``strongly nonuniform and magnetically structured''
(Marsch et al.\  2006) to the extent that one-dimensional
layered models may miss some important physical effects at
those heights.
However, the corona above the transition region seems to exist in
its hot ($T \sim 10^{6}$ K) state all the time.
Also, in~situ measurements of the solar wind plasma often can be
interpreted clearly as a superposition of a quasi-steady supersonic
flow (which depends mainly on the coronal source region of the
streams being measured or on stream-stream interactions) and a
rapidly varying turbulent or wavelike component.
We thus separate these two scales and model the fluctuations using
energy conservation equations derived from linear wave theory.

It is especially beneficial to treat waves and turbulent motions
statistically---rather than follow the oscillations explicitly
in time---when the dynamically important ranges of wavelength
and period span many orders of magnitude.
Alfv\'{e}n waves in the solar wind are measured to have periods
from seconds to days (e.g., Goldstein et al.\  1995;
Tu \& Marsch 1995) and there is evidence that the waves that
dominate the dissipation in the collisionless extended corona
have periods even below $10^{-3}$ s
(McKenzie et al.\  1995; Cranmer et al.\  1999).
If this full range of scales had to be resolved in order to track
the motions in a time-dependent numerical model, it would be
extremely difficult to simulate the macroscopic plasma properties
simultaneously from the photosphere to the heliosphere.

There are, however, some clear disadvantages in modeling
fluctuations as statistical ``wave energy fluxes'' rather
than as explicit variations.
Any nonlinear processes, such as shock steepening, turbulent
cascade, mode conversion, or ponderomotive forces, must
be inserted beforehand as source or sink terms in the
time-steady conservation equations.
Time-dependent MHD models have the benefit of producing such
effects naturally (e.g., Ofman \& Davila 1998; Ofman 2005;
Bogdan et al.\  2002, 2003; Suzuki \& Inutsuka 2005, 2006).
However, by resolving only certain temporal and spatial scales,
these models are limited in ways that a statistical treatment
is not.
For example, the relative importance of shock dissipation in the
time-dependent models of Suzuki \& Inutsuka (2005, 2006) may be an
artifact of either the inability to resolve small enough scales
(which could drive processes such as perpendicular cascade,
Landau damping, and collisionless particle energization at
shocks) or the waveguide-like trapping of fluctuations along
the model flux tube.
It is important to note, though, that neither extreme---i.e.,
neither coarse time-dependent simulations nor statistical
(and more approximate) time-independent models---is ideal.
There may be value in seeking some kind of hybrid methodology
between these two approaches.\footnote{%
Recent advances in modeling MHD turbulence in coronal loops
with a so-called ``shell model'' in wavenumber space may be
pointing the way (e.g., Giuliani \& Carbone 1998;
Nigro et al.\  2004).}

As indicated above, the only source of heating for the chromosphere
and corona that we consider is the dissipation of {\em waves
and turbulent motions.}
For open flux tubes, a substantial fraction of the energy appears
to be deposited at large heights in the wind's acceleration
region---i.e., at spatial scales much larger than the sizes of
low-lying closed loops in the quiet Sun and active regions.
This demands the energy be propagated for some distance,
presumably by waves or turbulent eddies, before it is dissipated.
However, the general phenomenology of turbulence is probably
not limited to the open-field regions.
Concepts from turbulence theory have been applied to the full
range of time scales for closed-field coronal energy input
as well, from the most rapid (AC) wavelike oscillations to the
slowest (DC) quasistatic stresses on magnetic footpoints
(see van Ballegooijen 1986; Hendrix \& van Hoven 1996;
Milano et al.\  1997; G\'{o}mez et al.\  2000; Chae et al.\  2002).
Indeed, some recent simulations of intermittent turbulent heating
in closed loops have been interpreted using similar cascade rate
expressions as we use in this paper (Rappazzo et al.\  2007).

The models presented below include the effects of both Alfv\'{e}n
waves and acoustic waves on the mean flow.
We assume implicitly that all waves propagate parallel to the
radially oriented flux tube, but this is not an essential feature.
In the magnetically dominated corona (where $\beta \ll 1$;
$\beta$ being the ratio of gas pressure to magnetic pressure) the
modeled Alfv\'{e}n waves can be considered essentially to be the
sum of true Alfv\'{e}nic wave power and fast-mode MHD wave power.
The acoustic waves can be considered equivalent to slow-mode
MHD waves.
As a starting point, we neglect all nonlinear couplings between
the Alfv\'{e}n and acoustic wave modes.
By ignoring the enhanced reflection and dissipation that may
arise because of these couplings
(e.g., Suzuki \& Inutsuka 2005, 2006) we essentially
{\em underestimate} the heat deposited in the extended corona
and heliosphere.
The fact that sufficient energy nonetheless exists to heat the
corona and accelerate the wind (mainly from the Alfv\'{e}n waves)
appears to deemphasize the need for such mode couplings.

It should be noted that the attention given to modeling the
steepening and shock dissipation of the acoustic waves ({\S}~4)
does not seem to have much of a ``payoff'' in the resulting
properties of the corona and solar wind.
The fully ionized outer atmosphere is extremely insensitive
to the magnitude (or sometimes even the presence) of acoustic
wave power that comes up from below the photosphere ({\S}~8.2).
Accordingly, many prior studies of the coupled chromosphere and
corona used a much simpler prescription for maintaining the
chromosphere.
We believe it is important, though, to treat both the chromospheric
and coronal heating at a comparable level of first-principles
modeling.
We also anticipate that the ZEPHYR code developed here will be
applied to the simulations of winds of other late-type stars
that may have acoustically heated coronae (e.g.,
Mullan \& Cheng 1993; Schrijver 1995), and thus the acoustic
waves should be treated as realistically as possible.

One final limitation of the models presented below is that the
energy conservation is treated in a {\em one-fluid} manner.
The protons, electrons, and heavy ions are modeled as having
a common flow speed $u$ and temperature $T$, with microscopic
velocity distributions that are simple isotropic Maxwellians.
This is an extreme simplification of reality, since it has been
known since the 1960s that the in~situ solar wind exhibits
significant departures from a thermalized equilibrium
(see reviews by Hundhausen 1972; Feldman \& Marsch 1997).
Many of these effects persist down into the extended corona
as well (e.g., Kohl et al.\  2006).
Therefore, theoretical models have long included a range of
attempts to deal with these features.
Fluid-based models have been extended to solve separate conservation
equations for each particle species, and they have been
reconstructed in various ways based on different parameterizations
for the anisotropic velocity distributions.
Several purely kinetic models have also been attempted (see
above-cited reviews, also Cranmer 2002; Hollweg \& Isenberg 2002;
Marsch 2005).

Despite the potentially clumsy ``averaging'' over real kinetic
effects, we believe the one-fluid approach is the {\em most
consistent with our present state of knowledge} about the primary
source of energy deposition: MHD turbulent cascade.
There are a number of competing ideas in the literature
regarding how the cascade proceeds to its smallest kinetic
scales and how either linear or nonlinear damping transfers
wave energy to the particles (for a recent summary see
{\S}~5.2.4 of Kohl et al.\  2006).
We thus do not yet know, from first principles, how to partition
the cascaded wave energy between particle species
($T_{e} \neq T_{p} \neq T_{\rm ion}$) and between
various directions in microscopic velocity space
($T_{\parallel} \neq T_{\perp}$).
Performing such partitioning at the present time would essentially
add new free parameters into the model.
Our adopted rate of Alfv\'{e}nic coronal heating (eq.~[\ref{eq:dmit}])
thus deals only with the total energy flux that cascades from
large to small scales and not the specific means of dissipation
once the energy reaches the small scales.

\section{Basic Physics}

The equations governing the expansion of a time-steady stellar
wind are derivable by taking successive velocity moments of the
Boltzmann equation, together with some assumption about the shape
of the velocity distribution function in order to close the
otherwise infinite chain of moment equations (see, e.g.,
Braginskii 1965; Collins 1989; Marsch 2005).
In this section we describe the conservation equations used in
the models ({\S}~3.1) and the adopted prescriptions for heat
transport due to radiation ({\S}~3.2) and conduction ({\S}~3.3).

\subsection{Conservation Equations}

We consider the flow of a pure hydrogen plasma along a radially
oriented magnetic field.
The goal is to solve for the time-steady radial dependence of the
mass density $\rho$, the bulk flow speed $u$,
and the Maxwellian temperature $T$.
The distance along the magnetic flux tube is denoted either as
$r$, measured from Sun-center, or $z$, measured from the lower
boundary of the model in the solar atmosphere (essentially the
photosphere).
For completeness, the equations below contain the dependence on
time $t$, though these terms are set to zero in the
time-independent solutions that we describe below.

The equation of mass conservation is
\begin{equation}
  \frac{\partial \rho}{\partial t} + \frac{1}{A}
  \frac{\partial}{\partial r} \left( \rho u A \right)
  \, = \, 0
  \label{eq:masscon}
\end{equation}
where $A$ is the cross-sectional area of the one-dimensional
flux tube along which the wind flows.
Magnetic flux conservation demands that the product $B_{0}A$ is
constant along the flux tube, where $B_0$ is the field strength
that we specify explicitly (see {\S}~8).
A time-steady one-dimensional flow thus constrains the product
$\rho u$ to be proportional to $B_{0}$.

The equation of momentum conservation is
\begin{equation}
  \frac{\partial u}{\partial t} + u \frac{\partial u}{\partial r}
  + \frac{1}{\rho} \frac{\partial P}{\partial r} \, = \,
  - \frac{GM_{\ast}}{r^2} + D
  \label{eq:momcon}
\end{equation}
where $P$ is the gas pressure, $G$ is the Newtonian gravitation
constant, and $M_{\ast}$ is the mass of the star.
The mass of the plasma in the modeled stellar wind region is
assumed to be negligible from a gravitational standpoint.
Also, $D$ is the bulk acceleration on the plasma due to
{\em wave pressure;} i.e., the nondissipative net ponderomotive
force due to the propagation of waves through an inhomogeneous
medium ({\S}~6).

The equation of internal energy conservation is
\begin{equation}
  \frac{\partial E}{\partial t} + u \frac{\partial E}{\partial r}
  + \left( \frac{E+P}{A} \right)
  \frac{\partial}{\partial r} \left( uA \right)
  \, = \, Q_{\rm rad} + Q_{\rm cond} + Q_{A} + Q_{S}
  \label{eq:dEdt}
\end{equation}
where $E$ is the internal energy density and the terms on the
right-hand side are volumetric heating/cooling rates due to
radiation, conduction, Alfv\'{e}n wave damping, and acoustic (sound)
wave damping.
The terms on the left-hand side that depend on $u$ are responsible
for enthalpy transport and adiabatic cooling in the accelerating wind.
The coupling of the above equations requires additional
constitutive relations to be specified:
\begin{equation}
  P \, \equiv \, n_{\rm tot} k_{\rm B} T \,\, ,
  \label{eq:Pdef}
\end{equation}
\begin{equation}
  E \, \equiv \, \frac{P}{\gamma - 1} + n_{p} I_{\rm H} \,\, ,
  \label{eq:Edef}
\end{equation}
\begin{equation}
  n_{\rm tot} \, \equiv \, n_{\rm H} + n_{e} \, = \,
  (n_{0} + n_{p}) + n_{e}  \,\, ,
  \label{eq:ntot}
\end{equation}
where $k_{\rm B}$ is Boltzmann's constant and a monatomic ratio
of specific heats $\gamma = 5/3$ is used.
The total particle number density $n_{\rm tot}$ is given by the
sum of the hydrogen number density $n_{\rm H}$ and the
electron density $n_e$, and the number densities of neutral
hydrogen and protons are denoted $n_{0}$ and $n_p$, respectively.
For the pure hydrogen plasma assumed in this paper, $n_{p} = n_{e}$.
The mass density is given by
$\rho = m_{\rm H}n_{\rm H} + m_{e}n_{e}$, although the second
term is safely considered to be negligible.

Note that equation~(\ref{eq:Edef}) becomes the ideal gas
relation $E = 3P/2 = \rho c_{v} T$ in the purely neutral limit,
where $c_{v}$ is the specific heat at constant volume.
Ionization is taken into account by the internal energy's
dependence on $I_{\rm H}$, the ionization potential of hydrogen
from the ground level (13.6 eV).
We use the same convention in the definition of $E$ as
Ulmschneider \& Muchmore (1986) and Mullan \& Cheng (1993).
An alternate definition of the internal energy is possible, though,
which is essentially given by $E - n_{\rm H} I_{\rm H}$.
This version reduces to the above ideal-gas relation in the limit
of a fully ionized plasma (e.g., McClymont \& Canfield 1983;
Fontenla et al.\  1990).
Either definition is consistent with the combined equations of
mass and energy conservation given above.

We adopted a relatively simple prescription to compute the ionization
fraction $x \equiv n_{p}/n_{\rm H}$.
This quantity is parameterized as a tabulated function of $T$ only,
where we have used the ionization balance from a recent
semi-empirical model of the solar photosphere, chromosphere, and
transition region (E.\  Avrett 2005, private communication; see
also Fontenla et al.\  1993, 2002, 2006).
The tabulated model we use is a modified version of the quiet-Sun
model C from the above papers, and it is the same one used by
Cranmer \& van Ballegooijen (2005) to model the magnetic structure
of the chromospheric network.
For convenience we call this model FAL-C$'$.

\begin{figure}
\epsscale{1.00}
\plotone{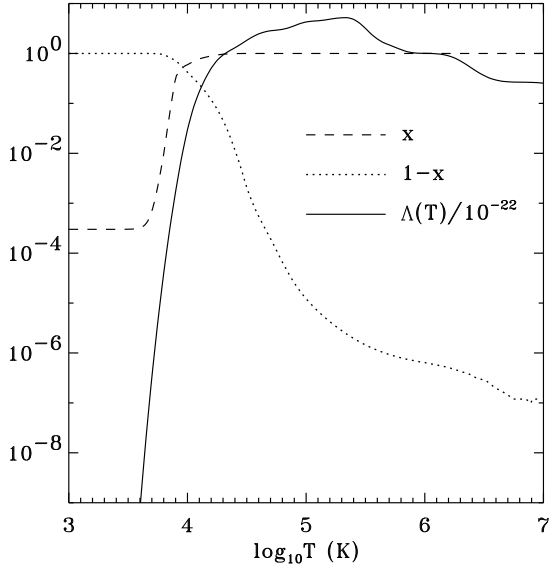}
\caption{
Temperature dependence of the hydrogen ionization state
$x$ ({\em dashed line}), the corresponding neutral hydrogen
fraction $1-x$ ({\em dotted line}), and the optically thin
radiative loss function $\Lambda (T)$ in units of $10^{-22}$
erg cm$^3$ s$^{-1}$ ({\em solid line}).}
\end{figure}

Figure 1 shows the adopted ionization fraction $x$, as well as the
corresponding neutral fraction $1-x$ (in order to see small
departures from total ionization at high temperatures) and the
radiative cooling function that was derived in part from the
FAL-C$'$ model (see {\S}~3.2).
For temperatures below the minimum tabulated value of about 4500 K,
we prevent $x$ from decreasing below a minimum value of
$3 \times 10^{-4}$, because photoionization is expected to
always keep some small fraction of metals ionized at and above
the photosphere.

We tested the ZEPHYR code with a more self-consistent model of
the hydrogen ionization balance.
The results were extremely similar, though, to those using the
tabulated fraction shown in Figure 1.
This model consisted of a three-level hydrogen atom, where the
$n=1$ and $n=2$ levels were assumed to remain in relative local
thermodynamic equilibrium (LTE) and the full rate equation between
$n=2$ and the continuum was solved iteratively with collisional
and radiative terms from Hartmann \& MacGregor (1980),
Vriens \& Smeets (1980), Ferland et al.\  (1992), and
Ferguson \& Ferland (1997).
Because the tabulated version was much faster in terms of computation
time, though, we decided to use it in the models shown below.
It will be important to include this kind of self-consistent
ionization balance when adapting this method to the winds of
other stars (e.g., Natta et al.\  1988).

\subsection{Radiative Heating and Cooling}

A complete treatment of the non-LTE transfer of energy between
radiation and matter in a partially ionized plasma is beyond the
scope of this paper.
Detailed computational efforts to model chromospheric and
coronal radiative transfer effects (e.g., Avrett \& Loeser 1992;
Carlsson \& Stein 1997, 2002; Rammacher et al.\  2005)
are important for reproducing the spectrum or studying time-dependent
or multidimensional dynamics, but for our purposes a simpler
treatment is warranted.
We use a similar general ``bridging'' approach as
Mullan \& Cheng (1993) to combine different limiting cases in
the optically thick lower atmosphere and the optically thin
upper atmosphere.

The adopted radiative heating/cooling rate is given by
\begin{equation}
  Q_{\rm rad} \, = \, e^{-\tau_{\rm R} / \tau_{0}} Q_{\rm thin}
  + (1 - e^{-\tau_{\rm R} / \tau_{0}}) Q_{\rm thick}
\end{equation}
where $\tau_{\rm R}$ is the Rosseland mean optical depth and
$\tau_{0} = 0.1$ is a constant that defines where the rate is
dominated by the optically thick or thin limits (see also
Ludwig et al.\  1994).
For a spherical stellar atmosphere, we use a definition for the
Rosseland optical depth,
\begin{equation}
  d\tau_{\rm R} \, = \, -\kappa_{\rm R} \rho
  \left( \frac{R_{\ast}}{r} \right)^{2} dr  \,\, ,
\end{equation}
that contains the geometrical correction factor suggested by
Lucy (1971, 1976); it is unimportant in the case of the Sun---where
$r \approx R_{\ast}$ in regions where
$\tau_{\rm R} \approx 1$---but we include it for later use in the
extended atmospheres of evolved stars.
The Rosseland mean opacity $\kappa_{\rm R}$ (in cm$^2$ g$^{-1}$)
is interpolated as a function of temperature and pressure from the
table of Kurucz (1992).
For a given one-dimensional model, we integrate downward from the
upper boundary far out in the supersonic wind (which has an assumed
optical depth of zero) to compute $\tau_{\rm R}(r)$.

In the optically thick photosphere and lower chromosphere we
assume that the heating and cooling is dominated by continuum
photons emitted and absorbed in LTE, with
\begin{equation}
  Q_{\rm thick} \, = \, 4\pi \rho \int
  \kappa_{\nu} (J_{\nu} - S_{\nu}) d\nu \, = \,
  4\pi \rho \kappa_{\rm R} (J-S)
  \label{eq:Qradthick}
\end{equation}
(e.g., Mihalas 1978; V\"{o}gler et al.\  2004).
Above, $S$ is the frequency-integrated source function, assumed
here in LTE to equal the local integrated Planck function,
$S=B=\sigma_{\rm R} T^{4}/\pi$,
with $\sigma_{\rm R}$ being the Stefan-Boltzmann constant.
The frequency-integrated mean intensity $J$ is given by the
gray atmosphere dependence on optical depth $\tau_{\rm R}$,
\begin{equation}
  J(\tau_{\rm R}) \, = \, \frac{3}{4\pi} \sigma_{\rm R} T_{\rm eff}^{4}
  \left[ \tau_{\rm R} + 2 q(\tau_{\rm R}) W(r) \right]  \,\, ,
  \label{eq:Jtau}
\end{equation}
where $T_{\rm eff} = 5800$ K is the solar effective temperature and
$q(\tau)$ is the Hopf (1930, 1932) function, for which we use
the following fit (accurate to better than 0.2\%),
\begin{equation}
  q(\tau) \, = \, 0.710 -
  \frac{0.133}{(1 + 0.15 \tau^{0.73})^{17.4}}  \,\, .
\end{equation}
Also, $W(r) = [1 - (1 - R_{\ast}^{2}/r^{2})^{1/2}]/2$
is the spherical dilution factor suggested for use in this context
by Chandrasekhar (1934) and Lucy (1971, 1976).
$Q_{\rm thick}$ is positive---giving net heating---when $T$ is
less than the radiative equilibrium temperature ($T_{\rm rad}$)
given by $J=S$, and it is negative---with net cooling---when
$T > T_{\rm rad}$.

In the optically thin upper chromosphere and corona, we use a
modified temperature-dependent radiative cooling function
that has been computed for photon losses due to a large collection
of spectral lines and continuum processes,
\begin{equation}
  Q_{\rm thin} \, = \, - n_{e} n_{\rm H} \Lambda(T)
  \left( 1 - \frac{T_{\rm rad}^4}{T^4} \right)
  \label{eq:Qradthin}
\end{equation}
(see also Cox \& Tucker 1969; Anderson \& Athay 1989b;
Schmutzler \& Tscharnuter 1993).
The radiative loss function $\Lambda (T)$ is shown in Figure 1,
and it has been assembled from three sources.
(1) For hydrogen and helium above 10$^4$ K, we used the output
from the PANDORA radiative transfer code which produced the
FAL-C$'$ semi-empirical model discussed above.
(2) For other elements above 10$^4$ K, we used a tabulated radiative
loss function from version 4.2 of the CHIANTI atomic database
(Young et al.\  2003) using a traditional solar abundance mixture
(Grevesse \& Sauval 1998) and collisional ionization balance
(Mazzotta et al.\  1998).
(3) For temperatures below 10$^4$ K, we used the fitting function
for partially ionized hydrogen given by Scholz \& Waters (1991).
For these low temperatures we also added a constant lower-limit
value of $10^{-34}$ erg cm$^3$ s$^{-1}$ to $\Lambda(T)$ to account
for photoionized metals, molecules, and dust, which may be
important contributors to radiative cooling at very low
temperatures in late-type stellar atmospheres (e.g.,
Schirrmacher et al.\  2003).

In order to ensure that the optically thin parts of the atmosphere
would smoothly approach radiative equilibrium (in the absence of
nonradiative heating) in the same way as in the optically thick
parts of the atmosphere, we multiplied the standard cooling
function by the term in parentheses in equation (\ref{eq:Qradthin}).
For temperatures above $\sim 2 T_{\rm rad}$ this correction factor
rapidly approaches unity.

Note from Figure 1 that there is no isolated ``Lyman alpha peak''
at temperatures of 1--2 $\times 10^4$ K.
For solar atmosphere models, this peak seems to be a spurious
feature that appears in optically thin radiative loss curves
computed without a radiation field.
The FAL-C$'$ model used here contained a full non-LTE treatment
of hydrogen as well as ambipolar diffusion that also provides
additional smearing of discrete features related to the strong
H~I Ly$\alpha$ transition (see also Kuin \& Poland 1991;
Fontenla et al.\  2002).

With regard to the extension to other stars, the computation of
the radiative cooling rate is in a similar situation as the
ionization fraction discussed in {\S}~3.1.
Our use of tabulated quantities from FAL-C$'$ seems to be
reasonable for modeling the solar atmosphere, but it will
eventually need to be replaced with a more self-consistent
procedure.
Specifically, evolved low-gravity stars with high mass loss
rates are expected to have more optically thick chromospheres.
Hartmann \& MacGregor (1980), Canfield \& Ricchiazzi (1980),
and Mullan \& Cheng (1993) computed approximate optically thick
radiation losses by taking account of reduced escape
probabilities in the damping wings of strong lines; it is possible
that these methods can be extended in a robust way to future
time-steady stellar wind models.

\subsection{Heat Conduction}

For the radially oriented flux tubes that we consider, the
conductive energy exchange rate is dominated by the divergence
of a parallel heat flux density, i.e.,
\begin{equation}
  Q_{\rm cond} \, = \, - \frac{1}{A} \frac{\partial}{\partial r}
  \left( q_{\parallel} A \right) \,\, .
\end{equation}
Because the solar atmosphere undergoes a transition from being
strongly collisionally coupled (at low heights and high densities)
to being nearly collisionless (at large heights and low densities),
it was realized long ago that the classical Spitzer-H\"{a}rm (SH)
prescription for thermal conductivity must break down somewhere
in the corona and solar wind.
We thus follow Wang (1993), and others, by using a
semi-empirical bridging law between the SH heat flux $q_{\rm SH}$
and a completely collisionless ``free-streaming'' heat flux
$q_{\rm FS}$,
\begin{equation}
  q_{\parallel} \, = \,
  \frac{\nu_{\rm coll} \, q_{\rm SH} + \nu_{\rm exp} \, q_{\rm FS}}
  {\nu_{\rm coll} + \nu_{\rm exp}}
  \label{eq:qpara}
\end{equation}
where $\nu_{\rm exp} = (u/\rho) |\partial \rho / \partial r|$
is the local wind expansion rate and $\nu_{\rm coll}$ is
the electron-electron Coulomb collision frequency,
\begin{equation}
  \nu_{\rm coll} \, = \, \frac{\ln\Lambda_{ee}}{275 \, \mbox{s}}
  \left( \frac{n_e}{10^{6} \, \mbox{cm}^{-3}} \right)
  \left( \frac{T}{10^{6} \, \mbox{K}} \right)^{-3/2}
\end{equation}
(Braginskii 1965; Olsen \& Leer 1996).
The electron Coulomb logarithm is approximated by
\begin{equation}
  \ln \Lambda_{ee} \, = \, 23.2 + \frac{3}{2} \ln \left(
  \frac{T}{10^{6} \, \mbox{K}} \right) - \frac{1}{2} \ln \left(
  \frac{n_e}{10^{6} \, \mbox{cm}^{-3}} \right)  \,\, .
\end{equation}
A more accurate version of equation~(\ref{eq:qpara}) was
derived by Cuperman \& Dryer (1985).
Under a number of simplifying approximations, though (such as
nearly Maxwellian distributions), their expression reduces to
something very close to the above bridging formula.

The radial distance where the heat flux undergoes the transition
from collisional to collisionless can be estimated by locating
the point at which $\nu_{\rm exp} = \nu_{\rm coll}$.
For the models of low-density high-speed solar wind streams
presented below, this occurs at about $r \approx 10$ solar
radii ($R_{\odot}$); for the models of high-density low-speed
solar wind, this occurs at $r \approx 50$--80 $R_{\odot}$.

In the collisionally dominated limit, we assume SH conduction,
\begin{equation}
  q_{\rm SH} \, = \, -\kappa \frac{\partial T}{\partial r}
  \label{eq:qSH}
\end{equation}
where the thermal conductivity is assumed to be dominated by
contributions from free electrons and neutral hydrogen,
$\kappa = F \kappa_{e} + \kappa_{H}$
(e.g., Nowak \& Ulmschneider 1977; McClymont \& Canfield 1983).
The effects of protons and heavy ions are neglected because they
tend to be overwhelmed by the electron conductivity in regions of
appreciable ionization (see also Ulmschneider 1970;
Hansteen \& Leer 1995).
The electron conductivity is given by
\begin{equation}
  \kappa_{e} = (1.84 \times 10^{-5} \,\, \mbox{erg}
  \,\, \mbox{cm}^{-1} \, \mbox{s}^{-1} \, \mbox{K}^{-7/2})
  \, \frac{T^{5/2}}{\ln \Lambda_{ee}}
\end{equation}
(Spitzer 1962; Braginskii 1965).
Partial ionization effects (i.e., $F$ and $\kappa_{\rm H}$)
are parameterized using the results of
Shmeleva \& Syrovatskii (1973) and McClymont \& Canfield (1983),
with
\begin{equation}
  F \, = \, \frac{1 + 4.49 w + 3.37 w^{2} + 0.59 w^{3}}
  {(1 + 3.86 w + 0.94 w^{2})^2}  \,\, ,
\end{equation}
where $w$ is the ratio of the electron-electron collision time
$\tau_{ee}$ to the electron-neutral hydrogen time $\tau_{e0}$, and
\begin{equation}
  w \, = \, \frac{\tau_{ee}}{\tau_{e0}} \, = \,
  \frac{n_{0}}{n_{e} \ln \Lambda_{ee}} \left(
  \frac{T}{52705 \, \mbox{K}} \right)^{2}  \,\, .
\end{equation}
Similarly, the neutral hydrogen conductivity is given by
\begin{equation}
  \kappa_{\rm H} \, = \, \frac{29.6 \, T} {1 +
  (n_{p}/n_{0}) (T/ 7.6 \times 10^{5} \, \mbox{K})^{-1/2}}
  \,\, .
\end{equation}
The above partial ionization effects are generally unimportant in
the solar atmosphere, but we include them for some attempt at
completeness, and with anticipation that the cooler and more
optically thick chromospheres of other stars may depend more
sensitively on them.

In the collisionless limit we use the free-streaming heat flux
derived for escaping electrons by Hollweg (1974, 1976), which was
based on empirical constraints from Forslund (1970) and
Perkins (1973) on the electron velocity distribution,
\begin{equation}
  q_{\rm FS} \, = \,
  \frac{3}{2} \alpha_{c} n_{e} u k_{\rm B} T \,\, ,
  \label{eq:qFS}
\end{equation}
where $\alpha_c$ is an order-unity correction factor that depends
on how the wings of the electron velocity distribution depart from
a Maxwellian shape.
We take a constant value of $\alpha_{c} = 4$ as has been often
used in solar wind modeling (see, e.g.,
Leer et al.\  1982; Scudder \& Olbert 1983; Withbroe 1988;
Canullo et al.\  1996; Landi \& Pantellini 2003).
Tests using Hollweg's (1974) more detailed prescription for how
$\alpha_c$ should depend on the solar wind speed and electron
temperature produced no more than a 10\% change in $T(r)$
from the $\alpha_{c} = 4$ model (and no more than a 2\% change in
the mass loss rate or terminal wind speed).
The simpler constant value was used in all subsequent ZEPHYR models.

Note that equations (\ref{eq:qpara}) and (\ref{eq:qFS}) are in
some ways similar to the heat flux relations used by
Smith \& Auer (1980) and others for flare plasmas, where the
classical heat flux is assumed to {\em saturate} at a value no
larger than a threshold flux proportional to $v_{e}^{3}$ (where
$v_e$ is the electron thermal speed).
The free-streaming heat flux given above is dimensionally similar
to the saturated heat flux, but the former is proportional to
$u v_{e}^{2}$ rather than $v_{e}^{3}$.
These two limits are related, though, because saturation occurs
when the electron-electron collisional mean free path becomes
larger than the local temperature scale height.
When heat can no longer be carried diffusively by classical
conduction, it may be advected directly at some characteristic
velocity $v_c$, and the heat flux then becomes proportional to
$v_{c} v_{e}^{2}$.
In the low-density supersonic solar wind, the characteristic
speed for, e.g., the total enthalpy flux is $v_{c} \approx u$.
To the extent that $\alpha_c$ is an order-unity correction
factor (that depends weakly on $v_e$), this applies also to
the heat flux carried in the non-Maxwellian tail of the
electron distribution.
For a high-density laboratory plasma, though,
Mannheimer \& Klein (1975) showed that $v_c$ scales directly
with $v_e$ (see also Smith \& Lilliequist 1979;
Craig \& Davys 1984).

\section{Acoustic Waves and Shocks}

We include the time-averaged effects of acoustic waves that
propagate parallel to the magnetic field.
The only source of these waves that we consider is the
convective motion below the photosphere which channels
compressive wave energy into the magnetic flux tubes.
Deep in the atmosphere these fluctuations take the form of
longitudinal, or sausage-mode tube waves (e.g., Spruit 1982;
Roberts 2000),
but because the individual flux tubes appear to merge together
somewhere in the low chromosphere into a region filled with
magnetic field (Cranmer \& van Ballegooijen 2005) we treat these
waves as standard acoustic oscillations and do not consider
thin-tube dispersive effects.
A separate source of compressive waves, which we do not model, may
be the gradual parametric decay of nonlinear MHD waves in the
outflowing stellar wind (see Sagdeev \& Galeev 1969; Goldstein 1978;
Jayanti \& Hollweg 1993).
It is usually assumed that these waves have very low frequencies
and thus are likely to have small rates of damping and steepening,
and thus a minimal impact on chromospheric or coronal
heating.\footnote{%
See, though, Suzuki \& Inutsuka (2005) for an example of naturally
produced compressive waves by similar nonlinear couplings.
This kind of process could account for a substantial fraction of the
low-frequency density fluctuations measured in interplanetary space
by in~situ spacecraft and radio scintillations.}

Below we describe how ZEPHYR models the propagation, steepening,
and dissipation of individual monochromatic ``packets'' of acoustic
wave energy ({\S}~4.1) and how the complete power spectrum is
specified ({\S}~4.2).

\subsection{Monochromatic Wave Train Evolution}

An arbitrarily steepened acoustic wave/shock train travels along
the field with constant frequency $\omega$ and a radially
varying wavenumber $k_{\parallel}$ determined by the dispersion
relation $\omega = (u + c_{s}) k_{\parallel}$.
The sound speed is given by $c_{s}^{2} = \gamma P/\rho$.
The energy density $U_S$ of linear acoustic fluctuations obeys
an equation of wave action conservation,
\begin{equation}
  \frac{\partial}{\partial t} \left( \frac{U_S}{\omega'}
  \right) + \frac{1}{A} \frac{\partial}{\partial r}
  \left[ \frac{(u + c_{s}) A U_S}{\omega'} \right] \, = \,
  - \frac{Q_S}{\omega'}
  \label{eq:J77ac}
\end{equation}
(e.g., Jacques 1977; Koninx 1992).
The Doppler shifted frequency in the frame of the accelerating
wind is given by $\omega' = \omega - u k_{\parallel}$ and
the wave energy density is given by
\begin{equation}
  U_{S} \, = \, \frac{1}{s} \rho v_{\parallel}^{2}
  \label{eq:Usdef}
\end{equation}
where $s$ is a dimensionless shape factor determined by the
spatial profile of the waves (see below).
For a small-amplitude sinusoid, $s = 2$, and for a fully
steepened sawtooth or N-wave, $s = 3$.
The parallel velocity variance, or squared wave amplitude,
is specified as $v_{\parallel}^{2}$.
Below we give the wave energy flux $F_S$ as a lower boundary
condition for the ZEPHYR code; this quantity is converted
into wave energy density using
\begin{equation}
  F_{S} \, = \, \left[ \frac{(\gamma + 3) u}{2} + c_{s}
  \right] U_{S}  \,\, .
  \label{eq:Fsdef}
\end{equation}
Note that in the limit of a static plane-parallel atmosphere
($u = 0$ and $A = \mbox{constant}$)
equations~(\ref{eq:J77ac})--(\ref{eq:Fsdef}) simplify into the
standard flux conservation quantities implicit in wave-heated
models of the solar atmosphere since the initial work of
Schwarzschild (1948), Biermann (1948), and Osterbrock (1961).

The right-hand side of equation~(\ref{eq:J77ac}) couples the
dissipation of acoustic wave energy to the $Q_S$ heating term
in equation~(\ref{eq:dEdt}).
We include two sources of wave damping: heat conduction
and entropy gain at shock discontinuities.
For the present models we ignore radiation damping of the waves
(which damps high frequencies at low heights, but does not
technically provide heat to the plasma), viscosity, and
ion-neutral friction.
The acoustic heating rate is given by
\begin{equation}
  Q_{S} \, = \, 2 \gamma_{\rm cond} U_{S} +
  \frac{\rho T \Delta S}{2\pi / \omega} \,\, ,
  \label{eq:Qs}
\end{equation}
where $\gamma_{\rm cond}$ is the linear damping
rate due to heat conduction and $\Delta S$ is the net
entropy jump across a shock (which is nonzero only above the height
where the wave train has steepened into shocks).

The damping rate due to heat conduction is given for
adiabatic waves ($\gamma = 5/3$) by
\begin{equation}
  \gamma_{\rm cond} \, = \, \frac{4 \omega^{2} \kappa}
  {15 k_{\rm B} c_{s}^{2} n_{\rm H}}
\end{equation}
(e.g., Landau \& Lifshitz 1959; Hung \& Barnes 1973; Whang 1997).
We use the same thermal conductivity $\kappa$ as is used in
the classical SH heat flux (eq.\  [\ref{eq:qSH}]).
For simplicity, the electron and neutral hydrogen conductivities
are assumed to dominate proton and heavy ion heat conduction as well
as other classical transport processes.
Note that $\gamma_{\rm cond}$ is the damping rate for the wave
amplitude; the damping rate for wave energy is twice that value
(see the factor of 2 in eq.~[\ref{eq:Qs}]).

The major contribution to the acoustic heating rate comes from
shock steepening and dissipation.
The gain in internal energy across an ideal inviscid shock is
given by
\begin{equation}
  T\Delta S \, = \, c_{v} \left[ T_{2} - T_{1} \left(
  \rho_{2} / \rho_{1} \right)^{\gamma-1} \right]
  \label{eq:TdS}
\end{equation}
where subscripts 1 and 2 denote quantities measured on the
upstream (supersonic) and downstream (subsonic) sides of the shock
(Landau \& Lifshitz 1959).
The above expression does not contain the internal energy
component from $P \Delta V$ work but only the energy that goes
into dissipation.
Equation (\ref{eq:Qs}) uses the approximation from so-called
``weak-shock theory'' that the volumetric heating rate is given
by the internal energy dissipated at one shock divided by the
mean time between shock passages in a periodic train.
This assumption breaks down for very strong shocks in the
chromosphere, which dissipate their energy in a relatively narrow
zone behind the shock (e.g., Carlsson \& Stein 1992, 1997),
but the models presented below do not develop such strong shocks.

To evaluate equation (\ref{eq:TdS}) we used the classical
Rankine-Hugoniot relations for a monatomic ($\gamma = 5/3$) gas.
These relations also are valid for a plasma with a constant
ionization state across the shock, and Carlsson \& Stein (2002)
found that shock trains in the solar chromosphere often approach
this nearly steady-state condition.
The density and temperature jump relations can be written
in terms of the upstream Mach number $M_{1}$,
\begin{equation}
  \frac{\rho_2}{\rho_1} \, = \,
  \frac{(\gamma + 1) M_{1}^2}{(\gamma - 1) M_{1}^{2} + 2}
  \label{eq:d2d1}
\end{equation}
\begin{equation}
  \frac{T_2}{T_1} \, = \,
  \frac{[2\gamma M_{1}^{2} - (\gamma - 1)]
  [(\gamma - 1) M_{1}^{2} + 2]}{(\gamma + 1)^{2} M_{1}^{2}}
  \label{eq:T2T1}
\end{equation}
Note, though, that for shocks of arbitrary strength,
equation (\ref{eq:TdS}) requires the absolute upstream and
downstream temperatures $T_1$ and $T_2$ to be computed, not just
their ratio.
(In the weak-shock limit, $T_{1} \approx T_{2} \approx T$,
the latter being the ``mean'' model atmospheric temperature.)
In many astrophysical models of shocks, $T_1$ is often assumed
to be the undisturbed equilibrium state and $T_2$ is computed
from equation (\ref{eq:T2T1}).
However, in time-dependent simulations of chromospheric shocks,
$T_1$ is often seen to fall {\em below} the time-averaged mean
temperature $T$ and often also below the radiative
equilibrium value $T_{\rm rad} \approx 4500$ K.
This is believed to arise from adiabatic expansion behind the
shock.

To compute $T_1$ and $T_2$, we first determine the upstream and
downstream densities $\rho_1$ and $\rho_2$ relative to the
known background model density $\rho$.
With respect to the propagating shock train, the background
density can be defined as that which occurs when the shock
passes through zero velocity in the reference frame of the
undisturbed atmosphere.
For an ideal sawtooth-shaped N-wave, this occurs for a given
height at a time halfway between shock passages.
Using this definition, the analytic results of
Bertschinger \& Chevalier (1985) can be used to estimate the
ratio of the minimum (preshock) density to the background value
to be
\begin{equation}
  \frac{\rho_1}{\rho} \, = \,
  \frac{1 + (\rho_{2} / \rho_{1})}{2 (\rho_{2} / \rho_{1})}
  \label{eq:d1d0}
\end{equation}
which then allows both $\rho_1$ and $\rho_2$ to be computed.
To convert densities into temperatures, some knowledge of the
thermodynamic cycle of the shock must be incorporated.
In other words, after the gas is heated and compressed, we need
to know what ``path'' it takes as it cools and expands back to
the preshock values, in order to be heated and compressed
again as the next shock in the train goes by.
Nearly all time-dependent models of periodic shocks in stellar
atmospheres have found that shocks first undergo rapid radiative
cooling at a roughly constant density, followed by nearly
adiabatic expansion back to the preshock density and temperature
(e.g., Weymann 1960; Osterbrock 1961; Ulmschneider et al.\  1978;
Bowen 1988).
This second phase seems to dominate the time between shock
passages---usually encompassing the halfway point used above
as the definition for the undisturbed density---so we assume
that the cooling from the mean temperature $T$ to the preshock,
or upstream temperature $T_1$ is adiabatic, and
\begin{equation}
  T_{1} / T \, = \, (\rho_{1} / \rho)^{\gamma - 1} \,\, .
  \label{eq:T1T0}
\end{equation}
This, in combination with equations (\ref{eq:d2d1})--(\ref{eq:d1d0}),
completes the specification of $T_1$ and $T_2$ needed to compute
$T \Delta S$.

In the limit of low-amplitude shocks (i.e., $M_{1}^{2} \approx 1+m$,
where $m \ll 1$), equation (\ref{eq:TdS}) reduces to the standard
weak-shock limit
\begin{equation}
  T \Delta S \, \approx \, \frac{2\gamma (\gamma - 1)}
  {3 (\gamma + 1)^2} \, c_{v} m^{3} T
  \label{eq:wst}
\end{equation}
(e.g., Ulmschneider 1970; Stein \& Schwartz 1972, 1973;
Mihalas \& Mihalas 1984).
In the strong-shock limit ($M_{1} \gg 1$ and
$\rho_{2}/\rho_{1} \rightarrow 4$), the ratio $T_{1} / T$
approaches a constant value of $(5/8)^{2/3} \approx 0.73$ and
$T_2$ grows without bound.
The first term of equation (\ref{eq:TdS}) dominates the second and
\begin{equation}
  T \Delta S \, \approx \, c_{v} T_{2} \, \approx \,
  0.228 \, c_{v} M_{1}^{2} T  \,\, .
\end{equation}
The weak and strong limiting expressions are valid to within
about 25\% of the exact result for $M_{1} < 1.2$ and
$M_{1} > 4.6$, respectively.
Because the peak Mach numbers of the shock trains in the ZEPHYR
models shown below are typically outside these ranges
(i.e., $M_{1} \approx 2$) we use the full procedure described by
equations (\ref{eq:TdS})--(\ref{eq:T1T0}), which is valid for
shocks of arbitrary strength.

In order to incorporate the shock dissipation into the acoustic
wave transport equation, the wave amplitude $v_{\parallel}$ must
be converted, where appropriate, into the Mach number $M_1$.
We must take account of how the initially sinusoidal wave
profile steepens into a sawtooth shock train.
Often shock heating is applied only above an estimated
shock formation height.
However, there exists a finite range of heights between the
first formation of the shock (i.e., where the acoustic wave train
first ``breaks'') and the height where the shock train
has evolved into a complete sawtooth shape.
Between these heights the velocity amplitude of the shock may
only be a fraction of the crest-to-trough velocity amplitude
of the wave.

To solve for the wave evolution and dissipation in this region,
we have computed numerical profiles of both the shape factor $s$
and a steepening efficiency factor $\varepsilon$.
These quantities are computed as a function of a dimensionless
steepening parameter $\zeta$, which measures by how much a
wave crest is approaching (or has overtaken) the zero-velocity
node immediately ahead of it.\footnote{%
The node propagates exactly at the linear phase speed, and the
crest propagates faster by a nonlinear factor proportional to
the wave amplitude $v_{\parallel}$.}
The efficiency $\varepsilon$ is defined as the ratio of the shock
velocity amplitude to the full velocity amplitude of the
(arbitrarily steepened) wave profile, and we define
\begin{equation}
  M_{1} \, = \, 1 + \left( \frac{\gamma + 1}{2} \right)
  \frac{\varepsilon \, v_{\parallel}}{c_s} \,\, .
\end{equation}
Where the wave is still a sinusoid, $\varepsilon = 0$ and
$M_{1} = 1$, which results in no heating.
As the wave steepens, $\varepsilon$ grows from 0 to 1.
At each radial grid zone, we compute the crest-to-node distance
factor $\zeta \equiv \Delta z / \lambda$, where $\lambda$ is the
local parallel wavelength and
\begin{equation}
  \Delta z \, = \, \frac{\lambda_0}{4} - \frac{\gamma + 1}{2}
  \int \frac{v_{\parallel} dz}{c_s} \,\, ,
  \label{eq:Delz}
\end{equation}
where the integration is taken from the lower boundary (at which
$\lambda = \lambda_0$) up to an arbitrary height.
By computing this quantity point-by-point along the grid, the
atmospheric stratification is taken into account accurately.
At the lower boundary, the wave profile is assumed to be a perfect
sinusoid, and $\zeta = 1/4$.
This quantity decreases steadily as the wave train propagates
upward and steepens.
When $\zeta$ reaches zero, the shock amplitude has grown to be
equal to the wave amplitude and the profile is assumed to
remain a sawtooth as it propagates upward.
We continue to follow the ever-decreasing $\zeta$ to values below
zero, though, because the profile only reaches the exact sawtooth
``N-wave'' shape in the limit of $\zeta \rightarrow -\infty$.

\begin{figure}
\epsscale{1.00}
\plotone{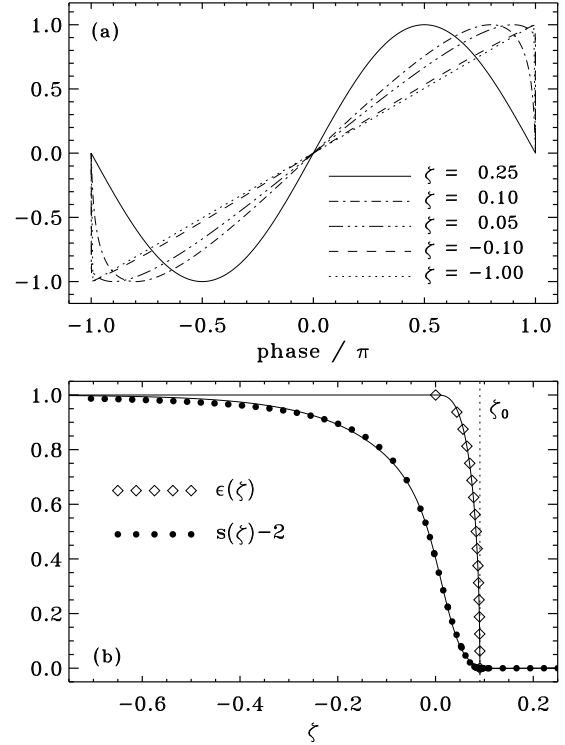}
\caption{
Properties of steepened acoustic waveforms.
({\em{a}}) Normalized phase function $\Phi$ as a function of the
dimensionless wave phase $\phi$ (in units of $\pi$) for a series of
steepening parameters $\zeta$ (see labels for values).
({\em{b}}) Steepening efficiency ratio $\varepsilon$
({\em open diamonds}) and scaled shape factor $s-2$ ({\em
filled circles}) computed numerically for a series of values of
$\zeta$.  Solid lines show the analytic fitting functions given in
the text.}
\end{figure}

Figure 2 shows the normalized phase function $\Phi (\phi)$ of a
gradually steepening acoustic wave, where the phase $\phi$ is
defined at a given (constant) height as $\pi - \omega t$;
it varies from $-\pi$ to $\pi$.
Figure 2 also plots the efficiency $\varepsilon$ and the shape
factor $s$ as a function of $\zeta$.
These have been derived from numerical simulations of this
steepening sinusoidal wave profile, and we give parameterized
fits (used by the ZEPHYR code) below.
The fit to the efficiency is given by
\begin{equation}
  \varepsilon (\zeta) \, = \, \left\{
    \begin{array}{ll}
      1 & \zeta \leq 0 \\
      \sqrt{1 - (\zeta / \zeta_{0})^{3.1}} &
        0 < \zeta < \zeta_{0} \\
      0 & \zeta \geq \zeta_{0}
    \end{array} \right.  \,\, .
\end{equation}
The constant $\zeta_{0}$ denotes the critical breakpoint at which
the wave train first steepens to infinite slope at the node ahead
of the crest; it is defined as
$\zeta_{0} \equiv (\pi - 2)/4\pi \approx 0.091$.
Note that as $\zeta$ decreases from 0.25 to $\zeta_0$, the
efficiency $\varepsilon$ remains zero because no shock
transition has yet formed.
Only when $\zeta$ decreases below $\zeta_0$ does there exist a
shock with a finite strength.

The shape factor $s$ is used to convert between wave
amplitude and energy density (eq.~[\ref{eq:Usdef}]), and it
is defined as the inverse of the phase-averaged square
of the normalized phase function, i.e.,
\begin{equation}
  \frac{1}{s} \, = \, \frac{1}{2\pi} \int_{-\pi}^{+\pi}
  d\phi \, \left[ \Phi (\phi) \right]^{2}
\end{equation}
(see also Koninx 1992; Suzuki 2004).
As an example, for a sinusoid the average of $\sin^{2}\phi$
over a full period is 1/2, and thus $s=2$.
We found the following fit from a series of simulated wave profiles
undergoing gradual steepening,
\begin{equation}
  s(\zeta) \, = \, 3 -
  \frac{1 + 1.32 e^{-37.5 \zeta}}{1 + 2.89 e^{-44.7 \zeta}}
\end{equation}
(see Figure 2 for a comparison between the numerically determined
values and the fit).
For most of the pre-break steepening (i.e., from $\zeta = 0.25$
down to $\zeta_0$) $s$ remains close to 2.
Note, though, that when the shock grows to full amplitude
at $\zeta = 0$ the shape factor has increased only to about 2.4.
The additional increase up to 3 occurs for negative values of
$\zeta$.

\subsection{Acoustic Power Spectrum}

The convection zone generates a continuous spectrum of acoustic
power, and we need to specify this distribution of energy as
a function of frequency at the photospheric base of the model.
The power spectrum at larger heights is determined implicitly as
a result of solving the monochromatic wave action conservation
equations (eq.~[\ref{eq:J77ac}]) for a range of frequencies.
The continuous power spectrum $P_{S} (\omega)$ is defined at the
base in units of the photospheric acoustic flux, with
\begin{equation}
  F_{S} \, = \, \int d\omega \, P_{S} (\omega)  \,\, .
  \label{eq:PintS}
\end{equation}
For waves that escape into the upper atmosphere, there should be
negligible power at frequencies below the acoustic cutoff,
\begin{equation}
  \omega_{\rm ac} \, \equiv \, \frac{\gamma g}{2 c_s}  \,\, ,
\end{equation}
because waves having $\omega < \omega_{\rm ac}$ are evanescent in
an ideal hydrostatic atmosphere (e.g., Mihalas \& Mihalas 1984;
see, however, Wang et al.\  1995; Schmitz \& Fleck 1998).
We used a constant value of $\omega_{\rm ac} = 0.03$ rad s$^{-1}$
(i.e., $f \approx 4.8$ mHz, or a period of 3.5 minutes), which
corresponds to temperature-minimum conditions in the upper photosphere.
From a theoretical standpoint, this value is slightly on the high side,
since the photospheric $T_{\rm eff}$ of 5800 K gives a cutoff
frequency about 13\% lower.
Our higher value for $\omega_{\rm ac}$ is an attempt to ensure that
the modeled frequencies are those that should be propagating
{\em everywhere} in the model---i.e., since we do not treat
evanescent energy loss, we want to include only waves that can make
it through both the photosphere and the temperature minimum region
without becoming evanescent.
Note, though, that from the standpoint of observational
helioseismology, our value is slightly on the low side (e.g.,
Jim\'{e}nez 2006), so it seems to be a satisfactory median value.

The shape of the power spectrum above the cutoff is a subject of
some ongoing controversy, which we do not attempt to address fully.
The presence of substantial power at high frequencies
($f \gtrsim 20$ mHz) is predicted by traditional theories of
sound generation from turbulent convection (e.g., Lighthill 1952;
Stein 1967; Musielak et al.\  1994; Ulmschneider et al.\  1996)
and also by observational inferences of time-steady chromospheric
heating (Kalkofen et al.\  1999; Ulmschneider et al.\  2005;
Cuntz et al.\  2007).
However, evidence also exists that there may be an extremely
steep decline in the acoustic power spectrum before frequencies
of order 20 mHz are reached, and thus that high frequencies
would not be important to atmospheric heating (e.g.,
Judge et al.\  2003; Fossum \& Carlsson 2005, 2006).
Recent advances in detecting high-frequency acoustic fluctuations
have been made by Wunnenberg et al.\  (2002), DeForest (2004),
Muglach (2006), and van Noort \& Rouppe van der Voort (2006),
but no firm conclusions yet exist regarding their impact on
chromospheric heating.
Future observations with higher spatial and temporal resolution
are definitely needed.

Provisionally, we model $P_{S}(\omega)$ with a high-frequency
tail reminiscent of the turbulent convection theories cited above.
The following parameterization
\begin{equation}
  P_{S} (\omega) \, \propto \, \left\{
    \begin{array}{ll}
      (\omega / \omega_{\rm max})^{\psi} /
      [1 + (\omega / \omega_{\rm max})^{2\psi}] &
      \omega \geq \omega_{\rm ac} \\
      0 & \omega < \omega_{\rm ac}
    \end{array} \right.
  \label{eq:PomegaS}
\end{equation}
has finite power at the cutoff, a peak value at
$\omega_{\rm max} > \omega_{\rm ac}$, and a declining tail with
$P_{S} \propto \omega^{-\psi}$ at high frequencies.
Typically, $\psi \approx 3$ and $\omega_{\rm max}$ is a factor
of 2 to 5 larger than $\omega_{\rm ac}$.
The normalization is given by specifying a known value of $F_S$ and
using equation (\ref{eq:PintS}).

\begin{figure}
\epsscale{1.00}
\plotone{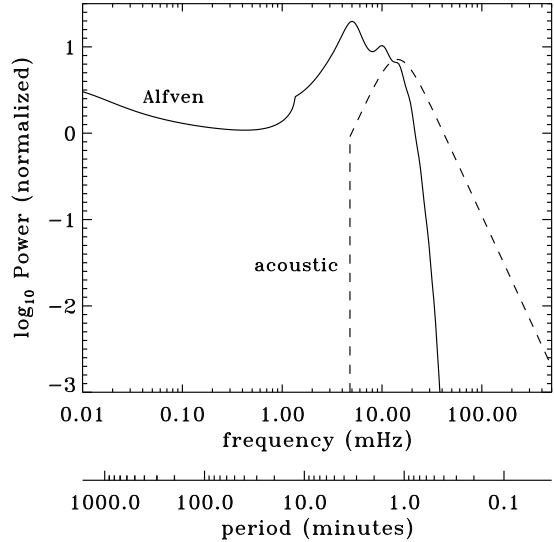}
\caption{
Normalized power spectra for Alfv\'{e}n waves $P_A$
({\em solid line}) and acoustic waves $P_S$ ({\em dashed line}).
The cyclic frequency $f = \omega / 2\pi$ is given in units of mHz,
and the period $p = 1/f$ is given in units of minutes.}
\end{figure}

Figure 3 illustrates the shape of the acoustic power spectrum
that is used in all of the solar models described in this paper.
We use constant values of $\omega_{\rm max}/ \omega_{\rm ac} = 3$
and $\psi = 2.5$.
The adopted value of $\psi$ is the most sensitive to the controversy
over high-frequency acoustic waves, and it deserves further discussion.
The standard Lagrangian treatment of the Lighthill-Stein sound
generation mechanism tends to give values of $\psi$ between 3 and
3.5 in the high-frequency limit for both acoustic and longitudinal
flux-tube waves (Ulmschneider et al.\  1996;
Musielak et al.\  2000).
However, an alternate Eulerian treatment of the turbulent
correlations (Rubinstein \& Zhou 2002) yielded a much shallower
decline with increasing frequency; $\psi \approx 1.3$--2.4.
We also note that the value $\psi = 3$ is a special case that
corresponds to frequency-independent shock steepening for the
discrete bins that we use (see below); i.e., when $\psi = 3$
the increase in frequency from one bin to the next is balanced
exactly by the power decrease, such that the wave train in each
bin steepens into a shock train at the same height.
We find that a realistic chromospheric temperature rise occurs only
when higher frequency waves steepen at successively lower
heights (which requires $\psi < 3$).
Our value of 2.5 satisfies this requirement while being only slightly
lower than the range predicted by traditional Lighthill-Stein theory.

In the ZEPHYR code, the continuous spectrum $P_{S}(\omega)$ is
modeled as a series of discrete frequency bins, each of which is
treated independently as described in {\S}~4.1.
We define the bins as ``octaves'' in frequency space; i.e.,
the first bin encompasses $\omega_{\rm ac}$ to $2 \omega_{\rm ac}$,
the second encompasses $2 \omega_{\rm ac}$ to $4 \omega_{\rm ac}$,
and so on.
Ideally, it would have been preferable to use even narrower bins
in order to more accurately represent the shape of the spectrum.
However, if the frequency bins were too narrow, each would
contain a vanishingly small amount of wave energy.
The calculation of nonlinear steepening (eq.~[\ref{eq:Delz}])
depends on the total amplitude that remains reasonably coherent
at a given frequency.
Narrow bins---treated independently as described above---would
essentially destroy this coherence and produce an unphysical delay
in the onset of steepening due to the low amplitudes in each bin.
The use of octaves is an attempt to balance the needs of frequency
resolution and realistic coherence for steepening.\footnote{%
Too much coherence would also be undesirable.  One-dimensional
simulations of chromospheric shocks often result in ``shock
cannibalization'' (i.e., overtaking and merging) and an
effective filtering out of high frequencies.  A realistic
multidimensional distribution of strong and weak acoustic
sources at and below the photosphere, though, is more likely
to result in an incoherent and randomized power spectrum such
as we assume here (see, e.g., Cadavid et al.\  2003;
Ulmschneider et al.\  2005).}

In the models computed for this paper, a series of 14 bins is used.
The power missed by ignoring frequencies above the maximum value of
$2^{14} \omega_{\rm ac}$ is only $\sim 10^{-5}$ of the total flux.
The monochromatic frequencies used in the wave action conservation
equation are computed as mean, or first-moment frequencies over
the power spectrum of each bin.
The first one ($\omega_{\rm ac}$ to $2 \omega_{\rm ac}$), for 
example, has a mean of $1.62 \, \omega_{\rm ac}$ (i.e.,
$f = 7.7$ mHz).

\section{Alfv\'{e}n Waves and Turbulence}

We model transverse MHD fluctuations in the radially oriented
magnetic flux tube as ideal incompressible Alfv\'{e}n waves.
The equations and assumptions we use generally follow
Cranmer \& van Ballegooijen (2005), who computed various
properties of Alfv\'{e}n waves from the photosphere to the
interplanetary medium assuming a known background plasma state
($u$, $\rho$, $B_{0}$).
Here we compute the wave properties and background plasma
properties together in a self-consistent manner.
The Alfv\'{e}n waves are believed to arise from transverse
jostling of magnetic flux tubes in the intergranular photosphere
by convective motions on granular ($\sim$1000 km) horizontal scales.
Note that, similar to the acoustic waves, the Alfv\'{e}n waves
are expected to have modified dispersive properties in the
photosphere and low chromosphere (e.g., they become {\em kink-mode}
tube waves in regions where the flux tubes are surrounded by
field-free regions; see Spruit 1981, 1982), but we ignore these
effects because they occur over a small range of heights.
Our treatment of Alfv\'{e}n wave reflection and turbulent
dissipation is complementary to the recent work of 
Verdini et al.\  (2005, 2006).

As in the previous section, we first discuss the overall wave energy
transport and dissipation mechanisms ({\S}~5.1) and then describe
the power spectrum of the fluctuations ({\S}~5.2).
The way that the frequency dependence is assimilated into the model
is slightly different from the acoustic-wave case, though.

\subsection{Alfv\'{e}n Wave Action Conservation}

For transverse incompressible waves, we solve a single wave action
conservation equation with a damping term that contains information
about the power spectrum and wave reflection.
We believe that a monochromatic treatment of the wave action
conservation (i.e., to treat each frequency bin independently from
the rest) is inappropriate to use when the dominant dissipation
mechanism is MHD turbulence.
A turbulent cascade inherently contains strong nonlocal ``mixing''
in frequency and wavenumber space.
Thus the turbulent damping rate is computed only using quantities
that have been integrated over the power spectrum (see below).

We assume the Alfv\'{e}n waves are dispersionless with a phase
speed given by $\omega / k_{\parallel} = u + V_{A}$ in the 
stationary reference frame of the Sun.
The Alfv\'{e}n speed is defined as
$V_{A} = B_{0} / (4\pi\rho)^{1/2}$.
Although we solve the full non-WKB transport equation to obtain
the {\em relative} distribution of outward and inward propagating
waves, we use the following simplified version of the wave
action conservation equation to compute the radial evolution
of the total (frequency-integrated) wave energy density $U_A$:
\begin{equation}
  \frac{\partial}{\partial t} \left( \frac{U_A}{\omega'}
  \right) + \frac{1}{A} \frac{\partial}{\partial r}
  \left[ \frac{(u + V_{A}) A U_A}{\omega'} \right] \, = \,
  - \frac{Q_A}{\omega'}
  \label{eq:J77alf}
\end{equation}
(see, e.g., Jacques 1977; Isenberg \& Hollweg 1982;
Tu \& Marsch 1995).
The presence of the Doppler shifted frequency $\omega'$
(measured in the solar wind frame) in the above equation is
deceptive, since the wave action conservation is essentially
{\em frequency-independent.}   Note that
$\omega' = \omega - u k_{\parallel} = \omega V_{A} / (u+V_{A})$,
and thus a constant factor of $\omega$ can be factored out of
each term in equation (\ref{eq:J77alf}).
The total Alfv\'{e}nic energy density $U_A$ is the sum of
the energy densities in kinetic and magnetic fluctuations,
\begin{equation}
  U_{K} \, = \, \frac{\rho v_{\perp}^{2}}{2} \,\, , \,\,\,\,
  U_{B} \, = \, \frac{B_{\perp}^{2}}{8 \pi}
\end{equation}
where $v_{\perp}$ and $B_{\perp}$ are the transverse velocity
and magnetic field oscillation amplitudes.
We assume equipartition between the two components
($U_{K} = U_{B}$), such that
$U_{A} = 2 U_{K} = \rho v_{\perp}^{2}$.

The use of equation (\ref{eq:J77alf}) contains the assumption that
the energy balance is dominated by outward-propagating waves, whose
energy density $U_{-}$ exceeds the energy density of
inward-propagating waves $U_{+}$.
By definition, $U_{A} = U_{-} + U_{+}$, and
\begin{equation}
  U_{\pm} \, = \, \frac{\rho Z_{\pm}^{2}}{4}
\end{equation}
where the Elsasser (1950) amplitudes are defined here as
$Z_{\pm} \, \equiv \, |v_{\perp} \pm B_{\perp}/(4\pi\rho)^{1/2}|$.
For future reference, we give the expression for the net outward
Alfv\'{e}n wave flux,
\begin{equation}
  F_{A} \, = \, u (U_{K} + 2 U_{B}) + V_{A} (U_{-} - U_{+})
  \label{eq:Fadef}
\end{equation}
(e.g., Heinemann \& Olbert 1980).
The two assumptions of kinetic/magnetic equipartition and
outward-wave dominance in equation (\ref{eq:J77alf}) tend to
break down in the lower atmosphere when there is strong non-WKB
wave reflection, but in the {\em corona}---where the dominant
Alfv\'{e}nic heating occurs---these assumptions seem to be
appropriate (see also Cranmer \& van Ballegooijen 2005).

The only physical source of Alfv\'{e}n wave damping that we include
is the turbulent cascade, which ultimately must terminate in an
irreversible conversion of wave energy into heat.
This is, in some sense, a controlled experiment to evaluate to what
degree turbulence may be the {\em dominant} cause of coronal heating
and solar wind acceleration, but other damping mechanisms for
Alfv\'{e}n waves have been suggested.
Collisional damping mechanisms for MHD waves have been studied
extensively in the context of the dense coronal base (e.g.,
Alfv\'{e}n 1947; Kuperus et al.\  1981;
Narain \& Ulmschneider 1990, 1996; Porter et al.\  1994;
Roberts 2000),
but the open field lines that feed the solar wind have lower
densities and are thus less collisionally dominated than
closed loops in the low corona.
Cranmer \& van Ballegooijen (2005) investigated linear viscous
dissipation of Alfv\'{e}n waves along an open magnetic flux tube
and found that viscosity should be negligible for waves having
periods longer than about 1 minute.

We adopted the following phenomenological form for the MHD
turbulence damping rate,
\begin{equation}
  Q_{A} \, = \, \rho \, {\cal E}_{\rm turb} \,
  \frac{Z_{-}^{2} Z_{+} + Z_{+}^{2} Z_{-}}{4 L_{\perp}}
  \label{eq:dmit}
\end{equation}
(Hossain et al.\  1995; Zhou \& Matthaeus 1990;
Matthaeus et al.\  1999; Dmitruk et al. 2001, 2002).
The transverse length scale $L_{\perp}(r)$ represents an effective
perpendicular correlation length of the turbulence for the
largest ``driving'' eddies.
We used the standard assumption that $L_{\perp}$ scales with the
transverse width of the open flux tube; i.e., that it remains
proportional to $B_{0}^{-1/2}$ (see also Hollweg 1986).
Ideally, the evolution of $L_{\perp}$ should be coupled to the
radial variation of the fluctuation energy (see, e.g., eq.~[3]
of Matthaeus et al.\  1999) as well as to the non-varying
field strength.
Future work should include both this effect and additional
tests of whether the above phenomenological form is an adequate
representation of the true anisotropic cascade.

The normalization of $L_{\perp}$ is one of the few free parameters
of our model.
We note that Cranmer \& van Ballegooijen (2005) found that
$L_{\perp}$ should be about 1100 km at the height in the low
chromosphere where thin flux tubes merge with one another.
Mapping this value down to the photosphere would yield a lower
boundary value $L_{\perp\odot} = 320$ km.
This spatial scale is intermediate between the probable horizontal
size an individual flux tube in the photosphere (50--100 km) and
the size of a convective granule (1000 km).
This value is also similar to the width of an intergranular lane
and also the mean separation between photospheric flux tubes in the
quiet-Sun supergranular network (350--700 km; see
Cranmer \& van Ballegooijen 2005).

The factor ${\cal E}_{\rm turb}$ in equation~(\ref{eq:dmit})
is a turbulent efficiency that accounts for regions where the
turbulence may not have time to develop before the waves or the
wind carry away the energy (see Dmitruk \& Matthaeus 2003).
We estimated this efficiency factor to be
\begin{equation}
  {\cal E}_{\rm turb} \, = \, \frac{1}{1 +
  ( t_{\rm eddy} / t_{\rm ref} )^{n}}
  \label{eq:Eturb}
\end{equation}
where the two time scales in this expression are $t_{\rm eddy}$,
a nonlinear outer-scale eddy cascade time, and $t_{\rm ref}$,
a timescale for macroscopic Alfv\'{e}n wave reflection.
In most of the models presented below we take $n = 1$, based on
analytic and numerical models of Dobrowolny et al.\  (1980),
Matthaeus \& Zhou (1989), and Oughton et al.\  (2006).
Dmitruk \& Matthaeus (2003) found that the turbulent cascade
has sufficient time to develop and heat the plasma only when
$t_{\rm eddy} \ll t_{\rm ref}$.
Thus, our efficiency factor above quenches the turbulent
heating when $t_{\rm eddy} \gg t_{\rm ref}$, i.e., when the
Alfv\'{e}n waves want to propagate away much faster than the
cascade can proceed ``locally.''
The reflection time is defined simply as
$t_{\rm ref} = 1/ |\nabla \cdot {\bf V}_{A}|$, and the
eddy cascade time is given by
\begin{equation}
  t_{\rm eddy} \, = \, \frac{L_{\perp} \sqrt{3\pi}}
  {(1 + M_{A}) \, v_{\perp}}  \,\, ,
\end{equation}
where the Alfv\'{e}nic Mach number $M_{A} = u/V_A$ and the
numerical factor of $\sqrt{3\pi}$ comes from the normalization
of an assumed shape of the turbulence spectrum (see
Appendix C of Cranmer \& van Ballegooijen 2005; see also
Higdon 1984; Shebalin et al.\  1983; Goldreich \& Sridhar 1995, 1997;
Bhattacharjee \& Ng 2001; Cho et al.\  2002).
When $n=1$, the efficiency factor provides an approximate
bridging between a Kolmogorov (1941) scaling, when
$t_{\rm eddy} \ll t_{\rm ref}$, and an IK-like
(Iroshnikov 1963; Kraichnan 1965) scaling,
when $t_{\rm eddy} \gg t_{\rm ref}$.
There is still some controversy, though, over which type
of cascade rate is appropriate for MHD turbulence in the
solar wind.

We separated the Alfv\'{e}n wave energy into outward ($Z_{-}$)
and inward ($Z_{+}$) components by solving a modified form of
the non-WKB transport equations of Heinemann \& Olbert (1980),
Barkhudarov (1991), Velli (1993), and
Orlando et al.\  (1996).\footnote{%
By WKB (Wentzel, Kramers, Brillouin) we do not refer to any
specific asymptotic expansion.
The WKB limit of pure outward propagation, with no reflection,
is amenable to the standard ``eikonal'' approximation by
defining a local wavenumber.
The treatment of non-WKB reflection is more general in that the
radial parts of the $Z_{\pm}$ eigenfunctions are computed
numerically without the use of a wavenumber.}
These frequency-dependent equations are discussed below in
{\S}~5.2 and their solution provides a spectrum-averaged value
of the effective local reflection coefficient
${\cal R} = Z_{+}/Z_{-}$.
Knowing ${\cal R}$ and $U_A$ (from eq.~[\ref{eq:J77alf}])
allows the Elsasser amplitudes to be computed at all heights,
\begin{equation}
  Z_{-} \, = \, \sqrt{\frac{4 U_A}{\rho (1 + {\cal R}^{2})}}
  \,\, , \,\,\,\,
  Z_{+} \, = \, {\cal R} |Z_{-}|  \,\, ,
\end{equation}
and the turbulent damping rate in equation (\ref{eq:dmit}) is
then specified.

\subsection{Alfv\'{e}n Wave Frequency Dependence}

For a series of Alfv\'{e}n wave frequencies $\omega$, we solved
the non-WKB wave transport equations in the dimensionless form
given by equations (24) and (33) of Barkhudarov (1991),
\begin{equation}
  \frac{d\Psi}{dr} \, = \, \frac{(\Psi^{2} - 1) \cos\Gamma}{2 H_A}
  \label{eq:dPsi}
\end{equation}
\begin{equation}
  \frac{d\Gamma}{dr} \, = \, \frac{(\Psi^{2} + 1) \sin\Gamma}
  {2 H_{A} \Psi} - \frac{2 \omega V_A}{u^{2} - V_{A}^{2}}
  \label{eq:dGam}
\end{equation}
where $\Psi$ is related to the frequency-dependent reflection
coefficient ${\cal R}_{\omega}$ via
\begin{equation}
  \Psi \, = \, \left( \frac{u - V_{A}}{u + V_{A}} \right)
  \, {\cal R}_{\omega}  \,\, ,
\end{equation}
$\Gamma$ is the angular phase shift between the $Z_{-}$ and $Z_{+}$
wave trains, and $H_A$ is the (signed) scale height for the
Alfv\'{e}n speed, or $V_{A} / (\partial V_{A} / \partial r)$.
Although the models of Barkhudarov (1991) were limited to
spherical geometry ($A \propto r^{2}$) the above relations are
valid for any $A(r)$ (see also Cranmer \& van Ballegooijen 2005).
We followed the general solution procedure outlined by
Barkhudarov (1991) for integrating across the Alfv\'{e}nic
singular point $r_A$, where $u = V_A$ and thus $\Psi = 0$.
The reflection coefficient remains finite at this point, and
it can be written exactly as
\begin{equation}
  {\cal R}_{\omega}(r_{A}) \, = \, \frac{|V_{A}'|}
  {\sqrt{\omega^{2} + (u' - V_{A}')^{2}}}
  \,\, ,
\end{equation}
where $u'$ and $V_{A}'$ are radial derivatives of the outflow speed
and Alfv\'{e}n speed taken at $r = r_A$.
The ZEPHYR code utilizes fourth-order Runge-Kutta numerical
integration to solve for ${\cal R}_{\omega}(r)$ above and below
$r_A$.
We ensured that $| {\cal R}_{\omega} | < 1$ at all heights and for
all frequencies.

The dissipation of Alfv\'{e}n waves is explicitly not included in
the equations given above for the reflection coefficient.
Although it is possible to include nonlinear damping consistent
with equation (\ref{eq:dmit}) in these transport equations (see,
e.g., Verdini et al.\  2005, 2006), we remain cautious about the
combination of strong turbulent damping and ``monochromatic''
wave quantities.
Additional simulations may be required in order to better guide
the use of phenomenological nonlinear terms when following the
development of a spectrum of Alfv\'{e}nic fluctuations.

The full frequency-averaged reflection coefficient is computed by
weighting ${\cal R}_{\omega}$ by the Alfv\'{e}n wave power spectrum
$P_{A}(\omega)$, with
\begin{equation}
  {\cal R}^{2}(r) \, = \, 
  \frac{\int d\omega \, P_{A}(\omega) \, {\cal R}_{\omega}^{2}(r)}
       {\int d\omega \, P_{A}(\omega)}
  \label{eq:Rweight}
\end{equation}
where the square of ${\cal R}$ is used because the power spectrum
is an energy density quantity and ${\cal R}$ is a ratio of
amplitudes.
The spectrum is used essentially as a weighting function, and we
assume it has a constant shape as a function of height (see, e.g.,
Figure 8 of Cranmer \& van Ballegooijen 2005).

Figure 3 shows the shape of the Alfv\'{e}n wave power spectrum that
we use in the ZEPHYR models discussed below.
Specifically, this is the power spectrum of total (kinetic plus
magnetic) energy from the model of Cranmer \& van Ballegooijen (2005)
taken at the height of the transition region.
It was computed from an empirically constrained {\em photospheric}
spectrum of the dynamics of thin flux-tubes, which is described as
a linear superposition of two types of motion.
First, isolated flux tubes undertake ``random walks'' in response to
convective granulation; they have a power spectrum that has been
constrained by observed G-band bright point motions (e.g.,
van Ballegooijen et al.\  1998; Nisenson et al.\  2003).
Second, flux tubes exhibit sporadic rapid horizontal ``jumps''
that probably represent merging, fragmenting, or reconnecting with
surrounding magnetic fields (e.g., Choudhuri et al.\  1993;
Berger \& Title 1996; Berger et al.\  1998; Hasan et al.\  2000)
which we modeled as a series of periodic impulsive motions.
High-resolution photospheric observations thus provided the
kinetic energy spectrum, and the partitioning between kinetic
and magnetic energy components (needed to compute the total energy
spectrum) was determined using the analytic linear theory of
kink-mode waves in a stratified atmosphere.
Finally, propagation effects between the photosphere and the
transition region ($z = 0.003 \, R_{\odot}$ in the model of
Cranmer \& van Ballegooijen 2005) were taken into account by
solving the combined non-WKB kink-mode and Alfv\'{e}n-mode 
transport equations for an empirically constrained (FAL-C$'$)
background plasma state.

There are several noteworthy features of the spectrum $P_{A}(\omega)$
in Figure 3.
There is power at the lowest frequencies, corresponding to periods
of hours to days, because of the assumed random-walk component of the
photospheric flux tube motion (with $P_{A} \propto e^{-\omega \tau}$,
where $\tau \approx 60$ s).
However, the relative amount of power at periods longer than 1 hour
is negligible compared to the total; this stands in contrast with
in~situ measurements that show the majority of power to be at
these long periods (e.g., Goldstein et al.\  1995;
Tu \& Marsch 1995).
There is growing evidence that the low-frequency fluctuations seen
in interplanetary space may be the result of the passage of multiple
uncorrelated flux tubes past the spacecraft and not intrinsic
turbulence within any one flux tube (see, e.g.,
McCracken \& Ness 1966; Jokipii \& Parker 1969;
Bruno et al.\  2001; Giacalone \& Jokipii 2004;
Giacalone et al.\  2006; Borovsky 2006).

The damping of evanescent kink-mode waves can be seen in Figure 3
from the mild discontinuity at the cutoff frequency of 1.4 mHz.
Only about 40\% of the energy of evanescent waves is lost, though,
because the flux tubes remain thin (and thus gravitationally buoyant)
only below a low-chromosphere ``merging height'' of 600 km.
Above this height the magnetic fields expand horizontally to fill
the volume above supergranular network lanes and the transverse
waves both below and above the cutoff can propagate freely as
Alfv\'{e}n waves.
The multiple peaks in the spectrum between about 3 and 20 mHz
are a result of propagation effects between the photosphere and
the merging height.
In the model of Cranmer \& van Ballegooijen (2005) the atmosphere
was modeled with the nonisothermal FAL-C$'$ temperature structure
and a radially varying filling factor for the flux tubes; both of
these factors resulted in a complicated frequency dependence for
the non-WKB transmission of kink-mode waves between $z=0$ and 600 km.
Coincidentally, the frequencies of the maxima in both $P_{A}$ and
$P_{S}$ are each about a factor three higher than their respective
cutoff frequencies.

It is important to note that the Alfv\'{e}n wave amplitude
$v_{\perp}$ (obtained from the WKB-like eq.\  [\ref{eq:J77alf}])
diverges from the actual transverse velocity of oscillating
magnetic flux tubes in the lower atmosphere.
The effects of evanescence and non-WKB wave reflection are not
{\em directly} included in equation (\ref{eq:J77alf}), but they
end up being of minimal importance in the corona and solar wind.
We thus can estimate the true velocity amplitude $w_{\perp}$
after the final iterated solar atmosphere parameters
have been determined, as
\begin{equation}
  w_{\perp}^{2} \, = \, v_{\perp}^{2} \left(
  \frac{1 + {\cal R}_{\omega}^2}{1 - {\cal R}_{\omega}^2}
  \right) \times \left\{
  \begin{array}{ll}
    1 \,\, , & \omega \geq \omega_{\rm kc} \\
    \exp \left[ 1 - \sqrt{1 - (\omega / \omega_{\rm kc})^2}
      \right] , & \omega < \omega_{\rm kc} \\
  \end{array}
  \right.  \, ,
  \label{eq:wperp}
\end{equation}
where $\omega_{\rm kc}$ is the kink-mode cutoff frequency.
Above the mid-chromosphere height where thin flux tubes merge
with one another, $\omega_{\rm kc}$ effectively goes to zero
and there is no evanescence.
The ${\cal R}_{\omega}$ factor above in parentheses corrects
for the approximation that equation (\ref{eq:J77alf}) follows 
only outward-going waves.
For the ZEPHYR models presented below, the spectrum-averaged
value of $w_{\perp}$ at the photosphere is typically a factor
of 2 to 5 times larger than $v_{\perp}$ at this lower boundary.
Above the transition region, though, $w_{\perp} \approx v_{\perp}$
(see also Figure 11 of Cranmer \& van Ballegooijen 2005).
Despite the fact that $v_{\perp}$ underestimates the actual
velocity amplitude, we believe it is more appropriate to use this
as an imposed lower boundary condition rather than to use
$w_{\perp}$.
The latter quantity depends on quantities that are computed
self-consistently along with the other plasma parameters and
are not known a~priori.

\section{Wave Pressure Acceleration}

Just as electromagnetic waves carry momentum and exert
pressure on matter, propagating acoustic and MHD waves can also
do work on the mean fluid via a similar kind of radiation stress.
(Bretherton \& Garrett 1968; Dewar 1970; Belcher 1971;
Alazraki and Couturier 1971).
For parallel-propagating acoustic and Alfv\'{e}n waves, the
time-averaged radial wave pressure acceleration was derived in
detail by Jacques (1977) to be
\begin{equation}
  D \, = \, -\frac{1}{2\rho}
  \frac{\partial U_A}{\partial r} - \left( \frac{\gamma + 1}{2\rho}
  \right) \frac{\partial U_S}{\partial r} - \frac{U_S}{A\rho}
  \frac{\partial A}{\partial r}
  \,\, .
  \label{eq:Ddef}
\end{equation}
As above, we ignore departures from kinetic-magnetic energy
equipartition for the Alfv\'{e}n waves; this ends up being
a good approximation for solar wind conditions (see, e.g.,
Heinemann \& Olbert 1980; Cranmer \& van Ballegooijen 2005).

To implement the above expression in the ZEPHYR code, we used
the wave action conservation equations (\ref{eq:J77ac}) and
(\ref{eq:J77alf}) to reformulate the momentum conservation
equation into a time-independent Parker (1958, 1963) critical
point equation with additional terms that affect the definition
of the critical point.
Assuming $\gamma = 5/3$ throughout, the modified momentum
equation becomes
\begin{displaymath}
  \left( u - \frac{u_{c}^2}{u} \right) \frac{du}{dr} \, = \,
  - \frac{GM_{\ast}}{r^2}
  + \left( u_{c}^{2} + V_{1}^{2} \right) \frac{d \ln A}{dr}
\end{displaymath}
\begin{equation}
  - \left( a^{2} + V_{2}^{2} \right) \frac{d \ln a^{2}}{dr}
  + \frac{1}{\rho} \left[ \frac{Q_A}{2 (u + V_{A})} +
  + \frac{4 Q_S}{3 (u + c_{s})} \right]
  \label{eq:critmod}
\end{equation}
where $a$ is the isothermal sound speed $(k_{\rm B}T/m_{\rm H})^{1/2}$
and the modified critical speed is given by
\begin{equation}
  u_{c}^{2} \, = \, a^{2} + \frac{U_A}{4\rho} \left(
  \frac{1 + 3 M_A}{1 + M_A} \right) + \frac{8 U_S}{3\rho}
  \left( \frac{M_S}{1 + M_S} \right)
  \label{eq:ucrit}
\end{equation}
with the bulk-flow Mach numbers $M_{A} = u/V_A$ and $M_{S} = u/c_s$.
In the presence of acoustic waves, the additional terms above are
given by
\begin{equation}
  V_{1}^{2}\, = \, \frac{U_S}{3 \rho} \left(
  \frac{1 - 7 M_S}{1 + M_S} \right)
\end{equation}
\begin{equation}
  V_{2}^{2}\, = \, \frac{4 U_S}{3 \rho} \left(
  \frac{M_{S} - 1}{M_{S} + 1} \right) \,\, .
\end{equation}
The modified critical radius $r_c$ is found by locating the point
where the right-hand side of equation (\ref{eq:critmod}) is zero,
and thus $u = u_c$.

For rapidly expanding superradial flux tubes,
Kopp \& Holzer (1976) pointed out that there may be more than one
possible location for the critical point.
Defining the right-hand side of equation (\ref{eq:critmod}) as the
radial derivative of a known function ${\cal F}(r)$, they essentially
found that the {\em global minimum} of ${\cal F}(r)$ specifies the
critical point location that allows for a stable and continuous
solution for $u(r)$ from the Sun to interplanetary space (see also
V\'{a}squez et al.\  2003; Cranmer 2005a).
The Kopp \& Holzer (1976) result, though, was derived for an
ideally polytropic corona without wave pressure acceleration.
It is not clear if this global minimum condition is applicable
to the models described in this paper.
For all of the ZEPHYR models discussed below in {\S}~8,
though, the global minimum of ${\cal F}(r)$ coincides with the
{\em largest} possible value of $r_c$.
For expediency, then, when there are multiple possibilities we
choose the largest value for the critical radius that also
exhibits a minimum in ${\cal F}$ (i.e., $d^{2}{\cal F}/dr^{2} > 0$).

Note that equation (\ref{eq:critmod}) includes the wave pressure
acceleration, but not the explicit derivatives of $U_S$ and $U_A$
that appear in equation (\ref{eq:Ddef}).
The modified momentum equation thus alleviates the need to perform
noisy numerical differentiation (see also
Jacques 1977; Hartmann \& MacGregor 1980; DeCampli 1981;
Holzer et al.\  1983; Wang \& Sheeley 1991).
The damping rates $Q_S$ and $Q_A$ appear explicitly in
the momentum equation, which provides additional nonlinear
coupling between the momentum and energy equations.

Wave pressure acceleration has been invoked by Laming (2004)
as a potential explanation for the relative enhancement of
low FIP (first ionization potential) elements relative to
high FIP elements in the corona and solar wind.\footnote{%
For other potential explanations of the FIP effect, see also
von Steiger \& Geiss (1989), Vauclair (1996),
Arge \& Mullan (1998), Schwadron et al.\  (1999),
and references therein and in Laming (2004).}
If Alfv\'{e}n waves exert an appreciable force on ions in
the chromosphere and transition region (where there are still
neutrals that do not feel this force) significant
``fractionation'' may occur.
We apply Laming's (2004) idea to the models presented below
by evaluating his integrated momentum equation for atoms and ions
of element $s$ undergoing fractionation.
A slightly simplified version of this equation (which assumes
that there is no explicit dependence of $D$ on the density
gradient) is
\begin{equation}
  \ln \frac{(\rho_{s} v_{s}^{2})_u}{(\rho_{s} v_{s}^{2})_l}
  \, = \, \int_{z_l}^{z_u} dz \,
  \frac{D}{v_{s}^2} \left[
  \frac{2 \xi_s}{\xi_{s} + (1 - \xi_{s})(\nu_{s,i} / \nu_{s,n})}
  \right]
  \label{eq:LamFIP}
\end{equation}
where $\rho_s$ and $v_{s} = (c_{s}^{2} + v_{\parallel}^{2})^{1/2}$
are the element's mass density and effective parallel turbulent
speed.
The ionization fraction of the element---essentially one minus
the neutral fraction---is given by $\xi_s$, and the collision
rates between ions and neutrals of element $s$ and the ambient
gas are given by $\nu_{s,i}$ and $\nu_{s,n}$ (see Laming 2004
for detailed expressions).
Subscripts $l$ and $u$ denote quantities computed at heights
$z_l$ (a lower boundary we take as the photosphere) and
$z_u$ (an upper boundary that can be anywhere in the corona
or solar wind), respectively.
The degree of FIP enhancement between these heights is obtained
by solving equation (\ref{eq:LamFIP}) twice: once for the
ratio $\rho_{s,u}/ \rho_{s,l}$ for a low FIP element, and once
for a high FIP element.
Dividing one ratio by the other cancels out the overall density
stratification between $z_l$ and $z_u$ and leaves only the
fractionated abundance difference.

\section{Solution Method}

We compute time-independent solutions to the conservation
equations given above by applying a new hybrid method of
iteration and relaxation.
Reasons for not following time-dependent fluctuations
explicitly were summarized in {\S}~2.
It has become common, though, to use time-dependent hydrodynamics
codes to find stable time-steady solar wind solutions (see
recent work by, e.g., Li et al.\  2004; Lionello et al.\  2005;
Lie-Svendsen \& Esser 2005, and references therein).
We break from this tradition for several reasons.
First, in the one-fluid case, the time-independent mass and
momentum conservation equations are relatively easy to solve,
and only the energy equation requires special treatment.
Second, when modeling the photosphere, chromosphere, corona,
and wind all in one grid, the energy equation changes its basic
character at different locations depending on which terms are
dominant.
In the corona, the Spitzer-H\"{a}rm conduction makes it a
second-order parabolic differential equation.
In the outer solar wind the gradual transition to collisionless
heat conduction reduces it to a first-order differential equation.
In the photosphere and chromosphere it is essentially a
zeroth-order differential equation---i.e., an algebraic balance
between heating and cooling.
These changes, combined with the huge dynamic range in quantities
such as density, temperature, and wave phase speeds, make it
difficult to implement and optimize a {\em robust} time-dependent
numerical scheme.
Even implicit methods, which are not necessarily limited by
propagation across the smallest grid zones, appear to be
prohibitively difficult to set up properly when the source terms
(i.e., the waves) have such a complex nonlinear dependence on the
primary plasma parameters.

The code developed for this work is called ZEPHYR.
After setting up an initial trial guess for the plasma parameters
$\rho$, $u$, and $T$, the code iterates a fixed number of times
(typically 100 to 200) alternately between solutions of the 
energy equation (eq.\  [\ref{eq:dEdt}]) and the other
constitutive and conservation equations (eqs.
[\ref{eq:masscon}], [\ref{eq:momcon}], [\ref{eq:Pdef}],
[\ref{eq:Edef}], [\ref{eq:ntot}], [\ref{eq:J77ac}],
[\ref{eq:J77alf}], [\ref{eq:dPsi}], [\ref{eq:dGam}])
to find a steady-state accelerating wind solution.
To stabilize the iteration process and avoid pathological
solutions, most of the plasma parameters are ``undercorrected''
using a scheme described below---i.e., rather than replacing
the old solution with the new one, a fractional step {\em toward}
the new solution is taken.

The spatial grid used by ZEPHYR has variable zone spacing
that depends on the height above the lower boundary.
The total number of grid zones $N$ is divided into two
subsets:  45\% of the zones are allocated to a fine mesh
with constant spacing in the lower atmosphere (i.e., between
$z=0$ and a fixed zone-midpoint height
$z_{\rm mid} = 0.005 \, R_{\odot}$)
and 55\% of the zones occupy the rest of the grid with
spacing that increases by 1\% per zone from the value at
$z_{\rm mid}$ up to the top of the grid.
All models described in this paper have $N = 1300$ and a fixed
grid spacing of $8.56 \times 10^{-6} \, R_{\odot} \approx 6$ km
in the lower atmosphere.
At the top of the grid ($z = 1200 \, R_{\odot} \approx 5.6$ AU)
the grid spacing has increased to 24 $R_{\odot}$.
The relative spacing $\Delta r / R_{\odot}$ thus increases 
from about $10^{-5}$ in the lower atmosphere up to 0.02 at the
top of the grid.

The initial condition for the temperature is a gray radiative
equilibrium atmosphere (i.e., $T = T_{\rm rad}$ as defined
in {\S}~3.2) that is perturbed by transitioning
(using a hyperbolic tangent function) to a corona/wind value
above a specified transition region height of $0.9 z_{\rm mid}$.
The temperature above this transition obeys a power law in
$r$, with a basal value of $1.5 \times 10^{6}$ K and
a very slow radial decline proportional to $r^{-0.05}$.
The lower boundary ($z=0$) is defined as $\tau_{\rm R} = 2$,
which corresponds to a base density $\rho_{\odot} =
2.45 \times 10^{-7}$ g cm$^{-3}$.
This choice for the lower boundary condition is used in order to
include the traditional photosphere (either $\tau_{\rm R} = 1$
or 2/3) in the interior of the grid.
The temperature at the lower boundary is defined by the gray
atmosphere condition (eq.\  [\ref{eq:Jtau}]).
The initial density distribution is assumed to be hydrostatic
(i.e., $u=0$).
Refined initial guesses for the density and outflow speed are
evaluated as the first steps in the iteration process
described below.
Tests have shown that finding the proper solution is
insensitive to the details of the initial guess, but the iteration
method converges faster when the initial guess has properties
closer to a realistic solar atmosphere.

The main ``outer'' iteration loop consists of two interior modules,
each of which undergoes a number of ``inner'' iterations.
The first module solves for the dynamics ($\rho$, $u$) and for
the various source terms in the momentum and energy equations.
The five steps taken in one inner iteration of this module are
as follows.
\begin{enumerate}
\item
The hydrogen ionization fraction $x$, the gas pressure $P$,
and the internal energy density $E$ are computed as described
in {\S}~3.1.
The optical depth scale $\tau_{\rm R}$ is integrated and the
radiative cooling/heating rate $Q_{\rm rad}$ is computed
({\S}~3.2).
The heat conduction rate $Q_{\rm cond}$ ({\S}~3.3) is also
determined using the four-point finite differencing scheme
discussed below for the radial derivatives $\partial T / \partial r$
and $\partial q_{\parallel} / \partial r$.
The heat conduction $Q_{\rm cond}$ is artificially suppressed
in the outermost 10 grid zones in order to produce a more
well-behaved upper boundary condition that is dominated by
the outward advection of all characteristics.
(These 10 zones are not considered to be part of the actual
solar wind solution.)
\item
The modified Parker critical point equation (\ref{eq:critmod}) is
solved for $u(r)$ by integrating up and down from $r_c$.
To step from the critical point (which generally falls between
grid zones) to the grid zones immediately above and below,
an analytic derivative $(\partial u / \partial r)_c$ computed
from L'H\^{o}pital's rule is used in order to avoid the well-known
instability at an X-type singular point.
At all other grid zones, fourth-order Runge-Kutta integration
is used (e.g., Press et al.\  1992).
The undercorrection scheme described below is used for $u(r)$.
\item
The time-steady mass conservation equation (\ref{eq:masscon})
is solved for $\rho(r)$ in a straightforward way.
The base density $\rho_{\odot}$ described above is kept fixed
and the prior step's solution for the basal outflow speed
$u_{\odot} = u(R_{\odot})$ is used to determine the constant
mass flux $\rho u A$.
The density is then computed exactly at each grid zone,
but the undercorrection scheme is also used for $\rho(r)$ in
order to prevent too rapid a change.
\item
The properties of the acoustic wave spectrum are computed at
each grid point and the spectrum-integrated values of
$U_S$ and $Q_S$ are determined ({\S}~4).
The steepening of each monochromatic wave train is computed
by integrating equation (\ref{eq:Delz}) up from the lower boundary
simultaneously with the wave action equation (\ref{eq:J77ac}).
Simple first-order Euler steps are used to perform both integrations.
\item
The non-WKB Alfv\'{e}n wave transport equations ({\S}~5.2) are
solved for a number of frequency dependent reflection coefficients
${\cal R}_{\omega}(r)$ which are summed over the power spectrum to
obtain ${\cal R}(r)$.
The adaptive fourth-order Runge-Kutta scheme developed by
Cranmer \& van Ballegooijen (2005) has been included in the
ZEPHYR code.
Once the $Z_{\pm}$ Elsasser variables have been determined,
the heating rate $Q_A$ is computed and the Alfv\'{e}n wave action
conservation equation is integrated up from the lower boundary to
obtain $U_A$ ({\S}~5.1).
\end{enumerate}
These steps are repeated for a fixed number of inner iterations
(typically 50 to 100) in order to reach internal consistency.
However, because the non-WKB wave reflection
equations (\ref{eq:dPsi}--\ref{eq:dGam}) dominate the computation
time, the reflection coefficient ${\cal R}$ is recomputed only for
the first two of these iterations.

The second main module of ZEPHYR solves the energy conservation
equation and obtains a time-independent temperature distribution
$T(r)$.
There are three main steps that are repeated until either a
certain degree of convergence has been attained
($\langle \delta E \rangle < 10^{-3}$) or a maximum number of
iterations (typically 1000) is exceeded.
The convergence quantity $\langle \delta E \rangle$ is defined to
be an average over the relative convergence at each grid zone.
For the radial grid with $N$ discrete zones,
\begin{equation}
  \langle \delta E \rangle \, \equiv \, \frac{1}{N}
  \sum_{i=1}^{N} \left(
  \frac{| \dot{E} |_i}{\mbox{max}|Q|_i} \right)
  \label{eq:delE}
\end{equation}
where $i$ denotes quantities computed at each grid zone.
The numerator is obtained by solving equation (\ref{eq:dEdt})
for the explicit time dependent term
$\dot{E} \equiv \partial E / \partial t$ at each grid zone
and taking the absolute magnitude.
This residual-like quantity is nonzero for solutions that have
not yet converged to a steady state.
The denominator is the absolute magnitude of the largest single
term in equation~(\ref{eq:dEdt});
this includes the two advection terms on the left-hand side,
and it also includes the separation of the optically thick
$J$ and $S$ terms in $Q_{\rm thick}$ (eq.~[\ref{eq:Qradthick}])
which balance one another in the photosphere.
The solution that satisfies exact time-independence (for the
energy equation) would have $\langle \delta E \rangle = 0$.
The following three steps repeated by the second module of ZEPHYR
are designed to hone in on this solution.
\begin{enumerate}
\item
The core procedure in solving the energy equation is
{\em relaxation} using $\dot{E}$ at each grid zone.
The sign of $\dot{E}$ is used to determine
whether the current solution for $T$ should be increased or
decreased, and the magnitude of the change is computed from a
positive-definite correction factor $c(r)$.\footnote{%
The fact that the magnitude of $\dot{E}$ is {\em not} used is the
main reason that this technique is called ``relaxation'' and
is not really a variety of time-dependent hydrodynamic evolution.}
This factor is initialized to a constant value of 0.16 at the
start of the inner iteration loop.
During each relaxation step it is either kept constant,
if $\dot{E}$ has kept the same sign from the last step to the
current step, or it is reduced in magnitude by 7\%, if $\dot{E}$
has switched signs from the last step to the current step.
The reduction in $c$ that occurs when $\dot{E}$ oscillates in
sign is a kind of ``annealing'' that allows the relaxation
method to focus in on the time-independent solution (i.e.,
the solution with $\dot{E} = 0$ everywhere).
The updated temperature at each grid zone is thus determined by
multiplying the old value by a factor of
$(1 + c \dot{E} / | \dot{E} |)$.
The optimized numerical values given above were determined from a
large number of tests with a range of values.
\item
Because the relaxation method above can lead to discontinuous
jumps in $T(r)$, we perform trial piecewise smoothing in order to
reduce unphysical fluctuations.
The radial grid with $N=1300$ is broken up into 130 pieces
each having 10 zones.
(At the start of each iteration, the offset point that defines
the start of the first 10-zone piece is shifted by one grid zone.)
If a piece contains more than one change of sign in the slope
$\partial T / \partial r$, it is smoothed using a Gaussian filter
(with a two zone half-width) until only 0 or 1 changes of sign
remain.
The piecewise nature of this smoothing is necessary so that the
entire grid does not ``suffer'' when just a small part of it
contains numerical noise.
\item
The convergence parameter $\langle \delta E \rangle$ for the
current inner iteration is compared to the best solution (i.e.,
the lowest value of $\langle \delta E \rangle$) that has been
found during this outer iteration loop.
If the solution has improved, the best solution is updated.
If the solution has gotten worse, we discard it and revert to
the saved best case.
Note, though, that this comparison is not performed at every
inner iteration step.
(Doing it every time could lead to an infinite loop with no
changes ever made to $T$.)
Tests showed that it is best to allow the solutions to
evolve for a while and only perform this comparison every
15 to 20 iteration steps (we use 17).
\end{enumerate}
The converged value of $T$ that emerges from this module is
undercorrected before starting the next outer iteration.

The undercorrection scheme that is used at various points in
the ZEPHYR code was motivated by globally convergent backtracking
methods for finding roots of nonlinear equations (e.g.,
Dennis \& Schnabel 1983).
Rather than taking the full suggested iteration step, which may
propel the solution away from the desired region of convergence,
it is sometimes best to take only a partial step.
For a scalar quantity $f_{i,j}$ at radial grid zone $i$ and
iteration step $j$, we specify this partial step as
\begin{equation}
  f_{i,j+1} = f_{i,j} \left| \frac{\tilde{f}_{i,j+1}}{f_{i,j}}
  \right|^{\epsilon}
\end{equation}
where $\tilde{f}_{i,j+1}$ is the next suggested iteration that
was obtained by solving one of the conservation equations.
The exponent $\epsilon$ describes the degree of undercorrection.
When the solutions are nearly converged, the full iteration step
should be taken ($\epsilon \approx 1$).
When the solutions are far from convergence, though, we require
substantial undercorrection ($0 < \epsilon \ll 1$).
To obtain this exponent, we use
\begin{equation}
  \epsilon = \epsilon_{0} + (1-\epsilon_{0}) \min \left(
  \left| \frac{\tilde{f}_{j+1}}{f_{j}} \right| ,
  \left| \frac{f_{j}}{\tilde{f}_{j+1}} \right| \right)
\end{equation}
where the minimum is taken over the entire radial grid (thus the
absence of $i$ subscripts) so that the worst agreement between
the current iteration and the next suggested iteration is
highlighted.
Because $\tilde{f}_{j+1}$ may be larger or smaller than $f_j$,
both the ratio and its reciprocal are used above, and the
largest value that the minimum can take is 1.
Tests have shown $\epsilon_{0} = 0.17$ to be a robust value;
it represents a practical lower limit to $\epsilon$ that
alleviates making infinitesimally small corrections.

When radial derivatives need to be taken in the ZEPHYR code,
we use the following four-point finite differencing scheme,
\begin{equation}
  \left( \frac{\partial f}{\partial r} \right)_{i} \, = \,
  C \left( \frac{f_{i+1}-f_{i-1}}{r_{i+1}-r_{i-1}} \right) +
  (1-C) \left( \frac{f_{i+2}-f_{i-2}}{r_{i+2}-r_{i-2}} \right)
\end{equation}
for discretized quantities $f_i$ on the radial grid that has
unequally spaced heights $r_i$.
The constant $C$ determines the weighting between two pairs of
centered differences.
The limit of $C = 1$ corresponds to standard two-point finite
differencing, which is generally accurate to second order.
Further Taylor-series expansion, assuming a constant-spacing grid
and a reasonably smooth function $f(r)$, would give a result that
is accurate to fourth order for $C = 4/3$.
Note, though, that quantities in the ZEPHYR code are tabulated on a
grid with variable spacing and they often exhibit numerical noise.
In this case, errors may be reduced by essentially averaging
between the $i \pm 1$ difference and the $i \pm 2$ difference
(i.e., using $C \approx 0.5$).
Tests with the ZEPHYR code found that $C = 0.2$ provides the
most accurate differentiation and noise reduction, and this
value is used in the models presented below.
For the four zones at the bottom ($i = 1, 2$) and top
($i = N-1, N$) of the grid we use a combination of the standard
two-point finite difference expression and linear extrapolation
to evaluate the derivatives.

\begin{figure}
\epsscale{1.00}
\plotone{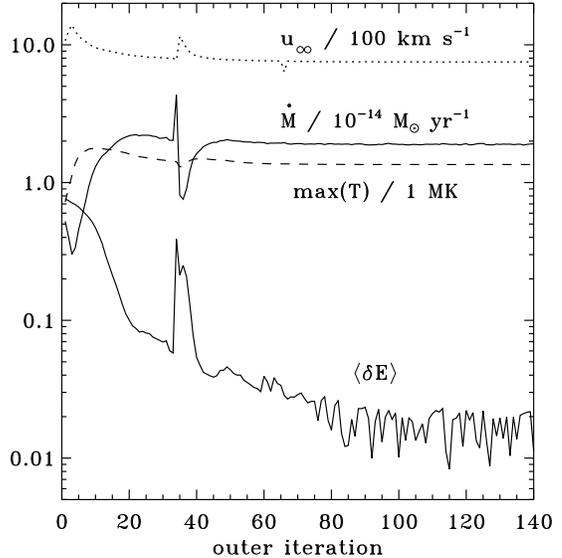}
\caption{
Convergence of the ZEPHYR outer iteration process toward a
steady-state solution.
From top to bottom, the plasma parameters shown include the solar
wind speed $u_{\infty}$ at the top of the grid ({\em dotted line}),
the spherical mass loss rate $\dot{M}$ ({\em upper solid line}),
the maximum coronal temperature ({\em dashed line})---all in
scaled units as listed in the captions---and the dimensionless
energy convergence parameter $\langle \delta E \rangle$
({\em lower solid line}).}
\end{figure}

Figure 4 illustrates the convergence of the outer iteration process
for the main polar coronal hole model described in {\S}~8.3.
The displayed plasma parameters reach reasonably steady final values
in about 70 outer iteration steps.
The energy convergence parameter $\langle \delta E \rangle$
decreases to its minimum range of variation (0.01--0.02) in
about 100 iterations.
Note, though, that $\langle \delta E \rangle$ is averaged over the
entire radial grid; in many parts of the grid the convergence
is much better than the average value indicates.
Below the chromosphere-corona transition region the convergence
is excellent (i.e., $\langle \delta E \rangle$ is always
less than $10^{-4}$).
In the narrow transition region itself, though, the piecewise
smoothing discussed above smears out the temperature distribution
so that the ideal ``conduction dominated'' solution cannot be
sustained exactly.\footnote{%
Note that the solar atmosphere should also contain some
degree of {\em ambipolar diffusion} between ions and neutrals
(Fontenla et al.\  1990, 1991, 1993).
This effectively provides additional ``smearing'' of the otherwise
extremely sharp transition region.  Our transition region
thicknesses resemble those of Fontenla et al.\  by pure coincidence.
None of the derived properties of the model either above or below
the transition region appear to depend on this smearing.}
Thus, the average value of the convergence parameter for
$10^{4} < T < 10^{5}$ K is about 0.2--0.3.
In the corona and solar wind acceleration region above the transition
region the average value of the convergence parameter is about
0.01, similar to the global average.
If the 30 or so grid zones of the transition region are excluded,
the global average $\langle \delta E \rangle$ is reduced by about
a factor of two to $\sim$0.005.

The full ZEPHYR code comprises approximately 2000 lines of Fortran.
With the typical iteration parameters given above, the code runs
in 2 to 4 hours of CPU time on various Sun Microsystems
(Ultra and SunFire) computers.

\section{Results}

In this section we present a series of solar atmosphere models
computed by the ZEPHYR code.
A series of tests was first performed to evaluate the sensitivity
of the dominant plasma properties to various input parameters
({\S\S}~8.1--8.2).
We used the grids of test models to determine the most likely
input parameters for a detailed model of a flux tube emerging from
a polar coronal hole at solar minimum ({\S}~8.3).
Then, using the same input parameters and varying only the
radial dependence of the magnetic field $B_{0} (r)$,
we modeled the pole-to-equator variation of solar wind conditions
at solar minimum ({\S}~8.4) and explored the properties of slow
wind streams that are connected to flux tubes emerging from
active regions ({\S}~8.5).

In addition to the global magnetic field strength (our primary
``control knob'') there are three key parameters that we varied
in the course of exploring the physics of atmospheric heating
and solar wind acceleration:
\begin{enumerate}
\item
The {\em photospheric acoustic flux} $F_{S \odot}$ injected at
the lower boundary mainly affects the chromospheric heating.
Probable values for $F_{S \odot}$ seem to range between
$10^7$ and $10^9$ erg s$^{-1}$ cm$^{-2}$ (see, e.g.,
Musielak et al.\  1994; Ulmschneider et al.\  1996, 2001;
Carlsson \& Stein 1997; Fawzy et al.\  2002).
For standard photospheric densities and sound speeds, this
gives a range for the photospheric acoustic velocity amplitude
$v_{\parallel \odot}$ of about 0.1 to 1 km s$^{-1}$.
These values are a bit smaller than traditional ``laminar''
granulation velocities of 1 to 2 km s$^{-1}$.
The sources of propagating waves are believed to be concentrated
in the dark intergranular lanes and thus are able to extract
only a fraction of the total kinetic energy of granulation
(Rimmele et al.\  1995; Nesis et al.\  1997, 1999;
Cadavid et al.\  2003).
\item
The {\em photospheric Alfv\'{e}n wave amplitude} $v_{\perp \odot}$
is specified instead of the basal flux $F_A$ or non-WKB velocity
amplitude $w_{\perp}$, since the latter quantities depend on
the cancellation between upward and downward propagating waves
that is determined as a part of the self-consistent solution.
Observational determinations of $w_{\perp}$ from the footpoint
motions of G-band bright points yield mean values around
1 km s$^{-1}$, with transient speeds up to 5 km s$^{-1}$
(e.g., Berger \& Title 1996).
Larger-scale analyses of the horizontal diffusion of magnetic
flux elements give smaller speeds of order 0.1 to 0.3 km s$^{-1}$
(Schrijver et al.\  1996).
Recall that $v_{\perp}$ is likely to be factors of 2 to 5 smaller
than $w_{\perp}$, so the range of possible values for
$v_{\perp \odot}$ may extend from below 0.1 up to 1 or 2 km s$^{-1}$.
\item
The {\em photospheric Alfv\'{e}n wave correlation length}
$L_{\perp \odot}$ sets the scale of the turbulent heating rate
$Q_A$ (eq.\  [\ref{eq:dmit}]).
Once this parameter is set, the value of $L_{\perp}$ at all larger
heights is determined by the adopted proportionality with
$B_{0}^{-1/2}$.
A practical lower limit for $L_{\perp \odot}$ seems to be about
10 km; i.e., the spatial scale over which radiative diffusion
may inhibit the collapse of strong fields into thin flux tubes
(e.g., Venkatakrishnan 1986; S\'{a}nchez Almeida 2001;
Cameron \& Galloway 2005).
The upper limit may be of the order of the size of photospheric
granules ($\sim$1000 km).
Possible intermediate length scales include the photospheric
radius of a of thin flux tube (50--100 km), the width of an
intergranular lane (about 300 km), and the mean separation between
the flux tubes in the supergranular network (350--700 km; see
Cranmer \& van Ballegooijen 2005).
\end{enumerate}
All other parameters have been fixed with the values given in
earlier sections.

\begin{figure}
\epsscale{1.00}
\plotone{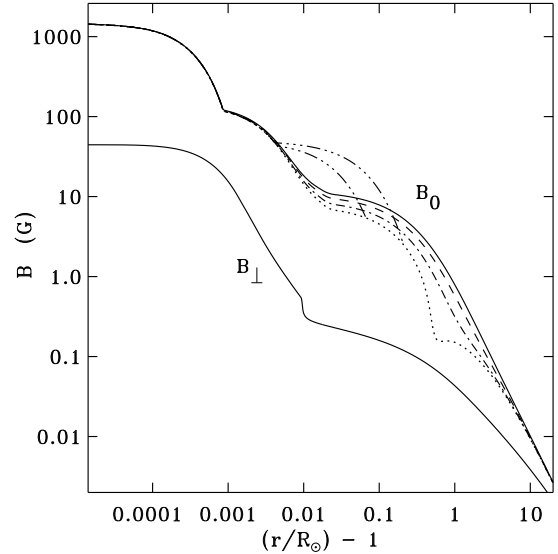}
\caption{
Radial dependence of the background magnetic field $B_{0}$.
The axisymmetric solar-minimum field of Banaszkiewicz et al.\  (1998)
is shown for field lines originating at
$\theta_{0} = 0\arcdeg$ ({\em upper solid line}),
{16\arcdeg} ({\em dashed line}),
{24\arcdeg} ({\em dot-dashed line}), and
{29.7\arcdeg} ({\em dotted line}).
All have been modified at low heights using the model of
Cranmer \& van Ballegooijen (2005).
Also shown are example active-region fields for
$h = 0.03$ and 0.07 $R_{\odot}$ ({\em dash-triple-dot lines}) and
the computed Alfv\'{e}n-wave magnetic amplitude $B_{\perp}$ for the
polar coronal hole model ({\em lower solid line}).
Above the largest height shown ($r \approx 20 \, R_{\odot}$)
the magnetic field is nearly exactly radial, with
$B_{0} \propto r^{-2}$.}
\end{figure}

\subsection{Coronal Parameter Study}

The test models discussed in this section all used the polar
coronal hole magnetic field model that was derived by
Cranmer \& van Ballegooijen (2005).
In the photosphere, chromosphere, and low corona (i.e.,
from $z=0$ to 12 Mm), this model was obtained by tracing the radial
magnetic field strength from the central axis of a two-dimensional
numerical model of the supergranular network.
This model contains thin intergranular flux tubes between the
photosphere and a ``merging height'' of 0.6 Mm where the flux
tubes have expanded laterally to the extent that the surrounding
field-free plasma disappears.
Above this height the merged network element undergoes further
funnel-like horizontal expansion to fill a supergranular canopy.
The field is directed completely vertically by the height of 12 Mm.
Above this height we applied a slightly modified version of the
empirically derived solar-minimum magnetic model of
Banaszkiewicz et al.\  (1998).
The radial dependence of $B_0$ is shown in Figure 5.

\begin{figure}
\epsscale{1.00}
\plotone{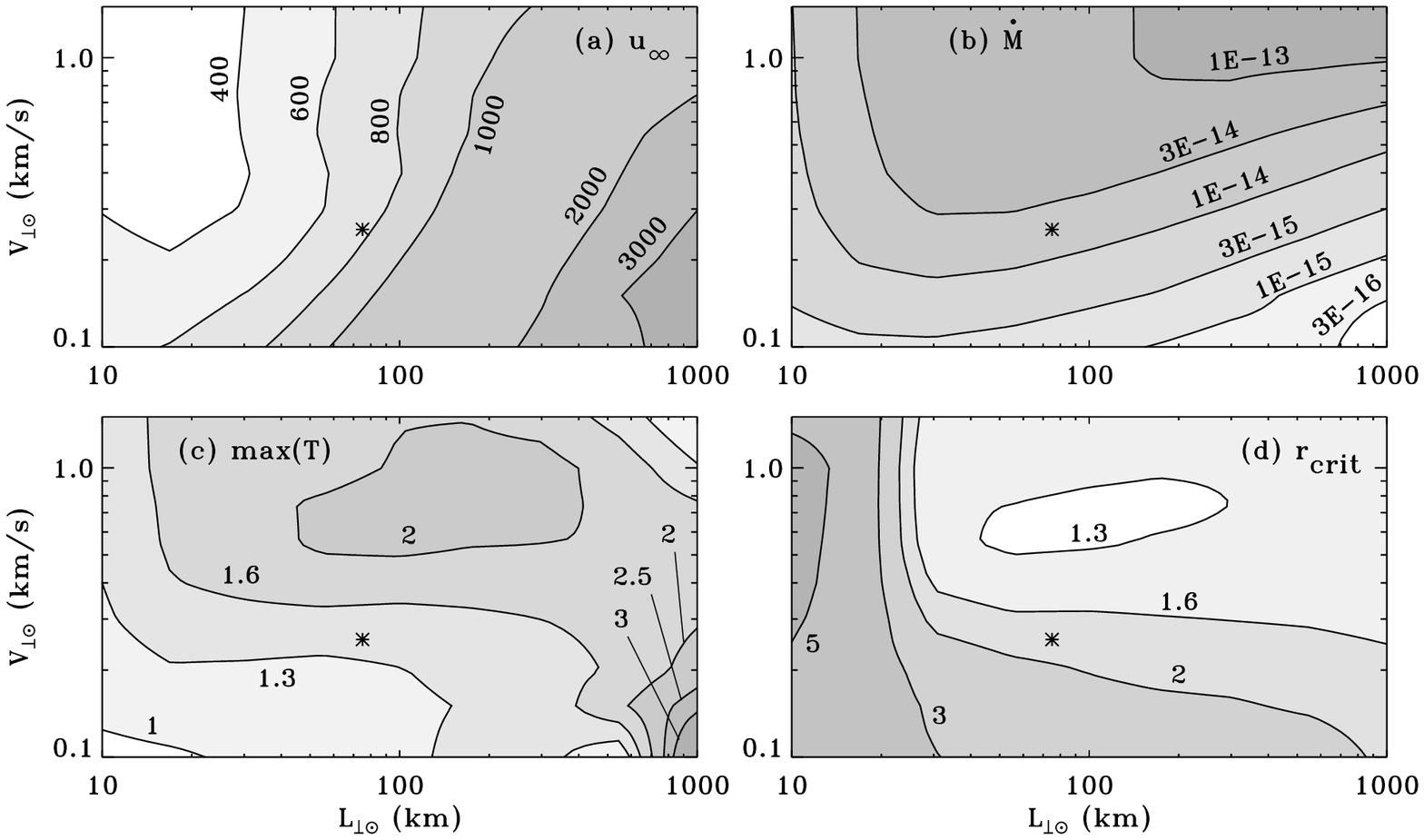}
\caption{
Contour plots of solar wind quantities resulting from varying the
coronal heating parameters $v_{\perp\odot}$ and $L_{\perp\odot}$:
({\em{a}}) terminal wind speed $u_{\infty}$ in units of km s$^{-1}$,
({\em{b}}) mass loss rate $\dot{M}$ in units of $M_{\odot}$ yr$^{-1}$,
({\em{c}}) maximum coronal temperature in units of MK,
({\em{d}}) heliocentric critical radius in units of $R_{\odot}$.
Also shown in each panel are the parameters chosen for the
model of fast wind from a polar coronal hole discussed in
{\S}~8.3 ({\em stars}). {\bf SEE LAST PAGE OF PAPER FOR LARGER VERSION.}}
\end{figure}

Figure 6 shows the result of producing a two-dimensional grid of
ZEPHYR models by varying $v_{\perp\odot}$ and $L_{\perp\odot}$.
The acoustic flux $F_{S \odot}$ was kept fixed at a median value of
$10^8$ erg s$^{-1}$ cm$^{-2}$ (see above).
The other two quantities were varied between the limits of
$0.1 \leq v_{\perp\odot} \leq 1.5$ km s$^{-1}$ and
$10 \leq L_{\perp\odot} \leq 1000$ km, with 9 points per quantity
spread logarithmically between those limits.
Figure 6a displays contours of the ``terminal'' outflow speed at
the upper edge of the spatial grid, which we call $u_{\infty}$.
This quantity is slightly larger than the outflow speed at 1 AU,
but never by more than 5\%.
Figure 6b shows the spherical mass loss rate, which is defined as
\begin{equation}
  \dot{M} \, = \, 4\pi \rho u r^{2} \,\, ,
\end{equation}
with the quantities on the right-hand side evaluated at the
top of the grid (where $A \propto r^{2}$).
Figures 6c and 6d show the maximum coronal temperature and the
heliocentric radius of the wave-modified critical point
(see {\S}~6), respectively.

Several general trends are evident in Figure 6.
The mass loss rate is primarily dependent on $v_{\perp\odot}$,
which represents the total amount of Alfv\'{e}nic wave energy
available to be dissipated in the corona.
The outflow speed, though, seems to depend mainly on
$L_{\perp\odot}$; this parameter tells us {\em where} the wave
energy is damped.
For large values of $L_{\perp\odot}$ the damping occurs over a
large range of heights, with increasingly more heating and
acceleration taking place above the critical point.
It has been known for some time that the relative heights of the
critical point and the dominant energy deposition are key in
determining the nature of a pressure-driven wind
(Leer \& Holzer 1980; Pneuman 1980; Leer et al.\  1982).
Heat that is deposited {\em above} the critical point is
converted nearly completely into kinetic energy of the wind
(and a higher value of $u_{\infty}$).
On the other hand, low values of $L_{\perp\odot}$ give a
more concentrated heat deposition mostly {\em below} the
critical point.
This energy raises the temperature in the subsonic part of the
corona, increases its scale height, and provides more downward
heat conduction into the upper transition region.
Less energy is thus available to accelerate the wind and
$u_{\infty}$ is lower.
The mass loss rate $\dot{M}$ is generally believed to be set
in the transition region by a balance between conduction,
radiative losses, and an upward enthalpy flux (e.g.,
Hammer 1982; Leer et al.\  1998).

Interestingly, for a large range of parameters in the lower-right
of Figure 6b ($L_{\perp\odot} \gtrsim 30$ km and
$v_{\perp\odot} \lesssim 0.4$ km/s), a combined power-law fit to
the parameter dependence of the mass loss rate yields good
agreement with the ZEPHYR models for
$\dot{M} \propto v_{\perp\odot}^{2.7}/L_{\perp\odot}$.
This resembles the classical Kolmogorov (1941) energy flux due
to isotropic hydrodynamic turbulence: $v^{3} / \ell$.
It seems to corroborate the general idea that the mass flux
of a thermally driven wind is proportional to the deposited
coronal energy flux.
The large variation in coronal heating, though, does {\em not}
seem to produce a large variation in the maximum coronal
temperature.
Figure 6c shows that this temperature varies only by about a
factor of two over most of the parameter space.
Owocki (2004) summarized how the energy losses due to both
conduction and the solar wind act as an effective ``thermostat''
to keep the coronal temperature from varying too widely.

\begin{figure}
\epsscale{1.00}
\plotone{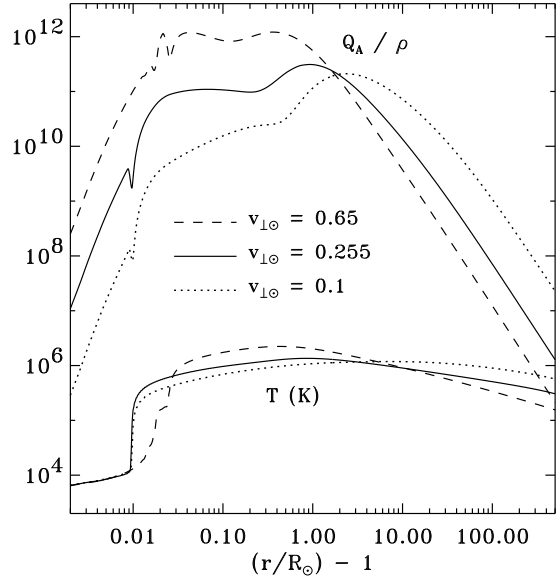}
\caption{
Coronal heating rates per unit mass ($Q_{A}/\rho$) in units
of erg s$^{-1}$ g$^{-1}$ and temperatures (in K) for three
models with constant values of $L_{\perp\odot} = 75$ km and
$F_{S \odot} = 10^{8}$ erg s$^{-1}$ cm$^{-2}$, and a range
of values for $v_{\perp\odot} = 0.1$ ({\em dotted lines}),
0.255 ({\em solid lines}), and 0.65 km s$^{-1}$
({\em dashed lines}).}
\end{figure}

Figure 7 illustrates how the coronal heating rate $Q_A$ changes
as $v_{\perp\odot}$ is varied and $L_{\perp\odot}$ is kept fixed.
We plot the heating rate per unit mass $Q_{A}/\rho$ in order to
more easily show which heights receive the most heating on a
particle-by-particle basis.
Note that larger values of $v_{\perp\odot}$ produce larger values
of both $Q_{A}/\rho$ and $T$ and move their local maxima
down to lower heights.
The presence of damping, though, leads to differences in how the
heating rate depends on $v_{\perp\odot}$ in various regions.
In the lower corona ($z \approx 0.05 \, R_{\odot}$), before
substantial Alfv\'{e}n-wave damping has had time to occur, the
three models show a power-law dependence of
$Q_{A}/\rho \propto v_{\perp\odot}^{2.58}$ that is similar to
the mass loss rate dependence of
$\dot{M} \propto v_{\perp\odot}^{2.11}$ for these models.
In the extended corona, though, the {\em peak value} of
$Q_{A}/\rho$ (above $z \approx 0.1 \, R_{\odot}$) varies more
weakly as $v_{\perp\odot}^{0.98}$.
The models with higher Alfv\'{e}n wave amplitudes have undergone
relatively more damping in the extended corona than the models
with lower $v_{\perp}$.

The above arguments do not explain all of the features of the
contours shown in Figure 6.
For example, although there is a general trend (in the center
and lower-right of Figure 6b) for a decreasing $L_{\perp\odot}$
to give a larger value of $\dot{M}$, the globally largest mass
loss rate occurs in the upper-right corner of the plot---i.e.,
for the largest values of {\em both} $v_{\perp\odot}$ and
$L_{\perp\odot}$.
In this region of the plot, added insight can be found in the
``cold'' wave-driven wind model of Holzer et al.\  (1983).
For undamped Alfv\'{e}n waves that dominate the critical
velocity (eq.\  [\ref{eq:ucrit}]) the terminal wind speed and
mass loss rate can be computed analytically.
Isolating the proportionality with $v_{\perp\odot}$ in
this model yields
\begin{equation}
  \dot{M} \propto v_{\perp\odot}^{4} \,\, ,
  \,\,\,\,\,
  u_{\infty}^{2} \propto (1 / v_{\perp\odot}^{2}) -
  \mbox{constant ,}
\end{equation}
where the constant term is typically negligible, thus giving
$u_{\infty} \propto v_{\perp\odot}^{-1}$
(see eqs.\  [39] and [41] of Holzer et al.\  1983).
In the upper-right corners of Figures 6a and 6b, these relations
come the closest to being satisfied.
Specifically, for $v_{\perp\odot} > 0.6$ km s$^{-1}$ and
$L_{\perp\odot} > 600$ km, the best power-law fits to the
model results are
$\dot{M} \propto v_{\perp\odot}^{3.42}$ and
$u_{\infty} \propto v_{\perp\odot}^{-0.89}$.
These parameters indeed correspond to a wave-dominated
critical velocity (i.e., large $v_{\perp\odot}$) without much
damping (large $L_{\perp\odot}$).

For the entire grid of values shown in Figure 6, a power-law
is a poor fit to the overall variations of $\dot{M}$ and
$u_{\infty}$ as a function of the two input parameters.
For completeness, though, we report the best-fitting exponents over
the full grid:
$\dot{M} \propto v_{\perp\odot}^{1.7} L_{\perp\odot}^{-0.07}$
and $u_{\infty} \propto v_{\perp\odot}^{-0.27} L_{\perp\odot}^{0.44}$.

\subsection{Chromospheric Parameter Study}

We produced a second grid of exploratory ZEPHYR models that kept
the coronal heating parameters ($v_{\perp\odot}$, $L_{\perp\odot}$)
fixed and varied only the basal acoustic wave flux $F_{S \odot}$.
As described further in {\S}~8.3, we chose optimal values of
$v_{\perp\odot} = 0.255$ km s$^{-1}$ and
$L_{\perp\odot} = 75$ km in order to model the fast wind that
emerges from a polar coronal hole (see also the stars in Figure 6).
For the median value of $F_{S \odot} = 10^{8}$
erg s$^{-1}$ cm$^{-2}$ used above, the transition region occurs
at a relatively large height of $z = 7200 \, \mbox{km}
\approx 0.01 \, R_{\odot}$.
A traditional view of chromospheric energy balance is that when
the acoustic heating is increased, the height of the transition
region should decrease (since the temperature would rise
more rapidly as a function of $z$).

\begin{figure}
\epsscale{1.00}
\plotone{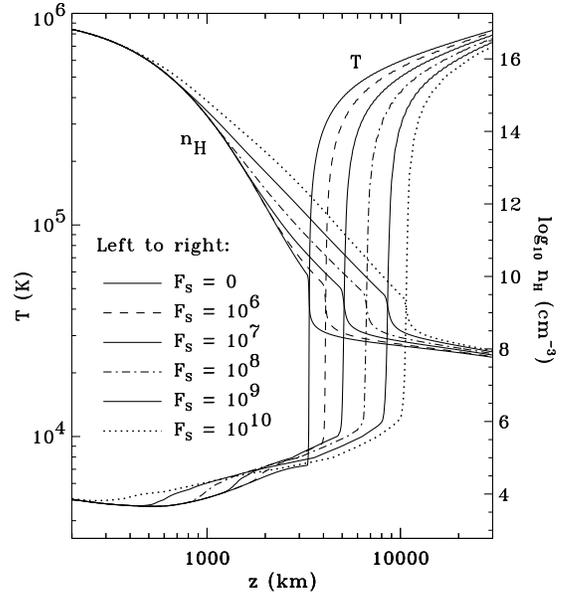}
\caption{
Temperature ({\em left axis}) and hydrogen number density
({\em right axis}) shown as a function of height above the
photosphere (in km) for models that vary the acoustic wave flux
$F_{S \odot}$ (in units of erg s$^{-1}$ cm$^{-2}$) and keep
the coronal heating parameters $v_{\perp\odot}$ and
$L_{\perp\odot}$ fixed (see captions).}
\end{figure}

Figure 8 shows that the {\em opposite} actually occurs when we
vary $F_{S \odot}$ and keep everything else fixed.
We produced a series of models with $F_{S \odot} = 10^{5}$,
$10^6$, $10^7$, $10^8$, $10^9$, and
$10^{10}$ erg s$^{-1}$ cm$^{-2}$, as well as a model with
$F_{S \odot} = 0$ (i.e., a model with Alfv\'{e}n wave heating only).
The models with $F_{S \odot} = 0$ and $10^5$ are virtually
identical and the latter is not shown.

Before discussing the chromospheric heating, we first note that
the coronal and solar wind parameters for these models are
remarkably constant.
This shows that varying the acoustic heating has relatively little
impact above the transition region.
As $F_{S \odot}$ is increased from zero to its maximum value, the
mass loss rate decreases by only 7\%, the terminal speed
increases by less than 1\%, and the peak coronal temperature
decreases by 5\%.

Why does the transition region height increase as additional
chromospheric heating is imposed?
The models having larger values of $F_{S \odot}$ have {\em larger
scale heights} because they have both larger chromospheric
temperatures and higher acoustic wave pressures.
The larger scale heights lead to a more shallow density
decrease in the approximately hydrostatic chromosphere.
The transition region tends to occur at a critical density
at which radiative cooling can no longer keep pace with the
imposed acoustic heating (i.e., because $\Lambda(T)$ has reached
its peak and can increase no further).
The ``shallower'' models with more acoustic heating
reach this critical density at a larger height.

The modeled transition region heights $z_{\rm TR}$ range from
3800 km (for $F_{S \odot} = 0$) to 11500 km
(for $F_{S \odot} = 10^{10}$ erg s$^{-1}$ cm$^{-2}$).
For consistency, we define $z_{\rm TR}$ as the height where
$T$ first reaches $2 \times 10^{5}$ K, which is the peak of the
radiative cooling curve; see Figure 1.
These heights are conspicuously larger than the standard values
of, e.g., 1700 to 2300 km from ``FAL'' semi-empirical solar
atmosphere models (Fontenla et al.\  1993).
Although there are some solar limb observations that suggest
large values similar to our modeled $z_{\rm TR}$
(Zhang et al.\  1998), these may be affected by the dynamics
of spicules and mass flows along closed magnetic loops, and
thus may not be comparable directly to the present models (see
also Filippov \& Koutchmy 2000).

The relative stretching of the chromospheres shown in Figure 8
can be understood further by examining the density dependence
of the total heating rate: $(Q_{S} + Q_{A}) \propto \rho^{\eta}$.
The exponent $\eta$ was computed for all of the models by
utilizing the radial dependence of both $\rho$ and the total
heating rate, with
\begin{equation}
  \eta \, = \, \frac{\partial \ln (Q_{S} + Q_{A}) / \partial r}
  {\partial \ln \rho / \partial r} \,\, .
\end{equation}
Because the heating is balanced by radiative cooling (which
usually has $Q_{\rm rad} \propto \rho^{2}$), we find that values
of $\eta$ closer to 2 give rise to a more extended ``matching''
between heating and cooling at chromospheric temperatures---and
thus a larger overall extent of the chromosphere.
Thus, since $\eta < 2$ for all of the models shown below, we
can understand how larger [smaller] values of $\eta$ correspond
to a higher [lower] transition region height.
For the shock heating described in {\S}~4.1, the possible
values of $\eta$ range between --0.5 and $+$1.
The lower limit would occur for an undamped weak shock train, for
which $T \Delta S \propto v_{\parallel}^3$ (eq.\  [\ref{eq:wst}])
and $\rho v_{\parallel}^{2} = \mbox{constant}$.
The upper limit occurs when the shock train strengthens to
the point where $v_{\parallel}$ saturates to a nearly constant
value and thus $T \Delta S \approx \mbox{constant}$.
(The peak Mach number $M_1$ tends to saturate at values between
2 and 3 for the majority of the $F_{S \odot}$ models.)

The ZEPHYR models shown in Figure 8 exhibit intermediate values
of the density exponent $\eta$, with the value of the exponent
varying slightly as a function of height in the chromosphere.
The lowest value of $z_{\rm TR}$ corresponds to the
$F_{S \odot} = 0$ model and $\eta = 0.4$--0.5.
The highest value of $z_{\rm TR}$ corresponds to the
$F_{S \odot} = 10^{10}$ model and $\eta = 0.8$--1.
It is clear that, in these models, an even lower (FAL-like)
transition region height would require $\eta \lesssim 0$, and
thus would require the heating rate to increase with height.
Such heating may occur for the following reasons.
\begin{enumerate}
\item
As described above, if the acoustic wave energy were to act more
like a weak undamped shock train, $\eta$ would approach --0.5.
The emerging picture of the chromosphere as a confluence of
multiple acoustic sources (each expanding quasi-spherically from
small regions at or below the photosphere) may be consistent
with the need for more low-amplitude acoustic wave energy
(see, e.g., Ulmschneider et al.\  2005).
A collection of weak and incoherent acoustic wave packets may
also exhibit effectively larger frequencies because of
their random phases and constructive interference.
\item
There could be additional important sources of linear wave
dissipation in the low chromosphere, similar in form to
$\gamma_{\rm cond}$ (which gives $\eta \approx -1$ to --0.5
depending on the magnetic geometry).
For example, Goodman (2000, 2004) suggested that the resistive
dissipation of currents driven by MHD waves may provide enough
energy to heat the chromosphere.
\item
Similarly, it has been argued on the basis of multiple lines of
observational evidence that both solar and stellar chromospheres
may be dominated by magnetic and not acoustic heating (e.g.,
Judge \& Carpenter 1998; Judge et al.\  2003;
Bercik et al.\  2005).
If these mechanisms grow stronger as one moves from the lower
to the upper chromosphere (the latter being more strongly
ionized and magnetized), this could produce the radial increase
of the total heating rate that would be needed to push down the
transition region.
\end{enumerate}
However, we should note that Anderson \& Athay (1989a) determined
empirically that the chromospheric heating rate seems to be roughly
proportional to $\rho$ (and thus $\eta \approx 1$) throughout most
of the chromosphere.
The mechanisms included in the ZEPHYR models may thus be reasonable
without the need for additional physics.

We experimented with several other variations on the acoustic
heating in order to test the assumptions described in {\S}~4.
A test model was created with the power spectrum $P_{S} (\omega)$
divided into a finer mesh of discrete frequency bins: separated by
factors of $\sqrt{2}$ rather than 2.
This model, which had $F_{S \odot} = 10^{8}$ erg s$^{-1}$ cm$^{-2}$,
produced a nearly identical chromosphere to the standard model
with the same value of $F_{S \odot}$.
The fine-mesh model's transition region height $z_{\rm TR}$ was
about 10\% higher than that of the standard model because slightly
more power was given to the lower frequencies (which form shocks at
larger heights).
Presumably, using a coarser frequency spectrum would lead to a
lower value of $z_{\rm TR}$, but this would start to become
unfaithful to the modeled spectral shape (eq.\  [\ref{eq:PomegaS}]).
Another model was run with just a single frequency bin;
i.e., a monochromatic low-frequency wave train with
$\omega = \omega_{\rm ac}$.
This model was designed to explore what would happen if the
high-frequency tail in $P_{S} (\omega)$ were absent (e.g.,
Fossum \& Carlsson 2005, 2006).
This model exhibited $z_{\rm TR} = 14300$ km, about a factor of
two larger than the standard model, and would certainly require
some additional kind of chromospheric heating to produce a
realistically low value of $z_{\rm TR}$.

Note that for all of the above models the minimum temperature
in the upper photosphere never dropped below the radiative
equilibrium value of $T_{\rm rad} \approx 4500$ K.
There is observational evidence, though, for a lower minimum
temperature that could extend intermittently down to values
between 3000 to 4000 K (see model A of
Fontenla et al.\  1993; Carlsson \& Stein 1997; as well as
recent work by Ayres et al.\  2006; Fontenla et al.\  2006).
Additional sources of cooling that could be included in ZEPHYR
include molecular opacity and dust formation (for cooler stars)
as well as adiabatic expansion effects due to waves and shocks
that may {\em not} be confined to the modeled flux tube.

\subsection{Polar Coronal Hole Model}

The exploratory models discussed above led to a choice
for the optimal set of parameters which would reproduce the
observed properties of high-speed solar wind streams that emerge
from polar coronal holes (mainly at the minimum of the solar cycle).
These parameters are
$v_{\perp\odot} = 0.255$ km s$^{-1}$, $L_{\perp\odot} = 75$ km,
and $F_{S \odot} = 10^{8}$ erg s$^{-1}$ cm$^{-2}$.
The terminal speed $u_{\infty}$ and mass loss rate $\dot{M}$
computed for this model are 753.5 km s$^{-1}$ and
$1.88 \times 10^{-14} \, M_{\odot} \, \mbox{yr}^{-1}$,
respectively.
These values give a total hydrogen number density of
2.9 cm$^{-3}$ at $r = 1$ AU.

The semi-empirical Alfv\'{e}nic turbulence model of
Cranmer \& van Ballegooijen (2005) determined a value for
$L_{\perp}$ at the mid-chromosphere merging height of about
1100 km, which corresponds to a photospheric value
$L_{\perp\odot} \approx 320$ km.
The Cranmer \& van Ballegooijen (2005) model also predicted a
photospheric non-WKB amplitude $w_{\perp\odot}$ of about
3.1 km s$^{-1}$, which translates into
$v_{\perp\odot} \approx 0.46$ km s$^{-1}$.
These values are both slightly higher than the ones chosen on the
basis of the self-consistent ZEPHYR models, though they certainly
fall within the range of plausibility.
The smaller value of the turbulence correlation length
(75 km) seems to be consistent with the observed horizontal size
of a thin flux tube in the photosphere (50--100 km).
This seems nicely consistent with being an outer
``stirring'' scale for the turbulence, since the horizontal
shaking and distortion of the flux tube can be expected to
take place mainly on the spatial scale of its own size.
(Earlier justifications for the larger [$\sim$ 300 km] correlation
length were based on this being either the mean separation
between flux tubes or the intergranular lane width.
These scales may be excited by convective driving as well,
but it makes sense for the primary {\em response} of each tube
to be on the smaller scale of its radius or diameter.)

For simplicity we continue to use the fiducial value for
$F_{S \odot}$ that was used in the two-dimensional grid shown
in Figure 6.
This energy flux density ($10^{8}$ erg s$^{-1}$ cm$^{-2}$)
corresponds to a photospheric acoustic wave amplitude
$v_{\parallel \odot}$ of about 0.29 km s$^{-1}$, and it is close
to that computed by Musielak et al.\  (2000) for linear
longitudinal flux-tube waves.
The flux computed in the Lighthill-Stein sound generation models
depends sensitively on the partitioning of gas and magnetic
pressure in the photosphere;
these models have $F_{S \odot} \propto (B_{0} / B_{\rm eq})^{-p}$,
where $B_{\rm eq} = (8 \pi P)^{1/2}$ is the field strength
consistent with equipartition between gas and magnetic pressures,
and $p$ takes on large values typically between 6 and 9.
Our polar coronal hole model has a photospheric ratio
$B_{0} / B_{\rm eq} = 0.77$.
Interpolating from Table 1 of Musielak et al.\  (2000) would yield
a flux for this model of about $1.1 \times 10^{8}$
erg s$^{-1}$ cm$^{-2}$, which is extremely close to what we use.

\begin{figure}
\epsscale{1.00}
\plotone{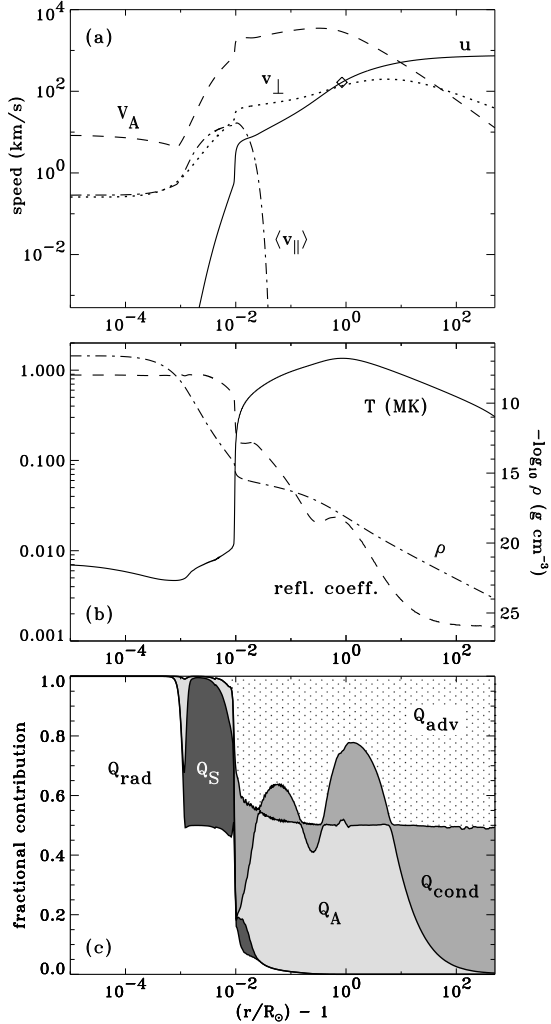}
\caption{
Radial dependence of various plasma parameters for a polar
coronal hole.
({\em{a}}) Velocity quantities in km s$^{-1}$:
wind speed $u$ ({\em solid line}), Alfv\'{e}n speed $V_A$
({\em dashed line}), Alfv\'{e}n wave amplitude $v_{\perp}$
({\em dotted line}), acoustic wave amplitude
$\langle v_{\parallel} \rangle$ ({\em dot-dashed line}),
and the wave-modified critical point ({\em diamond symbol}).
({\em{b}}) Temperature $T$ in MK ({\em solid line}),
spectrum-averaged reflection coefficient ${\cal R}$
({\em dashed line}), and mass density $\rho$
({\em dot-dashed line; right axis}).
({\em{c}}) Areas denote the relative contribution of terms to
the energy conservation equation; see labels.
($Q_{\rm adv}$ denotes the advection terms on the left-hand
side of equation (\ref{eq:dEdt}), other terms are defined
in the text.)}
\end{figure}

Figure 9 displays the radial dependence of several key plasma
parameters for the polar coronal hole model.
The critical point is denoted by a symbol, with
$r_{c} = 1.84 \, R_{\odot}$ and $u_{c} = 166$ km s$^{-1}$.
The wave-modified critical point is only
a small distance above the classical sonic point, which occurs
at 1.74 $R_{\odot}$ (where $u = a = 149$ km s$^{-1}$).
This is somewhat surprising because the fast solar wind is
typically assumed to be strongly ``wave dominated,'' and thus
one might have expected the modified critical point to be far
from the unmodified sonic point.
Both of these points are also near the so-called {\em turbopause}
radius at which the bulk solar wind speed begins to exceed the
turbulent fluctuation amplitude $v_{\perp}$
($r = 1.54 \, R_{\odot}$; see Veselovsky 2001).
In contrast, the Alfv\'{e}nic singular point (where $u = V_A$)
occurs at a substantially larger radius of
$r_{A} = 10.8 \, R_{\odot}$ and $u_{A} = 509$ km s$^{-1}$.
Note that Figure 9a displays a frequency-integrated acoustic
wave amplitude that we define for convenience as
\begin{equation}
  \langle v_{\parallel} \rangle \, = \, \left(
  \frac{2}{\rho} \sum_{\rm bins} U_{S} \right)^{1/2}
\end{equation}
(see eq.~[\ref{eq:Usdef}]), where we summed over the discrete
frequency bins discussed in {\S}~4.2 and assumed $s=2$ everywhere.
This acoustic wave amplitude peaks at about 17 km s$^{-1}$ in
the upper chromosphere (and is mildly nonlinear, with
$\langle v_{\parallel} \rangle / c_{s}$ peaking there at 0.96),
then is damped strongly in the low corona.

The temperature and density curves shown in Figure 9b appear to
be in reasonable agreement with prior expectations for the fast
solar wind (e.g., Kohl et al.\  2006).
When $\rho$ is converted into electron number density $n_e$, the
modeled values agree with the canonical polar coronal hole
measurements of Sittler \& Guhathakurta (1999) to within
$\pm$ 20\%.
These measurements come mainly from white-light polarization
brightness ($pB$) observations between $r = 1.5$ and 4 $R_{\odot}$.
Considering that the absolute $pB$ calibration uncertainties
are typically about 10\% and that the measured density depends
on the filling factor of polar plumes along the line of sight
(which introduces $pB$ variations up to a factor of two
in magnitude; see Fisher \& Guhathakurta 1995), this agreement
is good.

The Alfv\'{e}n wave amplitudes ($v_{\perp}$ in Figure 9a and
$B_{\perp}$ in Figure 5) compare favorably with the semi-empirical
models of Cranmer \& van Ballegooijen (2005), who compared those
models with various remote-sensing and in~situ measurements.
The spectrum-averaged reflection coefficient ${\cal R}$ shown
in Figure 9b is also similar to that of
Cranmer \& van Ballegooijen (2005); note that it need not be
monotonically decreasing with increasing distance.
The photospheric value of ${\cal R}$ in this model is 0.883,
which is smaller than the value of 0.974 from
Cranmer \& van Ballegooijen (2005).
The latter model utilized the FAL-C$'$ temperature structure
which had a sharper transition region than that produced by the
ZEPHYR code.
A more abrupt transition region has a larger
$V_A$ gradient and thus experiences stronger reflection.
Defining the transition region thickness $\delta z_{\rm TR}$
as the distance between temperatures of $2 \times 10^4$ and
$2 \times 10^5$ K, the FAL-C$'$ model has
$\delta z_{\rm TR} = 120$ km and the ZEPHYR polar coronal hole
model has a value of 560 km.
Although some of the computed thickness may be due to numerical
smoothing (see {\S}~7) not all of it is.
The ZEPHYR code did produce a sharper transition region
($\delta z_{\rm TR} = 270$ km) for the $F_{S \odot} = 0$ model
shown in Figure 8---which was also the model with the lowest
transition region height $z_{\rm TR}$.

The radial dependence of temperature in the extended corona and
heliosphere can be compared with various analytic limiting cases
(e.g., Hundhausen 1972).
At large distances where the wind speed is approximately
constant and $\rho \propto r^{-2}$, there are several cases
that give a power-law dependence $T(r) \propto r^{-\beta}$.
The polar coronal hole model shown in Figure 9 exhibits
$\beta \approx 0.30$ at 1 AU.
If the advection terms on the left-hand side of
equation (\ref{eq:dEdt}) are dominant, one obtains the
purely adiabatic exponent $\beta = 4/3$.
If classical heat conduction were to dominate the energy equation
out to 1 AU, $\beta = 2/7$.
However, we find that conduction and advection tend to
{\em balance} one another in the heliosphere and yield
intermediate values.
For the case that the advection terms are balanced by classical
Spitzer-H\"{a}rm conduction, there is no exact power-law
solution for $T(r)$.
For advection being balanced by collisionless conduction
(eq.\  [\ref{eq:qFS}]), one obtains
$\beta = 8 / (6 + 3 \alpha_{c})$, which for the adopted value
$\alpha_{c} = 4$ gives $\beta = 4/9$.
In the ZEPHYR models, the hybrid heat conduction
(eq.\  [\ref{eq:qpara}]) is mostly collisionless at 1 AU, but
it still contains some classical heat conduction.
The computed values of $\beta$ thus tend to fall between
the two above cases of 2/7 and 4/9.

Figure 9c shows the terms that dominate the energy conservation
equation as a function of distance.
The areas plotted here were computed by normalizing the absolute
values of the individual terms in equation (\ref{eq:dEdt}) by the
maximum value at each height (as in the denominator of
eq.\  [\ref{eq:delE}]) and then ``stacking'' them so that together
they fill the region between 0 and 1.
The photosphere is essentially definable as the region
where radiative heating balances radiative cooling (both
denoted as $Q_{\rm rad}$) and the chromosphere is the region
where acoustic heating ($Q_S$) balances $Q_{\rm rad}$.
The transition region and the very low corona
($T \lesssim 0.5$ MK) exhibit a complicated balance
of radiation, conduction, advection (i.e., enthalpy flux),
and some Alfv\'{e}nic and acoustic heating.
The extended corona is mainly a balance between the Alfv\'{e}n
wave heating ($Q_A$) and the advection terms on the left-hand
side of equation (\ref{eq:dEdt}), although conduction remains
nonnegligible.
In the heliosphere above 20--30 $R_{\odot}$ the direct heating
becomes less important and advection balances conduction.

In order to produce further comparisons with in~situ
solar wind measurements, we computed the nonequilibrium ionization
balance of oxygen as a function of distance in the ZEPHYR models.
Specifically, the ratio of number densities of O$^{7+}$ to
O$^{6+}$ is often used to aid in the identification of fast and
slow wind streams (e.g., Zurbuchen et al.\  2002).
Because of the steep decline in electron density with increasing
height, solar wind ions above a certain ``freezing-in radius''
encounter virtually no electrons, and thus are not sensitive to
ionization and recombination processes in interplanetary space
(Hundhausen et al.\  1968; Owocki et al.\  1983).
Interplanetary charge states thus carry information about the
plasma properties in the corona.
We adopted the nonequilibrium ionization code of
Gaetz et al.\  (1988), Esser et al.\  (1998), and
Esser \& Edgar (2000, 2001) to the ZEPHYR models at temperatures
above $10^4$ K.
Ionization and recombination are most sensitive to the electron
velocity distribution in the corona, and we relied on our basic
one-fluid assumption of a Maxwellian distribution with
$T = T_{e} = T_{p}$.
We also assumed that all oxygen ions flow with the bulk
wind speed $u$, independent of their charge.

\begin{figure}
\epsscale{1.00}
\plotone{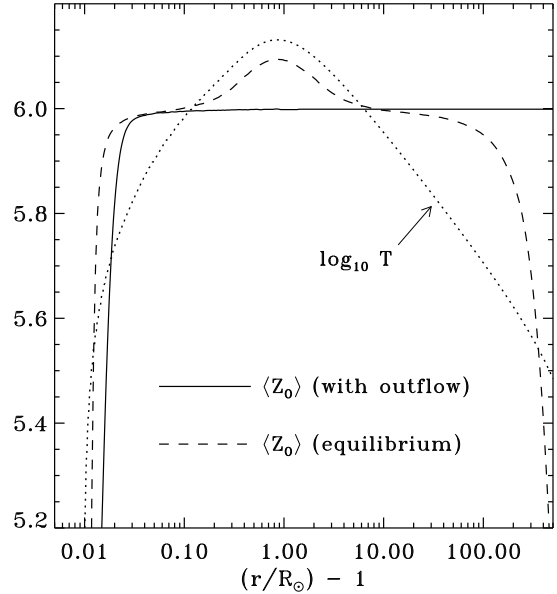}
\caption{
Oxygen ionization versus radial distance for the polar coronal
hole model.
The mean charge states for the nonequilibrium ``frozen in'' model
({\em solid line}) and the local coronal equilibrium
({\em dashed line}) are compared with the logarithm of the
temperature $T$, in K ({\em dotted line}).}
\end{figure}

Figure 10 shows the result of computing the oxygen ionization
state for the polar coronal hole model.
The mean charge state $\langle Z_{O} \rangle$ is computed as an
average of the net charge of the ions (in units of $e$) weighted by
the computed number density fractions of each stage of ionization.
The equilibrium solution assumes a local ``coronal'' balance
between collisional ionization, radiative recombination, and
dielectronic recombination, and is a strict function of
temperature.
The full nonequilibrium solution freezes in at a relatively low
height in the corona ($z \approx 0.05 \, R_{\odot}$) and 
is roughly constant at all larger heights.
Note, though, that the equilibrium and nonequilibrium solutions
are also different from one another in the upper transition
region ($z \approx 0.01$--0.02 $R_{\odot}$) where the wind
flow time over a scale height is beginning to approach the
relevant ionization and recombination times.
The dominant ionization state at 1 AU is O$^{6+}$ (i.e.,
$\langle Z_{O} \rangle \approx 6$) and the ratio of
O$^{7+}$ to O$^{6+}$ is discussed further below.

\subsection{Solar Minimum Axisymmetric Field}

There is a definite empirical relationship between the solar wind
speed measured in~situ and the inferred lateral expansion
of magnetic flux tubes near the Sun (Levine et al.\  1977;
Wang \& Sheeley 1990; Arge \& Pizzo 2000).
Specifically, flux tubes that expand more rapidly between the
coronal base and $r \approx 2.5 \, R_{\odot}$ tend to have
lower wind speeds at 1 AU.
There have been many theoretical attempts to explain this
observed anticorrelation.
Some models have treated the flux expansion as the fundamental
cause of the difference between fast and slow streams (e.g.,
Kovalenko 1978, 1981; Wang \& Sheeley 1991, 2006; Wang 1993;
Cranmer 2005a), some have treated it as a by-product of
varying boundary conditions in the low corona (e.g., Fisk 2003;
Schwadron \& McComas 2003), and some have considered a combination
of the two (Bravo \& Stewart 1997; Suzuki 2006).

Our goal is to test these ideas by varying {\em only} the
flux expansion rate and keeping the lower boundary parameters
($v_{\perp\odot}$, $L_{\perp\odot}$, $F_{S \odot}$) fixed.
The first way that we modify the flux expansion is to utilize the
full two-dimensional axisymmetric magnetic field model of
Banaszkiewicz et al.\  (1998).
This model utilizes a sum of dipole, quadrupole, and
current-sheet (effective monopole) terms to reproduce various
observed properties of the solar-minimum field.
Note that other assumptions lead to slightly different
solar-minimum field configurations (e.g.,
Sittler \& Guhathakurta 1999, 2002; V\'{a}squez et al.\  2003)
but the general trends of the polar and equatorial expansion
factors are not substantially different from the
Banaszkiewicz et al.\  (1998) case.

\begin{figure}
\epsscale{1.00}
\plotone{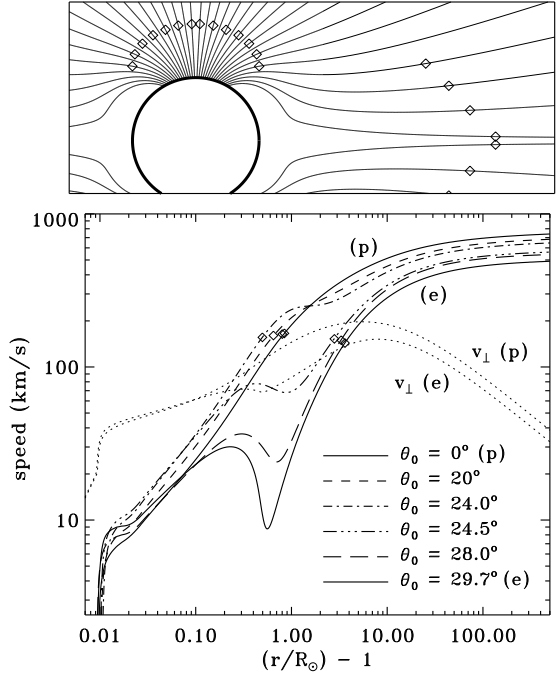}
\caption{
{\em Upper panel:} axisymmetric magnetic field geometry of
Banaszkiewicz et al.\  (1998) with the radii of wave-modified
critical points marked by open diamond symbols.
{\em Lower panel:}
outflow speeds for a series of axisymmetric flux tube models
(surface colatitudes $\theta_0$ given in captions) with critical
point heights denoted by open diamonds, and the Alfv\'{e}n wave
amplitude $v_{\perp}$ for the polar (p) and equatorial (e)
models ({\em dotted lines}).}
\end{figure}

Figure 11 illustrates a selection of the open field lines in the
Banaszkiewicz et al.\  (1998) model.
Figure 5 also shows the magnetic field $B_{0}(r)$ traced along
a subset of these field lines.
We define the surface colatitude $\theta_0$ as the main
identifier of each flux tube.
The polar flux tube that was used above in {\S}~8.3 has
$\theta_{0} = 0$.
We take the ``last'' open field line (at the outermost edge of the
open-field region) as $\theta_{0} = 29.7\arcdeg$.
Because the latter field line eventually stretches to a colatitude
of nearly {90\arcdeg} we also call this the ``equatorial'' flux
tube.
We assume that the differences in $B_0$ between flux tubes occur
only above $z \approx 0.01 \, R_{\odot}$ and that in the photosphere
and chromosphere all flux tubes are identical (see, e.g.,
Aiouaz \& Rast 2006).

Let us define an effective Wang \& Sheeley (1990) superradial
expansion factor for each flux tube as
\begin{equation}
  f_{\rm ss} \, = \, \frac{(B_{0} r^{2})_{\rm base}}
  {(B_{0} r^{2})_{\rm ss}}
  \label{eq:fss}
\end{equation}
where the coronal base is fixed at $r_{\rm base} = 1.04 \, R_{\odot}$
(i.e., a height equivalent to the size of one supergranule, well
above any ``canopy'' structure) and we use the traditional source
surface radius of $r_{\rm ss} = 2.5 \, R_{\odot}$
(Hoeksema \& Scherrer 1986).
It would not be proper to use the photosphere as the basal height
because the low-resolution magnetograms that typically are used
to determine $f_{\rm ss}$ resolve neither the individual
photospheric flux tubes nor the contrast between supergranular
network and cell interior.
For the Banaszkiewicz et al.\  (1998) field configuration,
$f_{\rm ss}$ ranges from 4.5 for the polar flux tube to 9.1
for the equatorial flux tube.
Note that this model does not produce the larger values of
$f_{\rm ss} \approx 20$--40 that have been found during more
active phases of the solar cycle (see {\S}~8.5).

A series of 20 ZEPHYR models was computed with the same photospheric
boundary conditions as the polar coronal hole model discussed
in {\S}~8.3, but varying $\theta_{0}$ from 0 to {29.7\arcdeg}.
Going from pole to equator, the asymptotic solar wind speed
$u_{\infty}$ was found to decrease and the mass loss rate
$\dot{M}$ was found to increase.
Figure 11 shows how the radial dependence of outflow speed changes
as $\theta_{0}$ increases from 0 to {29.7\arcdeg}.
There is an abrupt bifurcation between two classes of solar wind
solution: (1) polar solutions that have fast wind speeds, low
densities, and wave-modified critical points $r_{c} < 2 \, R_{\odot}$,
and (2) equatorial solutions that have slow wind speeds, high
densities, and $r_{c} \gtrsim 4 \, R_{\odot}$.
The colatitude at which this bifurcation occurs is
$\theta_{0} \approx 24.25\arcdeg$.
As discussed by V\'{a}squez et al.\  (2003) and in {\S}~6 above,
when there is more than one possible location for the critical
point, the most stable time-steady solution tends to be the one
with the largest value of $r_c$.
For the Banaszkiewicz et al.\  (1998) field geometry, the second
(large-$r_c$) solution appears only for
$\theta_{0} \gtrsim 24.25\arcdeg$.
The field line that divides the two classes of solutions
extends out to a heliospheric {\em latitude} (measured from the
ecliptic plane at $r \gg R_{\odot}$) of {16\arcdeg}.

Figure 11 shows that the solar wind solutions all exhibit
substantial wind speeds just above the transition region
($u \approx 10$ km s$^{-1}$).
Esser et al.\  (2005) found that such rapid flows at the
coronal base are a consequence of the rapid superradial
divergence in supergranular funnels, and that these flows seem
to be necessary to explain observed H~I Ly$\alpha$ emission
from the solar disk.
Note also that our modeled flux tubes nearest to the equator
exhibit a pronounced local minimum in the wind speed at the
cusp-radius of the current sheet in the
Banaszkiewicz et al.\  (1998) model.
Observational evidence for this kind of ``stagnation point'' in
equatorial streamers was reported by Strachan et al.\  (2002) and
was also modeled in a similar way as in this paper by
Chen et al.\  (2004) and Li et al.\  (2004).

It is also evident from Figure 11 that the polar model has a
larger Alfv\'{e}n wave amplitude than the equatorial model.
Some of this difference can be attributed directly to the varying
flux-tube area factors $A(r)$ (i.e., different manifestations of
wave action conservation).
However, the equatorial flux tube also exhibits relatively more
damping than the polar model.
In the corona, the Alfv\'{e}n waves essentially ``spend more time''
at low heights in the equatorial flux tube because of its lower
phase speed ($u + V_A$), and this allows a given damping rate
$Q_A$ to have more of an impact on diminishing the wave energy.
Thus, the equatorial models exhibit more heating at low heights
(due to the increased damping and the higher densities at the
coronal base) and less heating at large heights (due to the
lower wave amplitudes $v_{\perp}$ and larger turbulent correlation
lengths $L_{\perp}$ in the extended corona).
The peak temperature for the polar model, $T = 1.35$ MK, occurs
at $r = 1.86 \, R_{\odot}$, and the peak temperature for the
equatorial model, $T = 1.29$ MK, has a similar magnitude but
occurs at a lower height of $r = 1.26 \, R_{\odot}$ (see also
Figure 17 below).
This shift in the range of heights over which coronal heating
occurs seems to be an important factor in producing a fast
solar wind for the polar model (which is heated in the supersonic
region) and a slow wind for the equatorial model (which is
heated in the subsonic region); see {\S}~8.1.
The differences in temperature persist out to 1 AU, where
the polar temperature (0.41 MK) exceeds the equatorial
temperature (0.16 MK) by a larger relative amount than in
the extended corona.
The radial temperature exponent $\beta$ (i.e.,
$T \propto r^{-\beta}$) grows steeper from pole ($\beta \approx 0.30$)
to equator ($\beta \approx 0.47$).

Although the photospheric Alfv\'{e}n wave amplitude parameter
$v_{\perp\odot}$ was kept fixed in these models, the degree of
non-WKB reflection decreased slightly from the polar to the
equatorial flux tube.
This resulted in a decrease in the computed value of
$w_{\perp\odot}$ (eq.\  [\ref{eq:wperp}]) from its polar value
of 0.72 km s$^{-1}$ to an equatorial value of 0.59 km s$^{-1}$.
Interestingly, if we could have forced $w_{\perp\odot}$ to
remain fixed as a function of latitude, the resulting equatorial
value of $v_{\perp\odot}$ would have been about 20\% {\em larger}
than the polar value.
Figure 6 indicates that the effect would have been an even
larger equatorial mass loss rate and a slightly lower $u_{\infty}$.
Thus, the current set of models is in some sense ``robust''
because these effects arise even without such an extra
fine-tuning in $w_{\perp\odot}$.

\begin{figure}
\epsscale{1.00}
\plotone{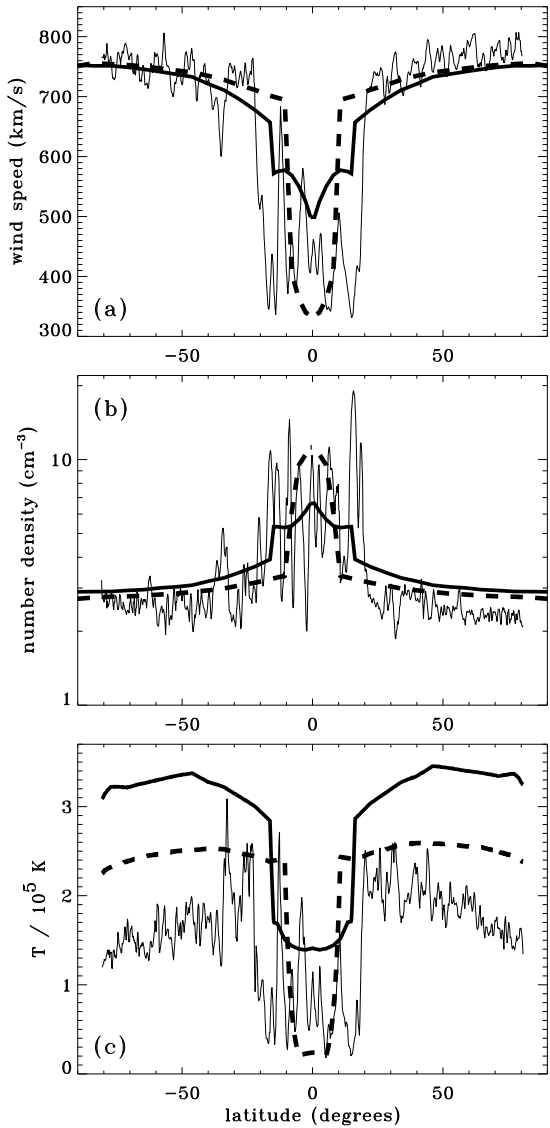}
\caption{
Latitudinal dependence of
({\em{a}}) outflow speed, ({\em{b}}) number density, and
({\em{c}}) temperature measured in interplanetary space
(see text for details).
Data from the {\em Ulysses} first polar pass ({\em thin solid lines})
are compared to the standard ZEPHYR models computed along axisymmetric
superradial flux tubes ({\em thick solid lines}) and the `Durham'
ZEPHYR models computed along the same flux tubes
({\em thick dashed lines}).}
\end{figure}

The detailed latitudinal dependence of the ZEPHYR models (for
an observer in interplanetary space) is shown in Figure 12.
We compare the model predictions with {\em Ulysses} SWOOPS
(Solar Wind Observations Over the Poles of the Sun;
Bame et al.\  1992) measurements during its first fast latitude
scan in 1994 and 1995 (see Goldstein et al.\  1996).
The measured values for flow speed, proton number density, and
proton temperature were obtained at heliocentric distances between
1.3 and 2.3 AU.
The densities were scaled to a common distance
of 1 AU by using an assumed $r^{-2}$ dependence.
For comparison we plot $u_{\infty}$, which does not vary
substantially between these distances, and the proton number density
($n_{p} = \rho/ m_{\rm H} / 1.2$) at 1 AU, scaled for an assumed
helium-to-hydrogen number density ratio of 0.05.
The model temperature $T(r)$ has been interpolated to varying
heights as a function of latitude to match the elliptical orbit
of the {\em Ulysses} probe during its polar pass.
Also, we used the Banaszkiewicz et al.\  (1998) model to map the
surface colatitude $\theta_0$ of each model flux tube to its
corresponding heliospheric latitude at $r \rightarrow \infty$.
The latter is the abscissa coordinate used in Figure 12.
To within about 10\% accuracy, one can estimate this latitude
as $| {90\arcdeg} - 3 \theta_{0} |$.

There are two sets of pole-to-equator ZEPHYR models shown in Figure 12.
We compare the standard model discussed above with a
preliminary model presented by Cranmer (2006), which we hereafter
call the ``Durham'' model.\footnote{%
This model was presented at the 37th meeting of the Solar Physics
Division of the American Astronomical Society in Durham, New Hampshire
(25--30 June 2006).}
The models differ in two specific ways.
(1) The standard model used the Alfv\'{e}n wave frequency spectrum
shown in Figure 3, but the Durham model assumed a single frequency
for the Alfv\'{e}n waves---corresponding to a period of 6 minutes.
(2) The standard model used an exponent $n=1$ in
equation (\ref{eq:Eturb}), but the Durham model used $n=2$.
The latter was an early guess for the functional form of
${\cal E}_{\rm turb}$ that we used prior to the incorporation of
the insights of Oughton et al.\  (2006).
The larger exponent resulted in more efficient quenching of the
MHD turbulence in the lower atmosphere, which thus required a
larger amount of Alfv\'{e}n wave energy to produce plasma conditions
appropriate for a polar coronal hole.
Specifically, the Durham model used $v_{\perp\odot} = 0.42$
km s$^{-1}$, $L_{\perp\odot} = 120$ km, and the same value of
$F_{S \odot}$ as was used in the standard model.
The Durham model's $u_{\infty}$ and $\dot{M}$ for the polar flux
tube were nearly identical to those of the standard model.
The peak coronal temperature, though, was substantially larger
(2.06 MK) and the temperature at 1 AU was slightly smaller (0.31 MK)
in comparison to the standard polar model.

There is reasonably good agreement between the modeled and
observed plasma properties shown in Figure 12.
We of course do not reproduce any of the measured north-south
asymmetry in the data, except for that in $T_p$ that arose because
the path of {\em Ulysses} was not perfectly north-south symmetric.
We also do not attempt to model the rapid near-ecliptic variability
that seems to result from longitudinal variations in $B_0$.
In some sense, the Durham model does better at achieving the
low equatorial wind speeds (down to 330 km s$^{-1}$) and
high number densities (up to about 10 cm$^{-3}$) that are often
observed in the plane of the ecliptic, but that model exhibits a
narrower range of slow-wind (large-$r_c$) latitudes than both the
standard model and the {\em Ulysses} data.
It is expected, though, that some fraction of the observed
$\pm 20\arcdeg$ latitudinal extent of the slow-wind region may be
due to a slight tilt in the main dipole component of the solar
magnetic field (i.e., a ``ballerina skirt'' phenomenon) because the
first {\em Ulysses} polar pass did not occur exactly at the
minimum of the activity cycle.

Note from Figure 12c that the ZEPHYR model temperatures between
1.3 and 2.3 AU are systematically higher than the measured values of
$T_p$.
The trend as a function of the coupled distance-latitude coordinate,
though, is similar to the measurements.
Our heliospheric temperatures are sensitive to the adopted method
of bridging between collisional and collisionless heat conduction
(eq.\  [\ref{eq:qpara}]) and the present method is admittedly
somewhat ad~hoc.
However, some of the discrepancy in Figure 12c may come from
the fact that $T_{p} \neq T_{e}$ in interplanetary space and that
we are essentially modeling their average.
In~situ measurements have shown that $T_{p} > T_{e}$ in
the fast solar wind and $T_{p} < T_{e}$ in the slow wind
(see, e.g., Feldman \& Marsch 1997).
The discrepancy is worst in the polar, fast-wind regions where
our modeled average temperature would exceed {\em both} the
observed $T_p$ and $T_e$.
However, the ZEPHYR code does not yet contain the dissipative
physical processes that would lead to temperature equilibration
between protons and electrons; these may alter subtly both the
net heating and the form of the conduction flux $q_{\parallel}$.

\begin{figure}
\epsscale{1.00}
\plotone{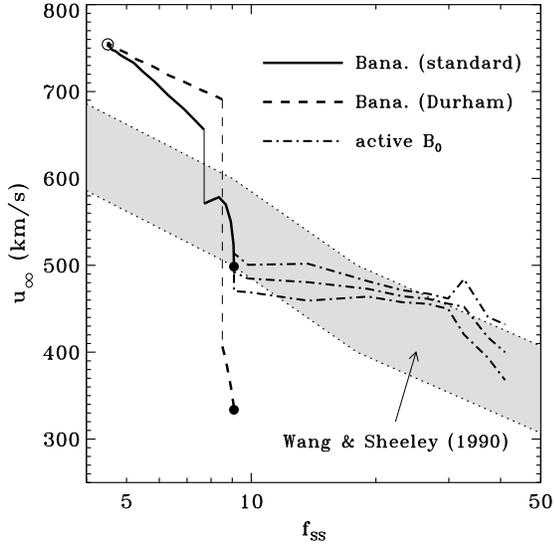}
\caption{
Terminal outflow speeds for the various sets of ZEPHYR models
plotted versus the flux-tube expansion factor $f_{\rm ss}$
(eq.\  [\ref{eq:fss}]), compared with the empirically
derived relationship from Table 2 of Wang \& Sheeley (1990).
The standard ({\em solid line}) and Durham ({\em dashed line})
pole-to-equator models computed using the
Banaszkiewicz magnetic field are shown along
with the active-region extensions to equatorial $B_{0}(r)$
discussed in {\S}~8.5 ({\em dot-dashed lines}).
The polar ({\em open circle}) and equatorial
({\em filled circles}) axisymmetric models are highlighted.}
\end{figure}

Figure 13 plots the terminal speeds of the standard and Durham
ZEPHYR models as a function of the Wang-Sheeley superradial
expansion factor $f_{\rm ss}$.
The abrupt changes in $u_{\infty}$ that occur because of the
emergence of the outer critical point are evident as vertical
discontinuities at the critical values of $\theta_0$.
Because of the relatively narrow range of $f_{\rm ss}$ values
subtended by the pole-to-equator (Banaszkiewicz et al.\  1998)
flux tubes, we see that the standard model actually follows the
the empirical Wang \& Sheeley (1990) relationship better than
the Durham model, despite the fact that the standard model
has a relatively high equatorial terminal speed of 500 km s$^{-1}$.

\begin{figure}
\epsscale{1.00}
\plotone{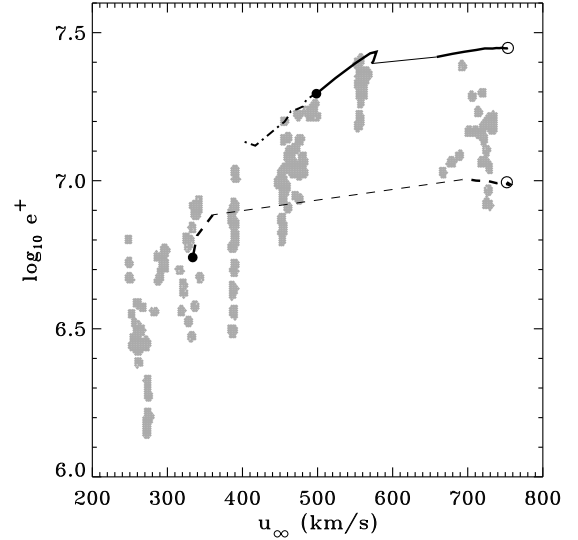}
\caption{
Comparison of {\em Helios} Elsasser spectra $e^{+}$ at
$r = 0.3$--0.5 AU with ZEPHYR model predictions at a
mean distance of $r = 0.4$ AU.
Data from Figure 3 of Tu et al.\  (1992) is shown
({\em gray points}).
Standard ({\em solid line}) and Durham ({\em dashed line})
pole-to-equator models are plotted together with active-region
flux tube models ({\em dot-dashed lines}) in a similar way as
in Figure 13.}
\end{figure}

Figure 14 compares modeled and measured Alfv\'{e}n wave
amplitudes in the heliosphere as a function of the in~situ
wind speed.
It is difficult to process the richly nonlinear turbulent
fluctuations (as measured in~situ) into a single ``amplitude,''
so we made an attempt to convert the ZEPHYR values of $v_{\perp}$
into something analogous to the frequency-averaged Elsasser
spectrum quantities reported by Tu et al.\  (1992).
Those data were obtained by the two {\em Helios} spacecraft
between 0.29 and 0.52 AU in the years 1979--1980.
(Measurements taken closest to the Sun are optimal if the goal
is to obtain the properties of the ``pristine'' Alfv\'{e}n waves.)
Individual time series of $Z_{-}^{2}/2$ (outward-propagating
kinetic energy) magnitudes for one-day periods were transformed
into power spectra, denoted $e^{+}(f)$.
The values shown in Figure 14 were averaged over frequencies
$f$ between $10^{-4}$ and $2 \times 10^{-4}$ Hz.
The plotted data points also correspond to time periods when the
solar wind exhibited its highest range of total energy flux
($F_{\rm tot} = \rho u [u^{2} + v_{\rm esc}^{2}]/2$), which
we verified to correspond most closely with the ZEPHYR model
results.

We extracted the Elsasser amplitudes $Z_{-}$ from the models
at a mean distance of $r = 0.4$ AU and converted them into the
average value of $e^{+}$ as follows.
First, representative spectral shapes for these fluctuations
were used in order to determine how much of the total energy
is contributed by the specified frequency band
(1--$2 \times 10^{-4}$ Hz).
We used the power-law exponents and breakpoints reported by
Tu et al.\  (1989), using the same {\em Helios} data, to
determine that fast and slow wind streams contain approximately
11\% and 21\%, respectively, of their power in this band.
We used a mean value of 16\% for all ZEPHYR models to avoid
introducing any possibly spurious correlations with wind speed.
To convert the (implicitly frequency-integrated) energy fraction
into an ``average'' spectrum quantity we divided by an effective
frequency $f_{\rm eff}$ that was defined to perform this conversion
exactly for the idealized spectra of Tu et al.\  (1989);
it turned out that $f_{\rm eff} \approx 9.5 \times 10^{-5}$ Hz
for both fast and slow wind spectra and we used just
$10^{-4}$ Hz.
Thus we plot the derived quantity
\begin{equation}
  e^{+} \, = \, \frac{0.16 \, (Z_{-}^{2}/2)}{10^{-4} \, \mbox{Hz}}
  \,\,\, \mbox{km}^{2} \, \mbox{s}^{-2} \, \mbox{Hz}^{-1}
\end{equation}
in Figure 14 in order to compare with the measured average
spectra from Tu et al.\  (1992).
The modeled and measured values---and trends with wind
speed---appear to agree quite nicely.

\begin{figure}
\epsscale{1.00}
\plotone{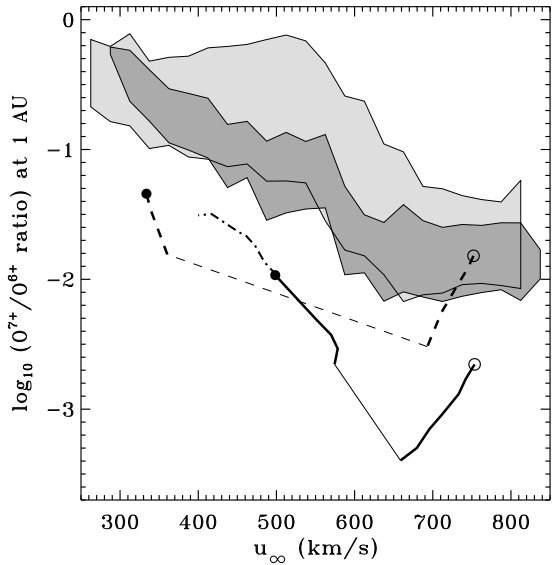}
\caption{
Oxygen freezing-in ratio (i.e., ratio of O$^{7+}$ to O$^{6+}$
number densities) at 1 AU plotted as a function of solar
wind speed in interplanetary space.
Symbols and line styles for the models are the same as
in Figure 14.
Binned {\em Ulysses} data from the 1990--1994 solar maximum
time period ({\em light gray region}) and the 1994--1995
fast latitude scan ({\em dark gray region}) are shown for
comparison.}
\end{figure}

In the remainder of this section we discuss two supplementary
calculations of in~situ quantities that are often used to
diagnose the properties of fast and slow wind streams: the
ionization fraction (Figure 15) and the FIP effect (Figure 16).

1. The {\em nonequilibrium oxygen ionization state} at 1 AU has
been computed for each of the standard and Durham model flux tubes.
Figure 15 shows the ratio of O$^{7+}$ to O$^{6+}$ number
densities as a function of the modeled terminal speed.
These are compared to statistical summaries of {\em Ulysses}
SWICS (Solar Wind Ion Composition Spectrometer;
Gloeckler et al.\  1992)
measurements of this ratio taken during two time periods:
the first polar pass discussed above (September 1994 to August
1995) and the initial phase of the mission that sampled
the peak of the solar cycle (December 1990 to September 1994).
The data points, obtained from the online database of
T.\  Zurbuchen and R.\  von Steiger (see also
von Steiger et al.\  2000), were grouped in sequential bins of
wind speed of size 25 km s$^{-1}$, and the mean and standard
deviation of the ionization ratio in each bin were computed.
The regions plotted in Figure 15 show the values within $\pm 1$
standard deviations of the mean for each bin in the two
data sets.

The general trend of the pole-to-equator ZEPHYR models is for
lower wind speeds to have higher ratios of O$^{7+}$ to O$^{6+}$
at 1 AU, in qualitative agreement with the observations.
The absolute values of the modeled ionization ratios, though,
are up to an order of magnitude lower than the {\em Ulysses}
measurements, with slightly better agreement for the hotter
Durham models.
This discrepancy may be the result of our simple assumption of
Maxwellian electron distributions and equal flow speeds for
all ion charge states (see, e.g., Esser \& Edgar 2000, 2001).
A ubiquitous weak ``halo'' of suprathermal electrons in the
low corona could be responsible for the larger overall degree
of ionization that is observed.

As one moves from the polar flux tube to the middle latitudes
(i.e., from $u_{\infty} \approx 750$ km s$^{-1}$ down to
about 650 km s$^{-1}$) the modeled ionization ratio decreases,
but then as one passes the bifurcation point between the low and
high critical radii, the ionization ratio increases for the rest
of the way from the mid-latitude region to the equator.
The overall trend for slow wind to have a high ionization ratio
is expected because of the higher temperatures in the narrow
zone just above the transition region.
In fact, the temperature at a constant height of
$z \approx 0.013 \, R_{\odot}$ has the same quantitative trend
as the modeled O$^{7+}$ to O$^{6+}$ ratios: there is an initial
decrease from the polar model (0.37 MK) to just before the
bifurcation latitude (0.28 MK), then an abrupt jump back to
nearly the polar value (0.34 MK) followed by a steady increase
toward the equatorial flux tube (0.46 MK).
The freezing in occurs at about this height, so it makes sense
that the ionization state tracks these variations.
Notice that this ionization ratio is sensitive only to the
temperature trends at low heights (and that the O$^{7+}$
to O$^{6+}$ ratio freezes in just above the transition region)
and is {\em not} sensitive to the temperatures at larger heights
in most of the extended corona.
Some interpretations of the in~situ charge states make
the implicit assumption that high electron temperatures in
the slow wind must persist all the way through the extended
corona into interplanetary space, and our models show that this
need not be the case.

\begin{figure}
\epsscale{1.00}
\plotone{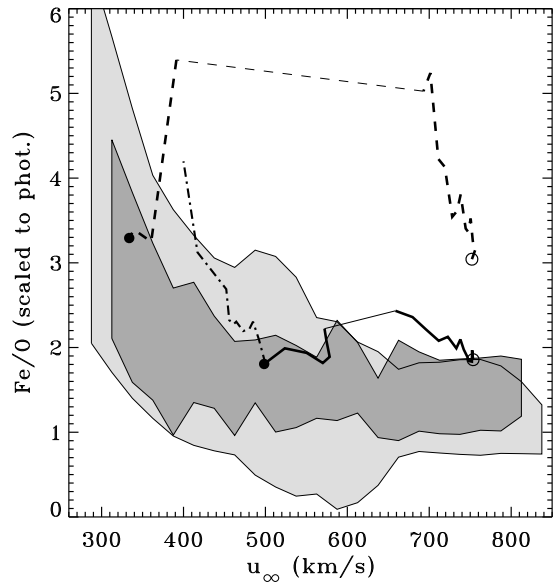}
\caption{
FIP fractionation ratio of Fe to O abundance in units of their
photospheric abundances.
Symbols, line styles, and grayscale regions are the same as
in Figure 15.}
\end{figure}

2. A specific diagnostic of {\em FIP fractionation} has been
computed for each of the standard and Durham model flux tubes.
Figure 16 shows a commonly measured ratio of the abundances of
a low-FIP element (Fe, 7.9 eV) and a moderately high-FIP element
(O, 13.6 eV) normalized to their photospheric abundance ratio.
We solved equation (\ref{eq:LamFIP}) as described in {\S}~6, and
by Laming (2004), for the various models and compared the ratios
to the {\em Ulysses} SWICS observations of the Fe/O ratio.
The SWICS ratio was normalized to an assumed photospheric
iron-to-oxygen abundance ratio of 0.0468 (Grevesse \& Sauval 1998)
and the modeled quantity is naturally in units of the
photospheric ratio.
Because the FIP fractionation takes place in the uppermost part
of the chromosphere (which is largely optically thin) and the
very low transition region, we assumed ``coronal'' ionization
equilibrium to hold for the modeled Fe and O ions.
We used the code developed by Cranmer (2000) to compute the 
Fe and O ionization fractions $\xi_s$ that are used in
equation (\ref{eq:LamFIP}); the rates were taken largely from
Mazzotta et al.\  (1998).

Figure 16 shows that our
standard pole-to-equator ZEPHYR model exhibits a roughly constant
Fe/O ratio of about 2 for outflow speeds between 500 and 750
km s$^{-1}$.
This is in reasonable agreement with the {\em Ulysses} data,
which have a roughly constant mean value of about 1.5 over
these speeds.
The Durham model varies nonmonotonically between ratios of about
3 and 5.5 over its wider range of outflow speeds and is clearly
inconsistent with most of the in~situ data.
The equatorial Durham model, with $u_{\infty} \approx 330$
km s$^{-1}$, does seem to fall into agreement with the
measured data, but the overall trend as a function of wind speed
is not the same.
The most promising models for matching the observed trend for
a steep enhancement of the Fe/O ratio (as the wind speed
decreases from 500 to 300 km s$^{-1}$) appear to be the
active-region flux tubes discussed below in {\S}~8.5.

We should note that an alternate theoretical explanation for
FIP fractionation may also be consistent with the ZEPHYR models.
Wang (1996) proposed that flux tubes with a stronger degree of
downward heat conduction in the low corona could give rise to a
larger amount of transient evaporative outflow of protons
relative to neutral hydrogen atoms.
These protons could collisionally dredge up singly ionized
species (which are preferentially low-FIP) from the transition
region and upper chromosphere.
In our models, the maximum value of $|Q_{\rm cond}|$ occurs within
the transition region (at $T \approx 10^{5}$ K), and in the
equatorial models this peak value is consistently {\em larger}
than in the polar models.
The ratio of equatorial to polar maximum conduction rates is
1.7 for the standard model and 2.8 for the Durham model.
Whether this is enough to produce a FIP effect with Wang's (1996)
mechanism is not known, but the variations are at least
in the correct qualitative direction to produce more of an
effect in the slow wind than in the fast wind.
Note, though, that Schwadron et al.\  (1999) disagree with
some of the basic assumptions of the Wang (1996) mechanism,
and a definitive explanation for the FIP effect remains elusive.

There are additional observational diagnostics that appear to be
correlated with fast and slow wind speeds.
Kojima et al.\  (2004) and Suzuki (2006) found that the ratio of
basal magnetic field strength to the superradial flux-tube
divergence factor (i.e., $B_{\rm base} / f_{\rm ss}$) may be a
better predictor of solar wind speed than just using
$f_{\rm ss}$ itself (see, however, Wang \& Sheeley 2006).
McIntosh \& Leamon (2005) found that the difference
in formation heights between the emission seen in the {\em TRACE}
1600 {\AA} and 1700 {\AA} filter bandpasses could be a useful
diagnostic of the eventual wind speed in flux tubes that are
associated with specific regions on the solar surface.
Although there have been numerous observational inferences of
blueshifts (i.e., outflow) in various spectral lines, it is
unclear to what degree these may be utilized as empirical
``lower boundary conditions'' for solar wind flux tubes (see
reviews by Peter et al.\  2004; Jones 2005).
Finally, there are definite fast/slow wind correlations in the
wealth of existing turbulence measurements---both in~situ
and remote sensing---that could be key tests of models that use
waves and turbulence for energy and momentum deposition.
Future development of the ZEPHYR models will encompass such
additional comparisons.

\subsection{Active Corona}

At times other than solar minimum, the coronal magnetic field is
substantially more complex than the axisymmetric configuration
modeled in the previous section.
During the approach to solar maximum the large polar coronal holes
disappear as magnetic flux of the opposite polarity is brought
toward the poles.
Active regions with strong fields emerge at all latitudes and
give rise to an interconnected web of closed loops and intermittent
open flux tubes.
In the extended corona, active regions often have the
appearance of cusp-like ``active streamers'' that seem like smaller,
more intense versions of the quiescent equatorial streamers that
dominate at solar minimum (Newkirk 1967; Liewer et al.\  2001;
Ko et al.\  2002).
There is increasing evidence that much of the slow solar wind is
connected somehow to open flux tubes that are rooted in or
near active regions (e.g., Neugebauer et al.\  2002;
Liewer et al.\  2004).

We investigated several different ways to model active-region
flux tubes with the ZEPHYR code.
First, we explored just changing the radial location of the cusp
in the equatorial Banaszkiewicz et al.\  (1998) flux tube.
Observationally, streamers exhibit a wide range of cusp heights
which gives rise to a large variation in their rates of
superradial expansion.
The cusp height was varied by first computing the ratio of the
equatorial ($\theta_{0} = 29.7\arcdeg$) to the polar
($\theta_{0} = 0$) magnetic fields $B_{0}(z)$, as shown in Figure 5,
then shifting the height coordinate $z$ either upward or downward
by a constant multiplier.
The ``new'' equatorial field was obtained by taking the product of
this shifted ratio and the polar field strength.
Changing the cusp height by up to a factor of two in either
direction did not have much of an effect on the resulting
solar wind properties.
The mass loss rates varied by about 18\% for these models (lower
cusps having lower $\dot{M}$), whereas the terminal speeds
$u_{\infty}$ varied by less than 3\%.
This near constancy in the solar wind speed occurred despite the
fact that $f_{\rm ss}$ was made to increase and decrease by factors
of $\sim$2 about the equatorial value of 9.1.

The above models cannot be considered true active-streamer
models because they were not given the significantly {\em stronger}
mean magnetic field strengths that active regions are observed to
have in the lower atmosphere.
On the smallest scales in the photosphere, plage and active regions
exhibit a much higher filling factor of strong-field flux
concentrations than quiet regions and coronal holes.
In the most tightly-packed active regions these flux concentrations
do not appear as isolated flux tubes but instead as fluted
ribbons, buckled flux sheets, and flower-like patterns (e.g.,
Berger et al. 2004; Rouppe van der Voort et al.\  2005).
We are concerned here only with the subset of magnetic flux that
extends out into interplanetary space, so we provisionally
retain the ``flux tube'' picture.
We thus modeled $B_{0}(r)$ in an active-region flux tube by
enhancing the field strength in the chromosphere and low corona.
This is done in general agreement with the larger observed magnetic
fields, but also to account for the smaller area $A(r)$ subtended
by open flux tubes that are ``crowded'' by neighboring closed
loops in active regions.

We added an active-field component to the standard equatorial
Banaszkiewicz et al.\  (1998) magnetic field by taking
\begin{equation}
  B_{0} (z) \, = \, \max \left[ B_{E} (z) \, ,
  \,\, B_{A} e^{-z/h} \right]
  \label{eq:Bactive}
\end{equation}
where $B_{E}(z)$ is the equatorial ($\theta_{0} = 29.7\arcdeg$)
model and $B_A$ is a constant that we set at 50 G.
We produced a grid of models that varied the scale height $h$
between 0 and 0.07 $R_{\odot}$.
The model with $h \rightarrow 0$ is equivalent to the standard
equatorial model with no active-field enhancement.
Equation (\ref{eq:Bactive}) takes the greater of the two magnetic
field quantities, which tends to retain the original $B_E$ in the
photosphere, low chromosphere, and outer corona and wind, but
produces the active-region enhancement in the upper chromosphere
and low corona (see Figure 5).
The smallest values of the scale height ($h < 0.008 \, R_{\odot}$)
do not produce any enhancement in $B_{0}$ because the quantity
$B_{A} e^{-z/h}$ is rapidly decreasing and it never exceeds $B_E$.
For larger values of $h$, the magnetic field strength at the
coronal base is enhanced and thus the Wang-Sheeley expansion
factor $f_{\rm ss}$ is enhanced as well.
The maximum value of $h = 0.07 \, R_{\odot}$ (i.e., 49 Mm)
corresponds to $f_{\rm ss} = 41$.
Even though we used the exact values of $f_{\rm ss}$ computed
for each model in the plots below, we found the following
approximate fit to be a useful illustration of how $f_{\rm ss}$
depends on $h$:
\begin{equation}
  f_{\rm ss} \, \approx \, \left\{
  \begin{array}{ll}
    9.1 \,\, , & h < 0.02 \, R_{\odot} \\
    9.1 + |3026 (h - 0.02)|^{0.7} \,\, , &
      h \geq 0.02 \, R_{\odot} \\
  \end{array} \right.
\end{equation}
where $h$ is in units of $R_{\odot}$.
The threshold value of 0.02 $R_{\odot}$ corresponds to the
point at which $B_{A} e^{-z/h}$ first produces a significant
enhancement at $z_{\rm base} = 0.04 \, R_{\odot}$.

ZEPHYR models were computed for active region field enhancements
with $h = 0.015$, 0.02, 0.024, 0.03, 0.035, 0.04, 0.045, 0.05,
0.06, and 0.07 $R_{\odot}$.
The active region models exhibited lower terminal speeds and
higher mass loss rates than the standard equatorial model.
As $h$ was increased from 0 to 0.07 $R_{\odot}$, $u_{\infty}$
decreased from 500 to about 390 km/s.
The mass loss rate increased from $2.9 \times 10^{-14}$ to
$3.5 \times 10^{-14}$ $M_{\odot} \, \mbox{yr}^{-1}$,
thus increasing the proton number density at 1 AU from
6.4 cm$^{-3}$ to about 10 cm$^{-3}$.
Figure 13 shows how $u_{\infty}$ decreases as
$f_{\rm ss}$ increases for these models.
There is reasonable agreement with the empirical
Wang \& Sheeley (1990) relationship, and it is important to
reiterate that all models in Figure 13 were produced by
fixing the basal parameters 
($v_{\perp\odot}$, $L_{\perp\odot}$, $F_{S \odot}$) at
values appropriate for the polar coronal hole and varying only
the magnetic field above the photosphere in order to produce
the fast-to-slow wind transitions.

Figures 15 and 16 illustrate the computed O$^{7+}$ to O$^{6+}$
freezing-in ratios and Fe/O (FIP fractionation) ratios for the
active region models.
These models mirror the observed trends for there to be larger
ratios for both quantities at lower wind speeds.
The strengthening of the FIP ratio, from $\sim$1.5 to 4 as the
wind speed decreases from 500 to 400 km s$^{-1}$, is
especially comparable to the observed increase toward the
lowest speeds.

\begin{figure}
\epsscale{1.00}
\plotone{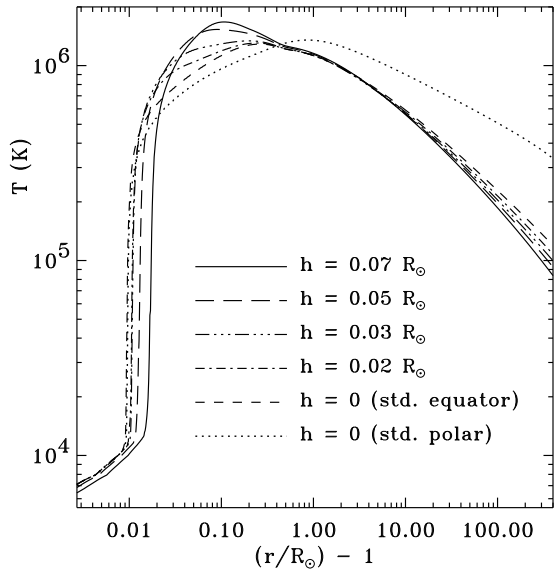}
\caption{
Radial dependence of temperature $T$ for a selection of ZEPHYR
models, highlighting the active-region extensions to the
equatorial $B_{0}(r)$ discussed in {\S}~8.5 (see captions).
For comparison the standard-model polar ({\em dotted line})
and equatorial ({\em dashed line}) model temperatures are also
shown.}
\end{figure}

Figure 17 compares the radial dependence of temperature for a
selection of active models to that for the standard polar
and equatorial models.
As $h$ increases from 0 to 0.07 $R_{\odot}$, the
peak coronal temperature increases from the equatorial
model's value of 1.29 MK up to 1.67 MK.
The heliocentric radius of this peak decreases from
1.26 to 1.11 $R_{\odot}$.
These trends appear to be a continuation of the variations
in coronal heating from the polar to the equatorial flux tubes.
As one progresses from pole to equator ($h=0$) to active region
($h > 0$), the models show a successively stronger decrease
in the Alfv\'{e}n speed with height, more reflection (and
damping) in the low corona, and thus less Alfv\'{e}n wave power
available to heights at and above the critical point.

The iterative convergence for the largest-$h$ active region
ZEPHYR models was slower and less robust than for the models
discussed in {\S\S}~8.1--8.4.
The models having $0.035 \leq h \leq 0.07 \, R_{\odot}$ were
run for 400 outer iterations in order to reach acceptable
convergence parameter values:
$\langle \delta E \rangle \approx 0.01$ to 0.02.
(Trial models with even larger values of $h \geq 0.09 \, R_{\odot}$
did not converge below $\langle \delta E \rangle \approx 0.04$
after 400 iterations and were not used.)
Unlike in Figure 4, where acceptable values of 0.01--0.02 were
retained in nearly every iteration (after 100 or so outer
iterations had passed), the active models exhibited these
values only sporadically.
Thus, the iteration with the lowest value of
$\langle \delta E \rangle$ did not necessarily represent the
``best'' single solution.
In order to evaluate the most likely time-steady plasma
parameters, we considered a collection of the
lowest-$\langle \delta E \rangle$ solutions sampled from the
iteration stream for each model run.
This was done by taking the distribution of values for a
given parameter, such as $u_{\infty}$, for all 400 outer
iterations and weighting it by a factor proportional to
$\langle \delta E \rangle^{-2}$ in order to deemphasize the
early iterations that are farthest from convergence.
Figure 13 shows both the mean and $\pm 1$ standard
deviation values for this weighted distribution of
well-converged terminal speed solutions.
As $h$ is increased, the standard deviation first decreases
from about 20 to 6 km s$^{-1}$, then increases again to
35 km s$^{-1}$.

It is unclear whether the nonconvergence of the large-$h$
active region models represents a failing of the numerical
technique or an intrinsic lack of a steady-state solution.
The coronal sources of slow solar wind are traditionally
viewed as more variable and filamentary than the source
regions of fast wind.
High-resolution coronagraphic observations of streamers
reveal an almost continual release of low-contrast density
inhomogeneities (Sheeley et al.\  1997; Wang et al.\  2000).
Multidimensional simulations of streamers have also exhibited
time-variable solutions (e.g., Suess et al.\  1996;
Endeve et al.\  2003, 2004).
For flux tubes in the vicinity of the boundary between open
and closed magnetic fields, the complex interplay between
heat conduction and transverse pressure balance can lead to
various instabilities that drive variability.
Certain open magnetic field configurations (possibly like
the one adopted in eq.~[\ref{eq:Bactive}]) may not be stable
when considered in the context of its neighbors in a
full-Sun model.

Other techniques for constructing active-region solar wind
models should be explored.
The specific functional form of the $B_0$ enhancement in the
low corona can be improved in order to agree with, e.g.,
force-free magnetic field reconstructions of active regions
(Schrijver et al.\  2006).
In addition, the photospheric field strength may also be
substantially higher in active regions than elsewhere.
It is unclear at first glance how the photospheric wave fluxes
would be modified in these regions, though.
For strong enough basal fields (i.e., in sunspots) convection
is quenched and the wave fluxes should be lower.
However, active regions exhibit a larger degree of magnetic
flux emergence (e.g., Abramenko et al.\  2006) which could
increase the frequency of impulsive horizontal ``jumps''
exhibited by flux-tube footpoints.
As summarized above, flux tubes are more more highly
concentrated in active-region intergranular lanes than
elsewhere.
The close presence of additional flux tubes may either
increase the basal wave amplitude in a given tube (via
``sympathetic'' motions from the neighbors) or
decrease it (because large horizontal displacements are
blocked by the neighbors).
Similarly, the correlation length $L_{\perp\odot}$ in these
regions may be either lower (because of the smaller domain
of motion when crowded) or higher (because the lane-filled
``ribbons'' may move together).
Comparisons between high-resolution observations of G-band
bright points in active regions (M\"{o}stl et al.\  2006) and
three-dimensional magnetoconvection simulations (e.g.,
Bushby \& Houghton 2005) are needed in order to clarify
these competing effects.

\section{Summary and Discussion}

The primary aim of this paper has been to construct
self-consistent models of chromospheric and coronal heating,
waves and turbulence, and wind acceleration in an open magnetic
flux tube rooted in the solar photosphere.
A key aspect of the ZEPHYR models presented above is that the
only true free parameters are:
(1) the properties of the waves injected at the base, and
(2) the background geometry and magnetic field strength along
the modeled flux tubes.
Everything else (e.g., the radial dependence of the rates of
chromospheric and coronal heating, the resulting temperature
structure of the atmosphere, and the solar wind speed and
mass flux) is an emergent property of the model and
not an ad~hoc input.
In the above sense, the ZEPHYR models are similar to those of
Suzuki \& Inutsuka (2005, 2006); the differences between the
two approaches involve mainly the specific physical processes
that are assumed to dissipate the Alfv\'{e}n waves and heat the
corona.

As discussed in {\S}~2, the ZEPHYR models presented here are
limited by being one-dimensional, time-independent, and one-fluid
solutions to the hydrodynamic conservation equations, with
heating that is derived only from a subset of all possible wave
and turbulent dissipation processes.
Even so, the results given in {\S}~8 show that a realistic
variation of asymptotic solar wind conditions can be produced
by varying only the background magnetic field geometry, as
predicted by Wang \& Sheeley (1990, 1991, 2003, 2006).
Specifically, our models show general agreement with some
well known empirical correlations:  i.e., a larger $f_{\rm ss}$
(expansion factor) gives rise to a slower and denser wind,
less intense Alfv\'{e}nic fluctuations at 1 AU, and larger
values of both the O$^{7+}$/O$^{6+}$ charge state ratio and
the FIP-sensitive Fe/O abundance ratio.
Satisfying these kinds of scalings are necessary but not
sufficient conditions for validating the idea that the
wind is driven by a combination of MHD turbulence
and non-WKB Alfv\'{e}n wave reflection.

Future work must involve more physical realism for the
models, expanded comparisons with existing observations, and
predictions of as-yet unobserved quantities that may be key
discriminators between competing theoretical models.
An important component of all three efforts will be to model
the divergent temperatures and flow speeds of protons,
electrons, and various heavy ion species in the extended
corona and heliosphere.
These mainly {\em collisionless} regions allow the kinetic
physics of wave dissipation to be probed to a level of detail
not possible in a collisionally coupled (i.e., essentially
one-fluid) plasma.
Even in a perfectly collisional plasma, there can be
{\em macroscopic} dynamical consequences depending on how
the energy is deposited into protons, electrons, and possibly
heavy ions as well.
For example, if all of the heat goes into electrons, there
can be substantially more downward conduction than in a
proton-heated model, which would affect the coronal temperature
distribution (Hansteen \& Leer 1995) and the stability of
helmet streamers (Endeve et al.\  2004).
Two-fluid effects also may affect the phenomenological form
of the MHD turbulent cascade assumed in the above models
(e.g., Galtier 2006).

The original motivation for this work was to understand the
preferential ion heating and acceleration in the corona
revealed by the UVCS (Ultraviolet Coronagraph Spectrometer)
instrument on {\em SOHO} (Kohl et al.\  1995, 1997, 2006).
If MHD turbulence gives rise to cyclotron-resonant Alfv\'{e}n
waves---which are likely to be responsible for the observed ion
properties---then we must first have a large-scale description
of the energy flux injected into the turbulent cascade in
order to further model the end-products of that cascade.
We hope to link the macroscopic properties modeled by ZEPHYR
with the microscopic kinetic processes studied by, e.g.,
Leamon et al.\  (1999, 2000), Cranmer \& van Ballegooijen (2003),
Voitenko \& Goossens (2003, 2004), Gary \& Nishimura (2004),
Dmitruk et al.\  (2004), Chandran (2005),
Gary et al.\  (2006), and Markovskii et al.\  (2006).

In addition to increasing the realism (and complexity) of the
models, attempts should be made to extract the dominant
physical processes of these models and create simpler
``scaling laws'' that can be used to estimate the wind
conditions for varying magnetic field geometries and
photospheric wave properties (see also
Leer et al.\  1982; Schwadron \& McComas 2003; Suzuki 2006).
An obstacle to accomplishing this for the present ZEPHYR
models is that the Alfv\'{e}n wave reflection is nontrivially
coupled to the plasma state via the Alfv\'{e}n speed $V_A$ and
solar wind speed $u$ (and their gradients).
The radial dependence of these quantities cannot be derived
a~priori from just the background magnetic field $B_{0}(r)$.
Approximations for $Z_{-}$ and $Z_{+}$ such as those given by
Dmitruk et al.\  (2002) should be tested further to evaluate
ways of estimating the amount of reflection without having to
solve the full non-WKB equations.

We also intend to use the methodology developed here to model
the winds of other kinds of stars on the cool side of the
Hertzsprung-Russell (H-R) diagram.
Similar ideas are beginning to be applied to younger and older
stars with solar-type winds (e.g., Airapetian et al.\  2000;
Falceta-Gon\c{c}alves et al.\  2006; Suzuki 2007;
Holzwarth \& Jardine 2007).
A benefit of computing the Alfv\'{e}n wave evolution with a
physically motivated damping rate (like eq.~[\ref{eq:dmit}])
is that the artificial ``damping lengths'' used in the past
are no longer needed.
Many cool stars that exhibit observational evidence for winds
have lower surface gravities than the Sun, stronger X-ray
emission (i.e., more coronal heating), and much larger mass
loss rates.
These properties are likely to be causally linked, but no
comprehensive and predictive models yet exist.

\begin{figure}
\epsscale{1.00}
\plotone{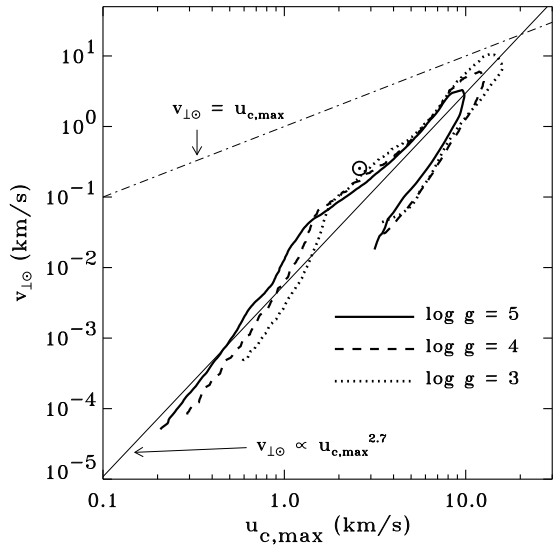}
\caption{
Photospheric transverse velocity amplitude $v_{\perp\odot}$
estimated from turbulent convection models of
Musielak et al.\  (2000) and Musielak \& Ulmschneider (2002),
plotted against the maximum convective velocity $u_{\rm c, max}$
from below the photosphere.
Models are for Population I stars with $T_{\rm eff}$ ranging from
2000 to 10000 K and $\log g = 5$ ({\em solid line}),
4 ({\em dashed line}), and 3 ({\em dotted line}).
The solar values used in this paper are denoted by the Sun
symbol ($\odot$).}
\end{figure}

In order to model cool-star winds with ZEPHYR we must have
values for the input parameters that drive the atmospheric
heating and wind acceleration.
For example, we must be able to predict the properties of
waves at the photospheric lower boundary for a given star.
Amplitudes can be estimated from existing models of
flux-tube wave generation from turbulent convection.
Figure 18 shows an example of how transverse kink-mode wave
amplitudes in stellar photospheres (for which we retain the
symbol $v_{\perp\odot}$) can be estimated from the peak value
of the subphotospheric convective velocity $u_{c, {\rm max}}$.
Musielak et al.\  (2000) computed a series of stellar interior
models with a range of effective temperatures (2000 to 10000 K)
and surface gravities ($\log g = 3$, 4, and 5, with $g$ in
units of cm s$^{-2}$) and presented the maximum values of
the ratio $u_{c, {\rm max}} / c_{s}$ in their Figure 1.
Also, Musielak \& Ulmschneider (2002) computed the photospheric
transverse wave fluxes for the same grid of stellar models.
We rescaled both quantities into velocity units using a
sound speed consistent with $T_{\rm eff}$, a magnetic field
strength consistent with the degree of pressure equipartition
assumed by Musielak et al., and a photospheric density
computed from the condition $\kappa_{\rm R} \rho H = 1$,
where $H$ is the photospheric scale height (proportional to
$T_{\rm eff}/g$) and $\kappa_{\rm R}$ is the Rosseland mean
opacity interpolated from the same tables used in other parts
of the ZEPHYR code (see also Cranmer 2005b).
Note that although the two velocities span many orders of
magnitude they ``collapse'' into something close to a
one-to-one power-law relation
(with $v_{\perp\odot} \propto u_{c, {\rm max}}^{2.7}$),
and that the solar value used in {\S}~8 coincides reasonably
closely with the model curves.

Another key input to cool-star wind models is an adequate
description of the large-scale magnetic field.
Observational determinations of the field geometries of
rapidly rotating young stars have been made possible via
Zeeman-Doppler imaging (e.g., Donati et al.\  1990, 2003),
time-resolved X-ray spectroscopy (Hussain et al.\  2005),
and combinations of these and related techniques.
However, not enough is known about how the magnetic flux
tubes break up---and penetrate the photosphere---on spatial
scales relevant to convective granulation and supergranulation.
The relationship between the star-averaged mean field
strength and the stronger fields inferred for thin flux
tubes is not yet well understood (e.g., Saar 2001).
Ongoing improvements in the observations make possible more
comprehensive and predictive models.
We anticipate that phenomenological approaches like those
described in this paper can contribute to substantial progress
in our understanding of magnetic activity, coronal heating,
and wind mass loss in a wide range of stars.

\acknowledgments

The authors would like to thank
Wolfgang Kalkofen, Jonathan Slavin, Eugene Avrett, Rudy Loeser,
and Takeru Suzuki for valuable discussions.
This work was supported by the National Aeronautics and
Space Administration (NASA) under grants {NNG\-04\-GE77G,}
{NNX\-06\-AG95G,} and NAG5-11913 to the
Smithsonian Astrophysical Observatory.
This research made extensive use of NASA's Astrophysics
Data System.

\vspace*{1.00in}
\begin{figure}[h]
\epsscale{1.13}
\figurenum{6}
\plotone{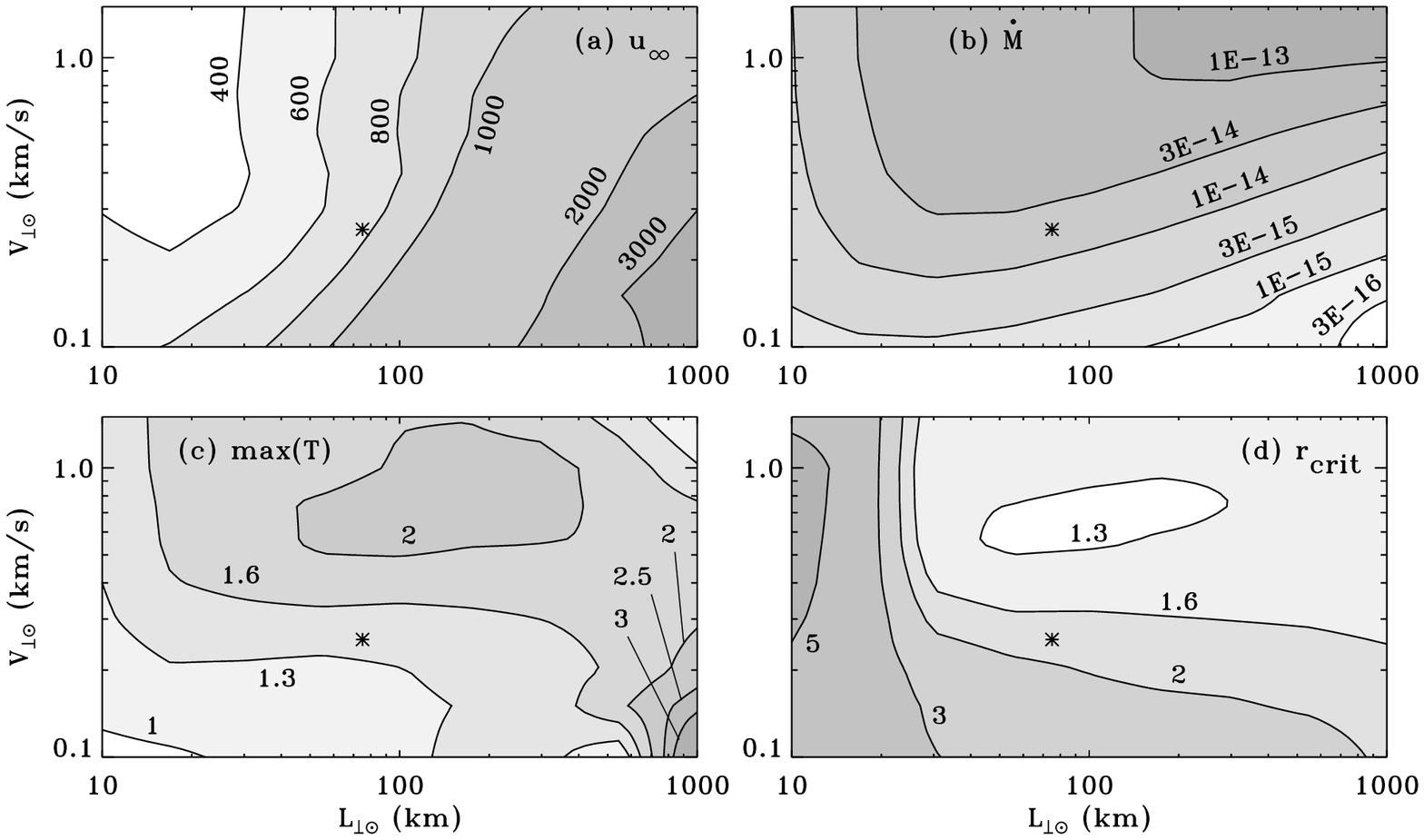}
\caption{
Contour plots of solar wind quantities resulting from varying the
coronal heating parameters $v_{\perp\odot}$ and $L_{\perp\odot}$:
({\em{a}}) terminal wind speed $u_{\infty}$ in units of km s$^{-1}$,
({\em{b}}) mass loss rate $\dot{M}$ in units of $M_{\odot}$ yr$^{-1}$,
({\em{c}}) maximum coronal temperature in units of MK,
({\em{d}}) heliocentric critical radius in units of $R_{\odot}$.
Also shown in each panel are the parameters chosen for the
model of fast wind from a polar coronal hole discussed in
{\S}~8.3 ({\em stars}).}
\end{figure}

\end{document}